\pdfoutput=1
\documentclass[12pt,onecolumn,oneside,letterpaper,peerreview]{IEEEtran}

\usepackage{cite}
\usepackage{graphicx}
\usepackage{amsmath}
\usepackage{times}
\usepackage{latexsym}
\usepackage{bm}
\usepackage{amssymb}
\usepackage[center]{caption2}
\usepackage{array}
\usepackage{fancyhdr}
\usepackage{cite,graphicx,amsmath,amssymb}
\usepackage{psfrag}
\usepackage{multirow}
\usepackage{longtable}

\ifCLASSINFOpdf

\else

\fi

\usepackage[dvipsnames,usenames]{color}
\usepackage[usenames,dvipsnames]{color}

\newtheorem{example}{Example}

\makeatother

\begin{document}
\title{A Survey on MIMO Transmission with Discrete Input Signals:
Technical Challenges, Advances, and Future Trends}

\author{Yongpeng Wu,~\IEEEmembership{Senior Member,~IEEE,}
Chengshan Xiao,~\IEEEmembership{Fellow,~IEEE,} \\ Zhi Ding,~\IEEEmembership{Fellow,~IEEE,}
 Xiqi Gao,~\IEEEmembership{Fellow,~IEEE,} and Shi Jin,~\IEEEmembership{Member,~IEEE}

\thanks{Y. Wu is with Institute for Communications Engineering,  Technical University of Munich,
Theresienstrasse 90, D-80333 Munich, Germany (Email:yongpeng.wu2016@gmail.com).}

\thanks{C. Xiao is with Department of Electrical and Computer Engineering,
Missouri University of Science and Technology, Rolla, MO 65409, USA. (Email: $\textmd{xiaoc}$@mst.edu).}

\thanks{Z. Ding is with Department of Electrical and Computer Engineering, University of California, Davis,
California 95616, USA. (Email: $\textmd{zding}$@ucdavis.edu).}

\thanks{X. Gao and S. Jin are with the National Mobile Communications Research Laboratory,
Southeast University, Nanjing, 210096, P. R. China. (Emails: xqgao@seu.edu.cn; jinshi@seu.edu.cn).}

}

\maketitle

\begin{abstract}
Multiple antennas have been exploited for
spatial multiplexing and diversity transmission in a wide range of
communication applications. However, most of the advances in the design of high speed
wireless multiple-input multiple output (MIMO) systems are based on information-theoretic principles
that demonstrate how to efficiently transmit signals conforming to Gaussian distribution.
Although the Gaussian signal is capacity-achieving,
signals conforming to discrete constellations are transmitted
in practical communication systems. The capacity-achieving design based on the Gaussian input signal can be quite suboptimal for practical MIMO
systems with discrete constellation input signals. As a result,
this paper is motivated to provide a comprehensive
overview on MIMO transmission design with discrete input  signals.
We first summarize the existing fundamental results for MIMO systems with
discrete input  signals. Then, focusing on
the basic point-to-point MIMO systems,
we examine transmission schemes based on
three most important
criteria for communication systems:  the mutual information driven designs,
the mean square error driven designs, and the diversity driven designs.
Particularly, a unified framework which designs low complexity transmission schemes applicable to massive MIMO systems in upcoming
5G wireless networks is provided in the first time.
Moreover, adaptive transmission designs  which switch
among these criteria based on the channel conditions to formulate the best transmission strategy
are discussed. Then, we provide a survey of the transmission designs with discrete input  signals for
multiuser MIMO scenarios, including MIMO uplink transmission, MIMO downlink transmission,
MIMO interference channel, and MIMO wiretap channel. Additionally, we discuss  the
transmission designs with discrete input  signals for other systems using MIMO technology.
Finally, technical challenges which remain unresolved at the time of writing are
summarized and the future trends of transmission designs with discrete input  signals
are addressed.

\end{abstract}

\section*{Glossary}
\begin{longtable}{ll}
3GPP &  3rd Generation Partnership Project \\
5G  &  fifth Generation  \\
ADC & analog-to-digital converter \\
AMC & adaptive modulation and coding \\
AF & amplify-and-forward \\
BER & bit error rate \\
BICM & bit-interleaved coded modulation  \\
BOSTBC & block-orthogonal property of space time block coding \\
BC &  broadcast channel \\
CQI & Channel Quality Indicator \\
CSI & channel state information \\
CSIT & channel state information at the transmitter \\
CDMA &  code-division multiple access systems  \\
DFE & decision-feedback equalization \\
DF & decode-and-forward \\
EPI & entropy power inequality  \\
FER & frame error rate \\
GSVD & generalized singular value decomposition \\
H-ARQ & hybrid automatic repeat request  \\
IPCSIR & imperfect channel state information at the receiver \\
IPCSIT & imperfect channel state information at the transmitter \\
KKT & Karush-Kuhn-Tucker \\
LCFE & linear complex-field encoder \\
LOS &  line-of-sight  \\
LTE & long term evolution  \\
LDPC & low density parity check codes \\
ML & maximum likelihood  \\
MRC & maximum ratio combining \\
MRT & maximum ratio transmission \\
MSE & mean squared error \\
MMSE & minimum mean squared error \\
MB & multichannel-beamforming \\
MAC & multiple access channel \\
MIDO & multiple-input double-output \\
MIMO & multiple-input multiple-output \\
MIMOME & multiple input multiple output multiple-antenna eavesdropper \\
MISO & multiple input single output \\
MISOSE & multiple input single output single-antenna eavesdropper \\
NVD & non-vanishing determinants  \\
OFDM & orthogonal frequency division multiplexing \\
OFDMA & orthogonal frequency division multiplexing access \\
OSTBC & orthogonal space-time block code \\
PER & packet error rate \\
PEP & pairwise error probability \\
PGP & Per-Group Precoding \\
PCSIT & perfect channel state information at the transmitter \\
PCSIR & perfect channel state information at the receiver \\
PSK & phase-shift keying \\
pdf & probability distribution function \\
PAM & pulse amplitude modulation \\
QAM & quadrature amplitude modulation \\
QOSTBC & quasi-orthogonal space-time block code \\
SNR & signal-to-noise ratio \\
SISO & single-input single-output \\
SVD & singular value decomposition \\
STBC & space-time block code \\
ST-LCFE & space time linear complex-field encoder \\
ST-LCP & space time linear constellation precoding \\
STTC & space time trellis codes \\
SDMA & spatial division multiple access \\
SCSI & statistical channel state information \\
SCSIT & statistical channel state information at the transmitter \\
SCSIR & statistical channel state information at the receiver \\
SER & symbol error rate \\
TAST & threaded algebraic space time \\
THP & Tomlinson-Harashima precoding \\
V-BLAST & vertical bell laboratories layer space-time \\
WSR & weighted sum-rate  \\
ZF & zero-forcing
\end{longtable}

\newpage

\section{Introduction}
The emergence of numerous smart mobile devices such as smart phones and wireless modems has led to an
exponentially increasing demand for wireless data services.
Accordingly, a large amount of effort has been invested on improving the spectral efficiency
and data throughput of wireless communication systems. In particular, multiple-input multiple-output (MIMO)
technology has been shown to be a promising means to provide multiplexing gains and/or diversity gains, leading
to improved performance. By increasing the number of antennas at the base stations  and mobile terminals,
communication systems can achieve better performance in terms of both system
capacity and/or link reliability \cite{Telatar1999,Foschini1998WPC,Zheng2003TIT}.
So far, MIMO technology has been adopted by third generation partnership project (3GPP) long-term evolution (LTE),
IEEE 802.16e, and other wireless communication standards.

\subsection{Future Wireless Communication Networks}
The unprecedented growth in the number of mobile data and connected machines ever-fast approaches
limits of 4G technologies to address this enormous data demand.  For example,
the mobile data traffic is expected to grow to 24.3 Exabytes per month by 2019 \cite{Cisco},
while the number of connected Internet of Things (IoT) is estimated to reach 50 Billion by 2020 \cite{UMTS}.
Also, emerging new services such as Ultra-High-Definition multimedia streaming and
cloud computing, storage, and retrieval require
the higher cell capacity/end-user data rate and extremely low latency, respectively.
It will become evident soon that current 4G technologies
will be stretched to near breaking points and
still can not meet the quality of
experience necessary for these emerging demands.
Therefore, the development of the fifth generation (5G)
wireless communication technologies is a priority issue currently for deploying future wireless communication
networks \cite{Huawei,Ericsson}.

\begin{figure*}[!ht]
\centering
\includegraphics[width=0.6\textwidth]{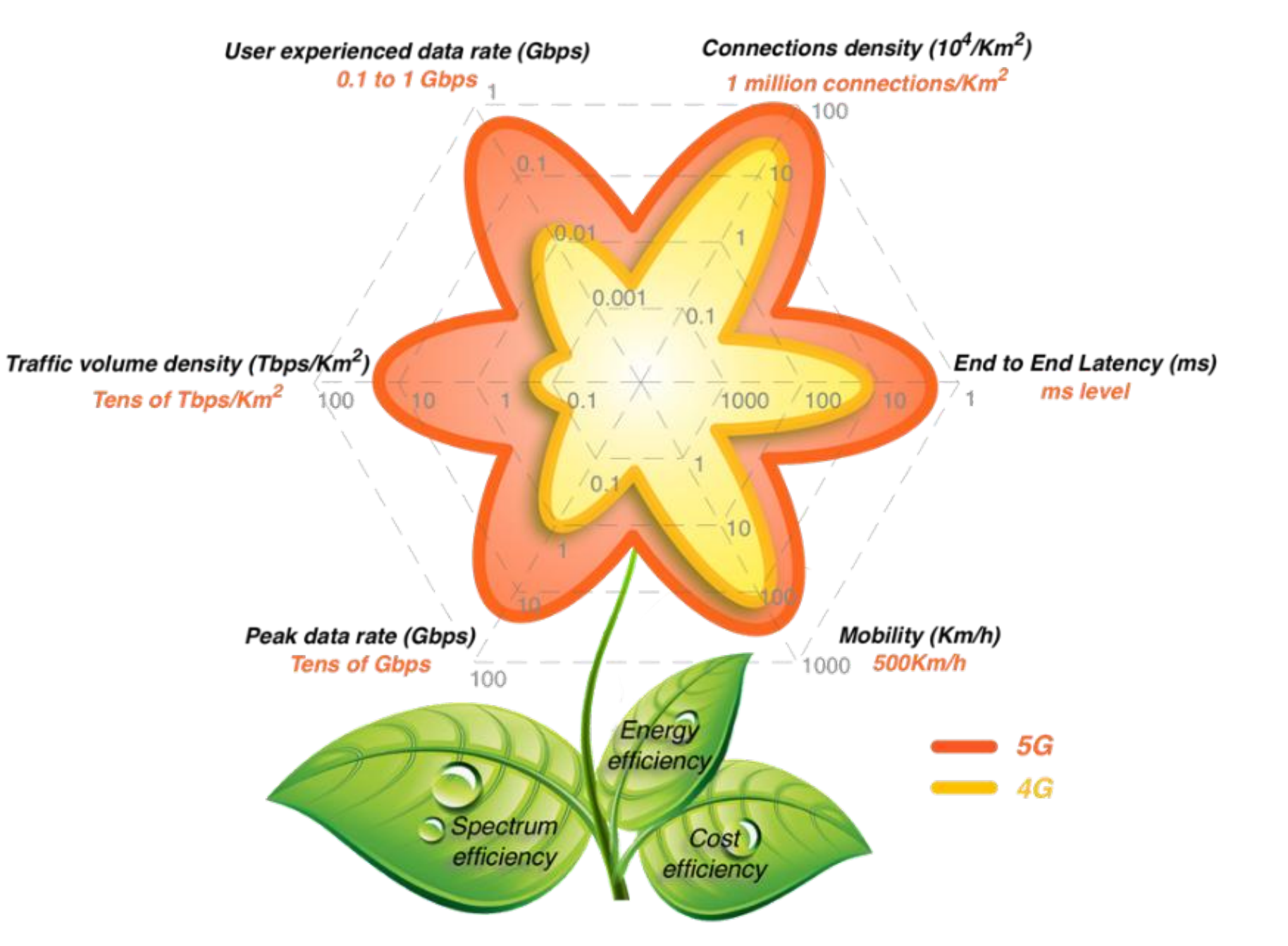}
\caption {\space\space Fundamental requirements of 5G: The 5G flower in \cite{IMT-2020}.}
\label{5G_Flower}
\end{figure*}

As illustrated in Figure \ref{5G_Flower}, the fundamental requirements for building 5G wireless networks are
clear. From system performance point of view,
5G wireless networks must support massive capacity (tens of Gbps peak data rate)
and massive connectivity (1 million /$K m^2$ connection density and tens of Tbps/$K m^2$ traffic volume density),
extremely diverging requirements for different user services
(user experienced data rate from 0.1 to 1 Gbps),
high mobility (user mobility up to 500 $Km/h$) and low latency communication (1 ms level end to end latency).
From transmission efficiency point of view, 5G wireless networks must simultaneously
realize spectrum efficiency, energy efficiency, and cost efficiency.
The evolution towards 5G wireless communication will be a cornerstone
for realizing the future human-centric and connected machine-centric
wireless networks, which achieve near-instantaneous, zero distance connectivity
for people and connected machines.

To achieve these requirements, the 5G air interface needs to
incorporate three basic technologies: massive MIMO, millimeter wave,
and small cell \cite{Andrews2014JSAC}. For millimeter wave transmission,
the millimeter wave beamforming formulated by
large antenna arrays, including both analog beamforming and digital beamforming,
is required to combat the higher propagation losses \cite{Rohde}.
For small cell, cell shrinking also needs the deployment of large and conformal antenna arrays
whose cost are inversely with the infrastructure density \cite{Puglielli2016PIEEE}.
Therefore,  MIMO technology will continue to play an important role in upcoming 5G and future wireless networks.

\subsection{MIMO Transmission Design}

While the benefits of MIMO can be realized
when only the receiver has the communication channel knowledge,
the performance gains can be further improved if the transmitter also obtains the channel
knowledge and performs certain processing operations on the signal before transmission.
For example, MIMO transmission
design can achieve double capacity at
-5dB signal-to-noise ratio (SNR) and
1.5 b/s/Hz additional
capacity at 5 dB SNR for a four-input two-output system over
independent identically distributed (i.i.d.) Rayleigh flat-fading channels \cite{Vu2007SPM}.
These are normal SNR ranges for practical applications
such as LTE, WiFi, and WiMax. The gains of MIMO transmission design
in correlated fading channels are even more obvious \cite{Wu2014TSP,Wu2016TIT}.
For the massive MIMO technology in future wireless communication networks,
it is shown in \cite{Marzetta2010TWC} that with perfect channel knowledge, linear precoding at the transmitters and simple signal detection at the
receivers can approach the optimum performance when the number
of the antennas at the base station tends to infinite.
A two step joint spatial division and multiplexing transmission scheme
is further proposed to achieve massive MIMO gains with reduced channel state information at the transmitter (CSIT)
feedback \cite{Adhikary2013TIT}. Therefore, based on some forms of CSIT,  MIMO transmission design
has a great practical interest in wireless communication systems.

The typical MIMO transmission designs can be mainly classified into three categories \cite{Xiao2011TSP}: (i)
designs based on mutual information; (ii) designs based on mean squared error (MSE);
(iii) diversity driven designs. The first category adopts mutual information
criterions from information theory such as ergodic or outage capacity
for the transmission optimization. The second category performs the transmission design
by optimizing the MSE related objective functions subject to various system constraints;
The third category maximizes the diversity of the communication system
based on pairwise error probability (PEP) analysis.

\subsection{MIMO Transmission Design with Discrete Input Signals}
\begin{figure*}[!ht]
\centering
\includegraphics[width=0.8\textwidth]{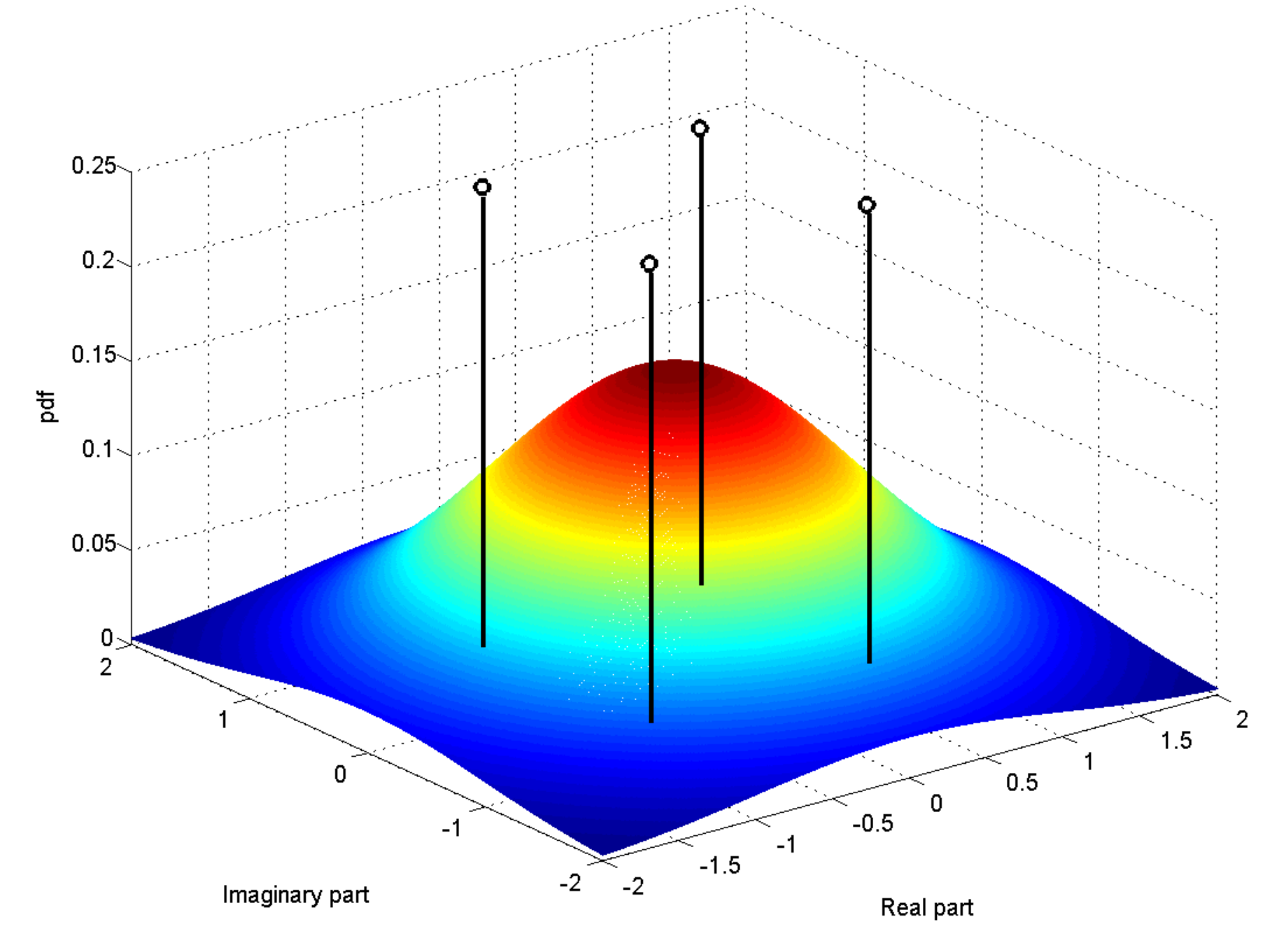}
\caption {\space\space The pdf of the Gaussian signal and the pmf of the QPSK signal.}
\label{pdf_finite_Gaussian}
\end{figure*}

From information theory point of view, the Gaussian distributed transmit signal achieves
the fundamental limit of MIMO communication \cite{Telatar1999}. However, the Gaussian transmit signal is
rarely used in practical communication systems. This is mainly because of two reasons: (i) The amplitude
of Gaussian transmit signal is unbounded, which may result in a substantial high power transmit
signal at the transmitter; (ii) The probability distribution  function (pdf) of Gaussian transmit signal is a continuous function,
which will significantly increase the detection complexity at the receiver. Therefore, the practical transmit signals are non-Gaussian input signals
taken from finite discrete
constellations, such as phase-shift keying (PSK), pulse amplitude
modulation (PAM), and/or quadrature amplitude
modulation (QAM). Figure \ref{pdf_finite_Gaussian} compares the pdf of the standard Gaussian distributed
signal and  the probability mass function (pmf) of the QPSK signal.   From Figure \ref{pdf_finite_Gaussian},
we observe that the QPSK signal is significantly different from the Gaussian signal.
Therefore, the communication based on discrete input signals
will normally be quite different from that of the Gaussian signal and needs specific transmission designs \cite{Lozano2006TIT,Perez-Cruz2010TIT,Xiao2011TSP,Wang2011TWC,Wu2012TWC_2}.

The main difficulties of MIMO transmission design with discrete input signals are threefold:
(i) The mutual information and the minimum
mean squared error (MMSE) of MIMO channels with finite discrete constellation signals usually lack explicit expressions. As a result,
the elegant water-filling type solutions for MIMO channels with Gaussian input signal
does not exist anymore.  Sophisticated numerical algorithms should  be designed to search for both
the power allocation matrix and the eigenvector matrix. (ii) The complexity of MIMO transmission
design with finite discrete constellation signals normally grows exponentially with the number of antenna dimension
and the number of the users, which often makes the implementation of the design prohibitive. (iii) Depending on
different channel conditions,  adaptive transmission is needed in practical systems which
determine the best transmission policy among various criterions such as  mutual information, MSE, diversity, etc.
This process is challenging since it involves a selection of a large number of parameters.

\subsection{The Contributions of This Paper}
\begin{figure*}[!ht]
\centering
\includegraphics[width=0.8\textwidth]{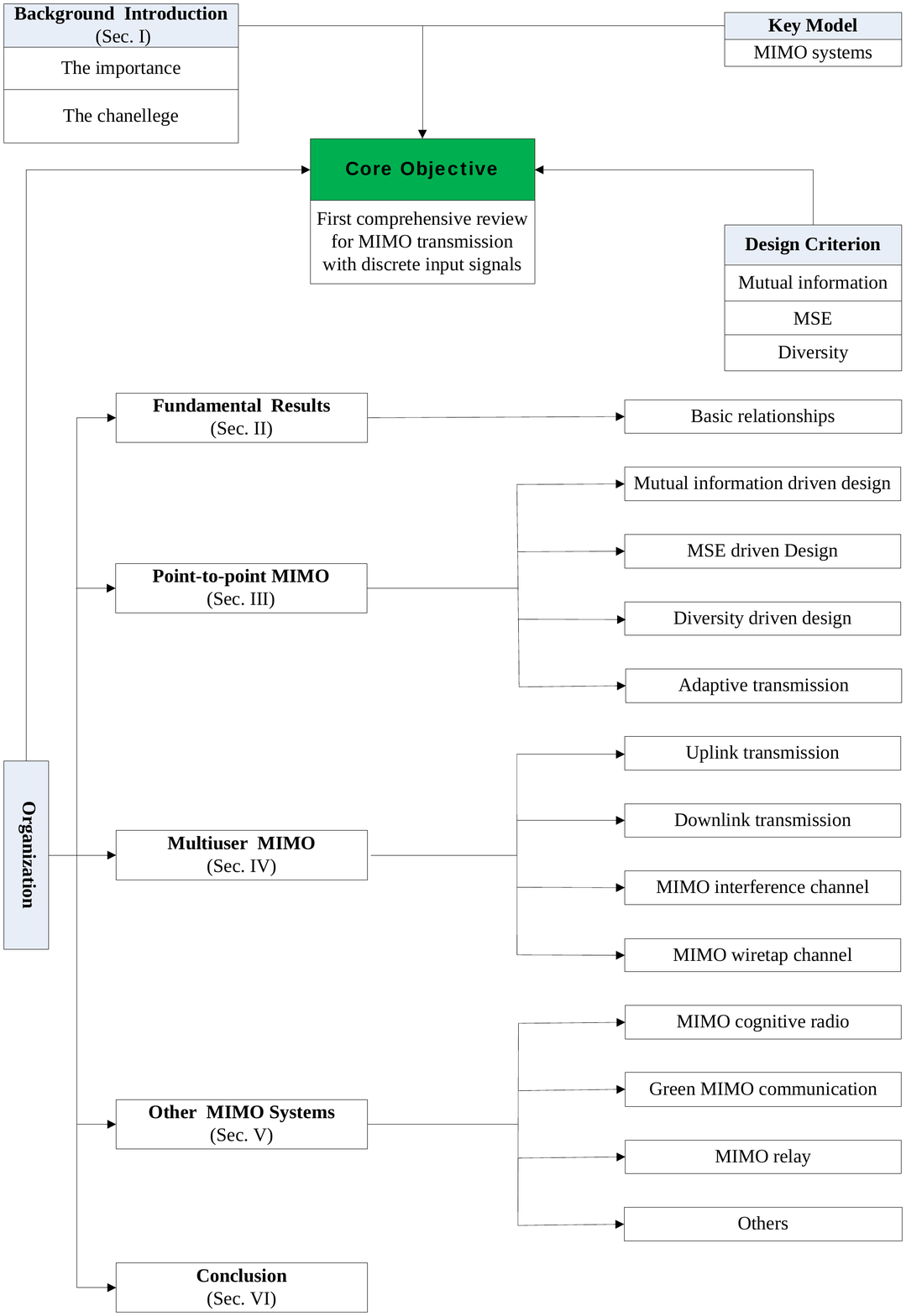}
\caption {\space\space The overall outline of this paper}
\label{fig:outline}
\end{figure*}

In this paper, a comprehensive review of MIMO transmission design with discrete input signals
is presented and summarized in the first time for various design criterions. We first introduce
the fundamental research results for discrete input signals, e.g.,
the mutual information and the MMSE relationships \cite{Guo2005TIT,Palomar2006TIT,Payar2009TIT},
the properties of the MMSE function\cite{Guo2011TIT,Wu2012TIT}, the definition of MMSE dimension \cite{Wu2012TIT_2}, etc.
Then, focusing on the basic point-to-point MIMO systems, we
provide in details the transmission designs with discrete input signals based on the criterions of mutual information (e.g., \cite{Lozano2008TCOM,Xiao2011TSP,Mohammed2011TIT,zeng2012linear,Ketseoglou2015TWC}), MSE (e.g., \cite{Palomar2003TSP,Serbetli2006TWC,Fischer2002}),
and diversity  (e.g., \cite{Alamouti2006JASC,Tarokh1999TIT,Zhou2003TIT,Jafarkhani2005book,Liu2008TIT,Vrigneau2008IJSTSP}), respectively.
In particular, a unified framework for the low complexity designs which can be implemented in large-scale MIMO antenna arrays for 5G wireless networks
is provided in the first time.    In light of this, the practical adaptive transmission schemes which switch among these criterions
corresponding to the channel variations are summarized for both uncoded systems (e.g., \cite{Zhou2005TVT,Zhou2004TWC,Chen2013TVT,Kuang2012TCOM})
and coded systems (e.g., \cite{McKay2007TVT,Choi2008JSAC,Tan2008JSAC,Zhou2011TWC}), respectively.
The specific adaptive transmission schemes defined in 3GPP LTE, IEEE 802.16e, and  IEEE 802.11n standards
are further introduced.  Moving towards
multiuser MIMO systems, we introduce the designs based on the above-mentioned criterions for multiuser
MIMO uplink transmission (e.g.,\cite{Wang2011TWC,Zhang2010TSP,Jorswieck2003TSP,Zhang2005TCOM}), multiuser MIMO downlink transmission (e.g.,\cite{Wu2012TWC_2,Chen2007TSP,Shi2007TSP,Tsai2008TWC}),
MIMO interference channel (e.g.,\cite{Wu2013TCom,He2010TWC,Shi2012TWC,Xie2013TWC}),
and MIMO wiretap channel (e.g.,\cite{Wu2012TVT,Li2016TWC,Reboredo2013TSP}). Finally, we overview transmission designs  with  discrete input signals
for other systems employing MIMO technology such as MIMO cognitive radio  systems (e.g, \cite{Zeng2012JSAC,Zhang2008TWC_Dec,Gharavol2011TWC}), green MIMO communication systems (e.g., \cite{Gregori2013TCOM,Xia2013TIT}),
and MIMO relay systems (e.g, \cite{Zeng2012TWC,Ding2007TSP,Mo2009TWC_2,Abualhaol2008}).
The overall outline of this paper is shown in Figure \ref{fig:outline}.

\section{Fundamental Results for Discrete Input Signals}
In this section, we briefly review the existing fundamental results
for discrete input signals.
To design transmission schemes for MIMO systems with discrete input signals
has to rely on these fundamental results.
\subsection{Fundamental Link between Information Theory and Estimation Theory}
The relationship between information theory and estimation theory is a classic
topic, which dates back to 50 years ago for a relationship
between the differential entropy of the Fisher information and the mean square score function in
\cite{Stam1959IC}. This relationship is called De Bruijn's identity.
The representation of mutual information as a function of causal
filtering error was given in \cite{Duncan1970TIT} as Duncan's theorem.
Other early connections between fundamental quantities in information
theory and estimation theory can be found in \cite{Ziv1969TIT,Kadota1971TIT,Seidler1971TIT,Bucy1979IC,Barron1986AP} and reference therein.
\subsubsection{Gaussian channel model}
Recently, D. Guo \textit{et al.} establish a fundamental relationship between the mutual information
and the MMSE \cite{Guo2005TIT}, which applies for discrete-time
and continuous-time, scalar and vector channels with additive Gaussian noise.
In particular, for the real-value linear vector
Gaussian channel $\mathbf{y} = \sqrt{snr} \mathbf{Hx} + \mathbf{n}$,
D. Guo \textit{et al.} find the following relationship regardless of the input distribution $\mathbf{x}$
\begin{align}\label{eq:i-mmse}
\frac{d}{{dsnr}}I\left( {{\bf{x}};\sqrt {snr} {\bf{Hx}} + {\bf{n}}} \right) = \frac{1}{2} {\rm mmse}(snr)
\end{align}
where the mutual information is in nats and
\begin{align}\label{eq:mmse}
{\rm mmse}(snr) = E\left[ {{{\left\| {{\bf{Hx}} - E\left[ {{\bf{Hx}}\left| {\sqrt {snr} {\bf{Hx}} + {\bf{n}}} \right.} \right]} \right\|}^2}} \right].
\end{align}
The right-hand side of Eq. (\ref{eq:mmse}) denotes the expectation of squared
Euclidean norm of the error corresponding to the best estimation of $\mathbf{Hx}$ upon
the observation $\mathbf{y}$ for a given SNR.

Eq. (\ref{eq:i-mmse}) is proved based on an incremental-channel method in \cite{Guo2005TIT}. The key idea of
this method is to model the change in the mutual information as the input-output mutual
information of an equivalent Gaussian channel with vanishing SNR. The mutual information
of this equivalent Gaussian channel is essentially linear with the estimation error.
Therefore, it is possible to establish a relationship between the incremental of the mutual information and
the MMSE.

For the continuous-time real-value scalar Gaussian channel $R_t = \sqrt{snr} X_t + N_t$, $t \in [0,T]$,
where $X_t$ is the input process and $N_t$ is a white Gaussian noise with a flat double-sided power
spectrum density of unit height, D. Guo \textit{et al.} establish another
fundamental relationship between the causal MMSE and the noncausal MMSE with any input process
\begin{align}\label{eq:cmmse-mmse}
{\rm{cmmse}}\left( {snr} \right) = \frac{1}{{snr}}\int_0^{snr} {mmse\left( \gamma  \right)d\gamma }
\end{align}
where
\begin{align}
{\rm{cmmse}}\left( {snr} \right) &= \frac{1}{T}\int_0^T {{\rm{cmmse}}\left( {t,snr} \right)dt} \\
{\rm{cmmse}}\left( {t,snr} \right) & = E\left[ {{{\left( {{X_t} - E\left[ {{X_t}\left| {Y_0^t;snr} \right.} \right]} \right)}^2}} \right] \\
{\rm{mmse}}\left( {snr} \right) &= \frac{1}{T}\int_0^T {{\rm{mmse}}\left( {t,snr} \right)dt} \\
{\rm{mmse}}\left( {t,snr} \right) & = E\left[ {{{\left( {{X_t} - E\left[ {{X_t}\left| {Y_0^T;snr} \right.} \right]} \right)}^2}} \right].
\end{align}
Eq. (\ref{eq:cmmse-mmse}) is proved in \cite{Guo2005TIT} by comparing Eq. (\ref{eq:i-mmse}) to Duncan's Theorem.
The fundamental relationships in Eq. (\ref{eq:i-mmse}) and (\ref{eq:cmmse-mmse})
and their generalizations in \cite{Guo2005TIT},  are referred to the I-MMSE relationships. These I-MMSE relationships
can be applied to provide a simple proof of Shannon's entropy power
inequality (EPI) \cite{Verdu2006}. The main idea is to use \cite[Eq. (182)]{Guo2005TIT}
to transform the EPI into the equivalent  MMSE inequality, which can be proved
much more easily. Similar approaches have been applied to prove the Costa's EPI
and the generalized EPI in \cite{Guo2006} and
a new inequality for independent random variables in \cite{Tulino2006TIT}.

I-MMSE relationships in \cite{Guo2005TIT} can be generalized.
For the complex-value linear vector Gaussian channel with arbitrary signaling,
D. P. Palomar \textit{et al.} derive the gradient of the mutual
information with respect to the channel matrix and other arbitrary
parameters of the system through the chain rule \cite{Palomar2006TIT}. It is shown
in \cite{Palomar2006TIT} that the partial derivative of the mutual information
with respect to the channel matrix is equal to the product of the channel matrix
and the error covariance matrix of the optimal estimation of the input given the output.
Moreover, the Hessian of the mutual information and the entropy with respect to various
parameters of the system are derived in \cite{Payar2009TIT}. In a special case where the left singular
vectors of the precoder matrix coincide with the eigenvectors of the channel correlation matrix,
it is proved in \cite{Payar2009TIT} that the mutual information
is a concave function of the squared singular values of the precoder matrix. Another
generalization direction is to extend the I-MMSE relationships to the abstract
Wiener space for the scalar Gaussian channel in \cite{Zakai2005TIT} and the vector Gaussian channel in \cite{Wolf2007AAP}.

D. Guo \textit{et al.} investigate the properties of the MMSE as a function
of the SNR and the input distribution for a real-value scalar Gaussian channel in \cite{Guo2011TIT}.
It is shown in \cite{Guo2011TIT} that the MMSE is
concave in the input distribution for a given SNR
and infinitely differentiable for all SNRs for a given input distribution.
An important conclusion in \cite{Guo2011TIT} is that the MMSE curve of a Gaussian input and the
MMSE curve of a non-Gaussian input intersect at most once throughout
the entire SNR regime. The MMSE properties are  extended to a real-value
 diagonal MIMO Gaussian channel in \cite{Bustin2010}, where each eigenvalue of the MMSE matrix
 with a Gaussian input is proved to
 have at most a single crossing point with that of the MMSE matrix with a non-Gaussian input
 for all SNR values.

 More results for the mutual information and the MMSE  properties for the
 scalar Gaussian channel are obtained.  Based on the Lipschitz continuity of the MMSE
 and I-MMSE relationship, Y. Wu \textit{et al.} prove that the mutual information with input
 cardinality constraints converges to the usual Gaussian channel capacity as the constellation cardinality
 tends to infinity \cite{Wu2010}. Y. Wu \textit{et al.} further provide a family of input constellation generated
 from the roots of the Hermite polynomials, which achieves this convergence exponentially \cite{Wu2010_2}.
 N. Merhav \textit{et al.} introduce techniques in statistical physics to evaluate
 the mutual information and the MMSE in \cite{Merhav2010TIT}.  N. Merhav \textit{et al.} establish several
 important relationships between the mutual infomration, the MMSE, the differential entropy,
 and the statistical-mechanical variables. Some application examples in \cite{Merhav2010TIT}
 shows how to use the statistical physics tools to provide useful analysis for the MMSE.
 K. Venkat \textit{et al.} investigate the statistic distribution of the difference
 between the input-output information density and half the causal estimation error \cite{Venkat2012TIT}.
 G. Han  \textit{et al.} propose a new proof of the I-MMSE relationship
 by choosing a probability space independent of SNR to evaluate the mutual information \cite{Han2014}.
 Based on this approach,  G. Han  \textit{et al.}
 extend the I-MMSE relationship to both discrete and
 continuous-time Gaussian channels with feedback in \cite{Han2014}.

A random variable with distribution $P$ is at the receiver.
The receiver performs the MSE estimation of this random variable
under the assumption that its distribution is $Q$. This is referred to
the mismatch estimation. S. Verd\'{u} initializes a new connection between the relative entropy and the mismatched estimation
for a real-value scalar Gaussian channel \cite{Verdu2010TIT}. The key relationship revealed in \cite{Verdu2010TIT}
is as follows:
\begin{align} \label{eq:D-MMSE}
D\left(P \parallel Q\right) = \frac{1}{2} \int_0^{\infty} {\rm mse}_Q\left(\gamma\right) - {\rm mse}_P\left(\gamma\right) d\gamma
\end{align}
where $D\left(P \parallel Q\right)$ is the relative entropy.
${\rm mse}_Q\left(\gamma\right)$ and   ${\rm mse}_P\left(\gamma\right)$
are MMSEs obtained by an estimator which assumes the input signal is distributed
according to $Q$ and $P$ for a real-value scalar Gaussian channel, respectively.
Eq. (\ref{eq:D-MMSE}) is a generalization of Eq. (\ref{eq:i-mmse}) for the scalar Gaussian channel
 with a mismatch estimation. Similarly, using the mutual information and the relative entropy as a bridge,
T. Weissman generalizes Eq. (\ref{eq:cmmse-mmse}) into the case of mismatch estimation in \cite{Weissman2010TIT}.
The statistic properties of the difference
 between the mismatched and matched estimation losses
 are studied in \cite{Venkat2012TIT}. The relationship between relative entropy and mismatched estimation
 is extended to the real-value vector Gaussian channel in \cite{Chen2013}. For a more complicated
 case of mismatched estimation where the channel distribution is mismatched at the estimator instead
 of the input signal distribution, W. Huleihel  \textit{et al.} derive the mismatched MSE for the
 vector Gaussian channel using the statistical physics tools \cite{Huleihel2014TIT}.
A brief summation
of recent research results
of the I-MMSE relationships for Gaussian channel
is given in Table \ref{table:I-MMSE-Gaussian}.

\begin{table}
\centering
  \renewcommand{\multirowsetup}{\centering}
 \captionstyle{center}
  {
\caption{Recent research results of the I-MMSE relationships for Gaussian channel}
\label{table:I-MMSE-Gaussian}
\newsavebox{\tablebox}
\begin{lrbox}{\tablebox}
\begin{tabular}{|c|c|c|}
\hline
  Paper   & Model &   Main Contribution \\ \hline
\multirow{2}{*} {D. Guo \textit{et al.} \cite{Guo2005TIT}}   & Real-value MIMO &  Reveal a fundamental relationship between mutual information and MMSE    \\
 &  Discrete-time, Continuous-time &  Reveal a fundamental relationship between causal and noncausal MMSEs
\\ \hline
\multirow{2}{*} {D. P. Palomar \textit{et al.}\cite{Palomar2006TIT} } & Complex-value MIMO &  Derive the gradient of the mutual information  \\
 &  Discrete-time &  with arbitrary parameters of the systems
\\ \hline
\multirow{2}{*} {M. Payar\'{o} \textit{et al.} \cite{Payar2009TIT}}   & Complex-value MIMO &  Derive the Hessian of the mutual inforamtion and entropy  \\
 &  Discrete-time &  with arbitrary parameters of the systems
\\ \hline
M. Zakai \cite{Zakai2005TIT}   & Real-value scalar &  Extend I-MMSE relationship in \cite{Guo2005TIT} to the abstract Wiener space (scalar version) \\ \hline
E. M.-Wolf \textit{et al.} \cite{Wolf2007AAP}   & Real-value MIMO &  Extend I-MMSE relationship in \cite{Guo2005TIT} to the abstract Wiener space (vector version)\\ \hline
\multirow{2}{*} {D. Guo \textit{et al.} \cite{Guo2011TIT}}   & Real-value MIMO &  Obtain various properties of MMSE  \\
 &  Discrete-time &  with respect to SNR and input distribution
\\ \hline
\multirow{2}{*} {R. Bustin \textit{et al.} \cite{Bustin2010}}   & Real-value diagonal MIMO & \multirow{2}{*} {Obtain properties of the MMSE matrix} \\
 &  Discrete-time &
\\ \hline
\multirow{2}{*} {Y. Wu \textit{et al.} \cite{Wu2010}}   & Real-value scalar & Prove the mutual information with input
 cardinality constraints converges  to \\
 &  Discrete-time &  the Gaussian channel capacity as the constellation cardinality
 tends to infinity
\\ \hline
\multirow{2}{*} {Y. Wu \textit{et al.} \cite{Wu2010_2}}   & Real-value scalar &  \multirow{2}{*} {Prove the convergence speed in \cite{Wu2010} is exponential} \\
 &  Discrete-time &
 \\ \hline
\multirow{2}{*} {N. Merhav \textit{et al.} \cite{Merhav2010TIT}}   & Real-value scalar &   {Introduce techniques in statistical physics to } \\
 &  Discrete-time &  evaluate
 the mutual information and the MMSE
 \\ \hline
 \multirow{3}{*} {K. Venkat \textit{et al.}  \cite{Venkat2012TIT}}   & Real-value scalar, mismatched case &  \multirow{2}{*} {Derive the statistical properties of difference
 between the  } \\
 &  with/without feedback &  \multirow{2}{*} {input-output information density and half the causal estimation error}  \\
 &  Continuous-time,  Discrete-time &
\\ \hline
\multirow{2}{*} {G. Han \textit{et al.} \cite{Han2014}}   & Real-value scalar, feedback &  \multirow{2}{*}  { Extend the I-MMSE relationship to channels with feedback  } \\
 &  Discrete-time, Continuous-time  &
\\ \hline
\multirow{2}{*} {S. Verd\'{u} \textit{et al.} \cite{Verdu2010TIT}}   & Real-value scalar & {Initializes a new connection between the relative entropy} \\
 &  Discrete-time, mismatched case & and the mismatched estimation
\\ \hline
\multirow{2}{*} {T. Weissman \cite{Weissman2010TIT}}   & Real-value scalar & {Establish a relationship between causal MMSE} \\
 &  Continuous-time, mismatched case &  and non-causal MMSE in mismatched case
\\ \hline
\multirow{2}{*} {M. Chen \textit{et al.}  \cite{Chen2013}}   & Real-value MIMO &  \multirow{2}{*} {Extend the relationship in \cite{Verdu2010TIT} to MIMO} \\
 & Discrete-time, mismatched case &
\\ \hline
\multirow{2}{*} {W. Huleihel \textit{et al.}  \cite{Huleihel2014TIT}}   & Real-value MIMO &  Derive the MSE when the mismatched estimation  \\
 & Discrete-time, mismatched case & occurs for the channel distribution
\\ \hline
\end{tabular}
\end{lrbox}
\scalebox{0.9}{\usebox{\tablebox}}
}
\end{table}

\subsubsection{Other Channel Models}
The I-MMSE relationships are investigated in other channel models.
For the scalar Poisson channel, D. Guo \textit{et al.} prove
that the derivative of the input-output
mutual information with respect to the intensity of the additive dark current
equals to the expected error between the logarithm of the actual input and the logarithmic
of its conditional mean estimate (noncausal in case of continuous-time) \cite{Guo2008TIT}.
The similar relationship holds for the derivative of the mutual information
 with respect to input-scaling, by replacing the logarithmic function
 with $x \log x$ \cite{Guo2008TIT}. Moreover, R. Atar \textit{et al.} prove that
 the I-MMSE relationships hold for Gaussian channel \cite{Guo2005TIT,Verdu2010TIT,Weissman2010TIT},
 including discrete-time and continuous-time channel, matched and mismatched filters, also
 hold for scalar Poisson channel by replacing the squared error loss
 with the corresponding loss function \cite{Atar2012TIT}.  L. Wang  \textit{et al.}
 derive the gradient of mutual information
 with respect to the channel matrix for vector Poisson channel \cite{Wang2014TIT}.
 In addition,  L. Wang  \textit{et al.} define a Bregman matrix based on the Bergman divergence in \cite{Wang2014TIT}
 to formulate a unified framework for the gradient of mutual information
for both vector Gaussian and Poisson channels. The Bergman divergence
is also used to characterize the derivative of the relative entropy
and the mutual information for the scalar binomial and negative binomial
channels in \cite{Taborda2014TIT}. The I-MMSE relationship for a scalar channel
with exponential-distributed additive noise is established in \cite{Raginsky2009}.
For continuous-time scalar channel with additive Gaussian/Possion noise in
the presence of feedback, the relationships between the directed information
and the causal estimation error are established in \cite{Weissman2013TIT}.

Some work consider more general channel models.
A scalar channel with arbitrary additive
noise is studied in \cite{Guo2005}.
It is found that the increase in the mutual information
due to the improvement in the channel quality equals
to the correlation of two conditional mean estimates
associated with the input and the noise respectively.
D. P. Palomar \textit{et al.} represent the derivative of mutual
information in terms of the conditional input estimation given
the output for general channels (not necessarily additive noise) \cite{Palomar2007TIT}.
The general results in \cite{Palomar2007TIT} embrace most of popular channel models including
the binary symmetric channel, the binary erasure channel,
the discrete memoryless channel, the scalar/vector Gaussian channel,
an arbitrary additive-noise channel, and the Poisson channel.
For a scalar channel with arbitrary additive noise,
D. Guo obtain the derivative of relative entropy with respect
to the energy of perturbation, which can be expressed as a mean
squared difference of the score functions of the two distributions \cite{Guo2009}.
Furthermore, Y. Wu  \textit{et al.} analyze the properties of the MMSE as a function of
the input-output joint distribution \cite{Wu2012TIT}. It is proved in \cite{Wu2012TIT}
that the MMSE is a concave function of the input-output joint distribution.
 N. Merhav analyze the MMSE for both cases of  matched and mismatched estimations
by using the statistical-mechanical techniques \cite{Merhav2011TIT}. The main advantage
of the results in \cite{Merhav2011TIT} is that they apply for very general vector channels
with arbitrary joint input-output probability distributions. By exploiting the statistical-mechanical techniques,
N. Merhav also establish relationships between rate-distortion function and the MMSE for a general
scalar channel \cite{Merhav2011TIT_2}. Then, simple lower and upper bounds on the rate-distortion function can
be obtained based on the bounds of the MMSE.  A brief summation
of recent research results
of the I-MMSE relationships for non-Gaussian channels
is given in Table \ref{table:I-MMSE-non-Gaussian}.

\begin{table}
\centering
  \renewcommand{\multirowsetup}{\centering}
 \captionstyle{center}
  {
\caption{Recent research results of the I-MMSE relationships for non-Gaussian channels}
\label{table:I-MMSE-non-Gaussian}
\begin{lrbox}{\tablebox}
\begin{tabular}{|c|c|c|}
\hline
  Paper   & Model &   Main Contribution \\ \hline
\multirow{2}{*} {D. Guo \textit{et al.} \cite{Guo2008TIT}}   &  Scalar Poisson   & Obtain the derivative of the
mutual information with respect to the additive dark current   \\
 &  Discrete-time, Continuous-time &  Obtain the derivative of the
mutual information with respect to the input-scaling
\\ \hline
\multirow{2}{*} {R. Atar \textit{et al.} \cite{Atar2012TIT}}   &  Scalar Poisson, mismatched case & \multirow{2}{*} {Obtain I-MMSE relationship}   \\
 &  Discrete-time, Continuous-time &
\\ \hline
\multirow{2}{*} {L. Wang \textit{et al.} \cite{Wang2014TIT}}   &  MIMO Poisson   &  Obtain gradient of mutual information
 with      \\
 &  Discrete-time  & respect to the channel matrix
\\ \hline
\multirow{2}{*} {C. G. Taborda \textit{et al.} \cite{Taborda2014TIT}}   &  Scalar binomial and negative binomial     & Obtain derivative of the relative entropy
and the mutual information     \\
 &  Discrete-time  &  with  respect to scaling factor
\\ \hline
\multirow{2}{*} {M. Raginsky  \textit{et al.} \cite{Raginsky2009}}   &  Scalar, exponential noise    &  \multirow{2}{*} {Obtain the I-MMSE relationship}
   \\
 &  Discrete-time  &
\\ \hline
\multirow{2}{*} {T. Weissman \textit{et al.} \cite{Weissman2013TIT}}   &  Scalar Gaussian and  Poisson   & Obtain the relationships between the directed information
   \\
 &  Continuous-time  with feedback &   and the causal estimation error
channels
\\ \hline
\multirow{2}{*} {D. Guo \textit{et al.} \cite{Guo2005}}   &  Scalar, arbitrary additive noise  & \multirow{2}{*} { Obtain the I-MMSE relationship}
   \\
 &   Discrete-time &
\\ \hline
\multirow{2}{*} {D. P. Palomar \textit{et al.} \cite{Palomar2007TIT}}   &   General channel  & Obtain the derivative of mutual
information in terms of
   \\
 &   Discrete-time &  the conditional input estimation given
the output
\\ \hline
\multirow{2}{*} {D. Guo \textit{et al.} \cite{Guo2009}}   &   Scalar, arbitrary additive noise   &  Obtain derivative of relative entropy
 with respect  \\
 &   Discrete-time &
to the energy of perturbation
the output
\\ \hline
\multirow{2}{*} {Y. Wu \textit{et al.} \cite{Wu2012TIT}}   &   Scalar,  arbitrary additive noise   &  Analyze the properties of the MMSE  as a function  \\
 &   Discrete-time &
 of the input-output joint distribution
\\ \hline
\multirow{2}{*} {N. Merhav\textit{et al.} \cite{Merhav2011TIT}}   &   MIMO,  general channel   &   \multirow{2}{*} {Analyze the MMSE } \\
 &   Discrete-time, mismatched case &
\\ \hline
\multirow{2}{*} {N. Merhav\textit{et al.} \cite{Merhav2011TIT_2}}   &   Scalar,  general channel    &   Establish relationships between rate-distortion  \\
 &   Discrete-time & function and the MMSE
\\ \hline

\end{tabular}
\end{lrbox}
\scalebox{0.85}{\usebox{\tablebox}}
}
\end{table}

\subsection{Other Analytical Results with Discrete Input Signals}

\subsubsection{Asymptotic Results in Low and High SNR Regime}
Shannon's pioneering work \cite{Shannon1948BSTJ} have shown that the binary antipodal
inputs are as good as the Gaussian input in low SNR regime. Recently,
S. Verd\'{u} investigate the spectral efficiency in low SNR regime  for a general
class of signal inputs and channels \cite{Verdu2002TIT}. In \cite{Verdu2002TIT},
S. Verd\'{u} provides two important criterions to characterize
the spectral efficiency in low SNR regime: the \textit{minimum energy
per information bit} required for reliable communication and the
\textit{wideband slope} of the spectral efficiency. S. Verd\'{u} further
define that any input signals which achieve the same minimum energy
per information bit achieved  by the Gaussian input are first-order optimal.
If the perfect channel knowledge is available at the receiver,
 both BPSK and QPSK are first-order optimal for a general MIMO channel.
Also,  S. Verd\'{u} define that any first-order optimal input signals which achieves
the wideband slope achieved by Gaussian input are second-order optimal.
For a MIMO Gaussian channel with perfect channel knowledge at the receiver,
it is proved that equal-power QPSK is second-order optimal.
Moreover, V. V. Prelov \textit{et al.} provide the second order
expansion of the mutual information in low SNR regime for
very general classes of inputs and channel distributions \cite{Prelov2004TIT}.

For a parallel Gaussian channels with $m$-ary constellation inputs,
A. Lozano \textit{et al.} provide both lower and upper bounds of the
MMSE in high SNR regime \cite{Lozano2006TIT}. For an arbitrary MIMO
Gaussian channel,  F. P\'{e}rez-Cruz \textit{et al.} further derive
high SNR lower and upper bounds of the MMSE and the mutual information \cite{Perez-Cruz2010TIT}.
For a scalar channel with arbitrary inputs and additive noise,
 Y. Wu \textit{et al.} define the MMSE dimension when SNR tends to be infinity in \cite{Wu2012TIT_2}.
The MMSE dimension represents the gain of  the non-linear estimation error over the linear estimation
error in high SNR regime. For MIMO Gaussian block Rayleigh fading channels with $m$-ary constellation inputs, it is proved that
the constellation with a higher minimum Euclidean distance achieves a higher mutual information
in high SNR regime \cite[Appendix E]{Duyck2013TIT}.  For a scalar Gaussian channel with arbitrary input distributions independent of SNR,
the exact high SNR expansion of the mutual information is obtained in \cite{Alvarado2014TIT} in the first time.
The obtained asymptotic expression validates the optimality of the Gray code for bit-interleaved coded modulation (BICM)
in high SNR regime. By using Mellin transform,
A. C. P. Ramos \textit{et al.}  derive low and high SNR asymptotic expansions of the ergodic mutual information
and the average MMSE for a scalar  Gaussian channels with arbitrary discrete  inputs
over Rician and Nakagami fading \cite{Ramos2014TIT}.
High SNR expansions are derived for a MIMO Gaussian channel over Kronecker
fading with the line-of-sight (LOS) \cite{Rodrigues2014TIT}.  A brief summation
of the recent analytic results for discrete input signals
in asymptotic SNR regime is given in Table \ref{table:asy-SNR-non-Gaussian}.

\begin{table}
\centering
  \renewcommand{\multirowsetup}{\centering}
 \captionstyle{center}
  {
\caption{Recent analytic results for discrete input signals in asymptotic SNR regime}
\label{table:asy-SNR-non-Gaussian}
\begin{lrbox}{\tablebox}
\begin{tabular}{|c|c|c|}
\hline
  Paper   & Model &   Main results in asymptotic SNR regime \\ \hline
\multirow{2}{*} {S. Verd\'{u} \cite{Verdu2002TIT}}   &  MIMO, low SNR   &  Analyze the  minimum energy
per information bit   \\
 &  A general class of channels &  Analyze the wideband slope
\\ \hline
\multirow{2}{*} {V. V. Prelov \textit{et al.} \cite{Prelov2004TIT}}   &  MIMO, low SNR   & \multirow{2}{*} {Obtain the second order
expansion of the mutual information }   \\
 &  A general class of channels &
\\ \hline
\multirow{2}{*} {A. Lozano \textit{et al.} \cite{Lozano2006TIT}}   &  Parallel MIMO Gaussian channels & \multirow{2}{*} {Provide lower and upper bounds of the
MMSE}   \\
 &  high SNR &
\\ \hline
\multirow{2}{*} {F. P.-Cruz \textit{et al.} \cite{Perez-Cruz2010TIT}}   &   MIMO Gaussian channels & {Provide lower and upper bounds of}   \\
 &  high SNR &  the
MMSE and the mutual information
\\ \hline
\multirow{2}{*} {Y. Wu \textit{et al.} \cite{Wu2012TIT_2}}   &   Scalar,  arbitrary additive noise   &   \multirow{2}{*} {Define the MMSE dimension} \\
 &    high SNR &
 \\ \hline
\multirow{2}{*} {D. Duyck \textit{et al.} \cite{Duyck2013TIT}}   &   MIMO Gaussian channels & {Prove the mutual information is a monotonously}   \\
 &  block Rayleigh fading, high SNR &  {increasing function of the minimum Euclidean distance}
\\ \hline
\multirow{2}{*} {A. Alvarado \textit{et al.} \cite{Alvarado2014TIT}}   &   Scalar Gaussian channels & {Obtain the exact high SNR expansion }   \\
 &  high SNR &  {of the mutual information}
\\ \hline
\multirow{2}{*} {A. C. P. Ramos \textit{et al.} \cite{Ramos2014TIT}}   &   Scalar Gaussian channels, Rician fading & {Obtain asymptotic expansions of the }   \\
 & Nakagami fading, low SNR, high SNR &  {ergodic mutual information and the average MMSE}
\\ \hline
\multirow{2}{*} {M. R. D. Rodrigues \cite{Rodrigues2014TIT}}   &   MIMO Gaussian channels, high SNR & {Obtain asymptotic expansions of  }   \\
 & Kronecker fading with LOS &  {ergodic mutual information and the average MMSE}
\\ \hline
\end{tabular}
\end{lrbox}
\scalebox{1}{\usebox{\tablebox}}
}
\end{table}

\subsubsection{Asymptotic Results in Large System Limits}
Some work provide the analytic results for discrete input signals
in large system limit.
A technique which has been widely used in statistical physics,
referred to replica method, is initially  exploited to analyze the asymptotic performance
of various detection schemes of code-division multiple access systems (CDMA) in large dimension \cite{Tanaka2002TIT,Guo2005book,Muller2004TIT,Guo2005TIT_2}.
By using this replica method, R. R. M\"{u}ller derive an asymptotic mutual information expression for
a spatially correlated MIMO channel with BPSK inputs in the first time  \cite{Muller2003TSP}. This expression is
extended to arbitrary input at the transmitter in \cite{Wen2006TCOM}.
C.-K. Wen \textit{et al.}  obtain the asymptotic sum rate expression for the Kronecker fading channels for the centralized MIMO
multiple access channel (MAC) and the distributed MIMO MAC  in \cite{Wen2007TCOM} and \cite{Wen2007TIT}, respectively.
The asymptotic sum-rate expression including the LOS  effect
for the centralized MIMO MAC is derived in \cite{Wen2010TCOM}.  C.-K. Wen \textit{et al.} further obtain
the asymptotic sum-rate expression for the Kronecker fading channels
for the centralized MIMO MAC with amplify-and-forward (AF) relay \cite{Wen2010JSAC}.  Also, M. A. Girnyk \textit{et al.} derive an
asymptotic mutual information expression for i.i.d. Rayleigh fading for  K-hop AF relay MIMO channels \cite{Girnyk2010arxiv}.
For other performance criterions, R. R. M\"{u}ller  \textit{et al.} obtain an asymptotic expression for
the average minimum transmit power for i.i.d. Rayleigh fading MIMO channels with a non-linear vector precoding \cite{Muller2008JSAC}.
Similar problem is also considered for MIMO broadcast channel (BC) in \cite{Zaidel2012TIT}.  A more general analysis, referred to one step replica symmetry
breaking analysis, is used in \cite{Zaidel2012TIT} to reduce the asymptotic approximation errors in \cite{Muller2008JSAC} for
 non-convex alphabet sets.
 The effect of the transmit-side noise caused by hardware impairments is investigated in \cite{Vehkaperaa2015TCOM}
for point-to-point MIMO systems with i.i.d. discrete input signals and i.i.d. MIMO Rayleigh fading channels.
It is assumed that the receiver has perfect channel state information (CSI) but the transmitter does not know the CSI.
Also, the transmit-side noise is generally unknown at both transmitter and the receiver. As a result,
the receiver employs a mismatched decoding without taking the consideration of the transmit-side noise.
The general mutual information in \cite{Ganti2000TIT} is adopted as the metric to evaluate the performance. An
asymptotic expression of the general mutual information is derived when both the numbers of transmit
and receive antennas go to infinity with a fixed ratio. Numerical results indicate that
the derived asymptotic expression provides an accurate approximation for the exact general mutual information
and the effects of transmit-side noise and mismatched decoding become significant only at high modulation orders.
A brief summation
of recent research results
of the analytic results (transmitter side) for discrete input signals
in large system limit  is given in Table \ref{table:I-MMSE-non-Gaussian_system}.

\begin{table}
\centering
  \renewcommand{\multirowsetup}{\centering}

 \captionstyle{center}
  {
\caption{Analytic results (transmitter side) for discrete input signals in large system limit }
\label{table:I-MMSE-non-Gaussian_system}
  \begin{lrbox}{\tablebox}
\begin{tabular}{|c|c|c|}
\hline
  Paper   & Channel &   Criterion \\ \hline
R. R. M\"{u}ller \cite{Muller2003TSP}  &  Point-to-point MIMO, spatially correlated Rayleigh &   Asymptotic mutual information (BPSK) \\  \hline
C.-K. Wen  \textit{et al.}  \cite{Wen2006TCOM}  &  Point-to-point MIMO, Kronecker fading  &   Asymptotic mutual information  \\ \hline
C.-K. Wen  \textit{et al.}  \cite{Wen2007TCOM}  &  Distributed MIMO MAC, Kronecker fading  &   Asymptotic sum-rate  \\ \hline
C.-K. Wen  \textit{et al.}  \cite{Wen2007TIT}  &  Centralized MIMO MAC,  Kronecker fading  &   Asymptotic sum-rate  \\ \hline
C.-K. Wen  \textit{et al.}  \cite{Wen2010TCOM}  &  Centralized MIMO MAC,  Kronecker fading with LOS  &   Asymptotic sum-rate  \\ \hline
C.-K. Wen  \textit{et al.}  \cite{Wen2010JSAC}  &  Centralized MIMO MAC with AF relay,  Kronecker fading &   Asymptotic sum-rate  \\ \hline
M. A. Girnyk  \textit{et al.}  \cite{Girnyk2010arxiv}  &  K-hop AF relay MIMO, i.i.d. Rayleigh fading  &   Asymptotic mutual information  \\ \hline
R. R. M\"{u}ller \textit{et al.}  \cite{Muller2008JSAC}  & Point-to-point MIMO,  i.i.d. Rayleigh fading, vector precoding &   Asymptotic average minimum transmit power  \\ \hline
B. M. Zaidel \textit{et al.}  \cite{Zaidel2012TIT}  & MIMO BC,  i.i.d. Rayleigh fading, vector precoding (1RSB ansatz) &   Asymptotic average minimum transmit power  \\ \hline
\multirow{2}{*} {M. Vehkaper\"{a} \textit{et al.} \cite{Vehkaperaa2015TCOM} } & Point-to-point MIMO,  i.i.d. Rayleigh fading, i.i.d. input signals &   \multirow{2}{*} {General mutual information}   \\
 &  Transmit-side noise and mismatched decoding &    \\ \hline
\end{tabular}
\end{lrbox}
\scalebox{0.9}{\usebox{\tablebox}}
}
\end{table}

\subsubsection{Other Results}
There are some other analytical research results for discrete input signals.
For a scalar Gaussian channel with a finite size input signal constraint,
N. Varnica \textit{et al.} propose a modified Arimoto-Blahut algorithm
to search for the optimal input distribution which maximizes the
information rate \cite{Varnica2002}. The corresponding maximum information rate
is referred to constellation constrained capacity.
J. Bellorado \textit{et al.} study the constellation constrained capacity of MIMO Rayleigh fading channel.
It is shown in \cite{Bellorado2006TWC} that for a given information rate, the SNR gap between
the uniformly distributed signal and the constellation constrained capacity when the
constellation size tends to infinity is 1.53 dB. This is the same with
the QAM asymptotic shaping gap for the scalar Gaussian channel in \cite{Forney1989JSAC}.
Also, it is conjectured in \cite{Bellorado2006TWC} that when the elements of input signal vector are chosen from
a Cartesian product of one dimension PAM constellations, the
 constellation constrained capacity can be achieved by optimizing the distribution of each element
of the signal vector independently. This reduces the computational complexity significantly.
This conjecture is proved in \cite{Yankov2014}. M. P. Yankov  \textit{et al.}
derive a low-computational complexity  low bound of the constellation constrained capacity
for MIMO Gaussian channel based on QR decomposition \cite{Yankov2015}.
For one dimension channel with memory and finite input state,
D. M. Arnold \textit{et al.} propose a practical algorithm
to compute the information rate \cite{Arnold2006TIT}. For two dimension channel,
tight low and upper bounds of the information rate
are derived in \cite{Chen2006TIT}. O. Shental \textit{et al.} propose
a more accurate computation  method for the information rate of two dimension channel with memory and finite input state
based on generalized belief propagation \cite{Shental2008TIT}.

\section{Point-to-Point MIMO System}
In this section, we focus on the transmission designs for point-to-point MIMO systems
with discrete input signals. We classify these designs into four  categories:
i) design based on mutual information; ii) design based on MSE; iii) diversity driven
design; iv) adaptive transmission.

\subsection{Design Based on Mutual Information}
The mutual information maximization yields
the maximal achievable rate of the communication link with the implementation of
reliable error control codes \cite{Cover}. For point-to-point
MIMO system, the maximization of the mutual information has been obtained
in \cite{Telatar1999}. It is proved in \cite{Telatar1999} that the mutual information is maximized
by a Gaussian distributed signal with a covariance structure satisfying
two conditions. First, the signal must be transmitted along the right eigenvectors
of the channel, which decomposes the original channel into a set
of parallel subchannels. Second, the power allocated for each
subchannel is based on a water-filling strategy, which depends on
the strength of each subchannel and the SNR.

Although the Gaussian signal is capacity-achieving, it is too idealistic to be implemented in
practical communication systems. The discrete input signals are taken from a finite discrete constellation set, such
as PSK, PAM, or QAM. Therefore, some work begin to investigate MIMO transmission designs based on
the discrete constellation input mutual information.

\subsubsection{Mutual Information Maximization}
By exploiting the I-MMSE relationship in \cite{Guo2005TIT},
 A. Lozano  \textit{et al.} propose a mercury water-filling power allocation policy which maximizes the input-output mutual
information for  diagonal MIMO Gaussian channels with arbitrary inputs \cite{Lozano2006TIT}.
The essential difference between the mercury water-filling  policy and
the conventional  water-filling policy is that it will pour mercury  onto each of the
subchannel to a certain height before filling the water (allocated power) to
the given threshold.  Because of the poured mercury, the classical conclusion for the Gaussian input signal that,
the stronger subchannel receives more filling water (allocated power) does not hold any more.
The asymptotic expansions of the MMSE function and the mutual information in low and
high SNR regime are obtained in \cite{Lozano2006TIT}.
Based on these expansions, it is proved that allocating power only to the strongest channel
is asymptotically optimal in low SNR regime, which is the same with the conventional water-filling
allocation.  For the optimal power allocation policy in high SNR regime, in contrast,
the less power should be allocated to the stronger
subchannel for equal constellation inputs. This is significant different from the conventional water-filling
allocation. Also, it is revealed that the more power should be allocated to the richer constellation inputs
for equal channel strength at high SNR.
For the non-diagonal MIMO Gaussian channel with arbitrary inputs,
an iterative numerical to search for the optimal linear precoder based on
a gradient decent update of the precoder matrix is proposed in \cite{Palomar2006TIT}.
Moreover, necessary conditions for the optimal linear precoding matrix maximizing the mutual information of
the non-diagonal MIMO Gaussian channel with finite alphabet inputs are presented in \cite{Xiao2008GlobalCom},
from which a numerical algorithm is formulated to search for the optimal precoder.
Simulations indicate that the obtained non-diagonal and non-unitary precoder
in \cite{Xiao2008GlobalCom} outperforms the mercury water-filling design in non-diagonal MIMO Gaussian channel.
The design in \cite{Xiao2008GlobalCom} has been extended to the case where only statistical CSI (SCSI) is available
 at the transmitter \cite{Xiao2009ICC}. In the mean time, F. P\'{e}rez-Cruz \textit{et al.} also
 establish necessary conditions that the optimal precoder should satisfy for
 the non-diagonal MIMO Gaussian channel with finite alphabet inputs
through  Karush-Kuhn-Tucker (KKT) analysis \cite{Perez-Cruz2010TIT}.
The  MMSE function and the mutual information of the non-diagonal MIMO Gaussian channel with finite alphabet inputs
are expanded in asymptotic low and high SNR regime. It is proved
that in low SNR regime, the optimal precoder design for finite alphabet
inputs is exactly the same with the optimal design for Gaussian input.
In high SNR regime, low and upper bounds of the mutual information
are derived. Base on the bounds, it is shown that the precoder that
maximizes the mutual information is equivalent to the precoder that
maximizes the minimum Euclidean distance
of the receive signal points in high SNR regime.
This result indicates that the three important criterions in MIMO transmission
design: maximizing the mutual information, minimizing the symbol error rate (SER),
and minimizing the MSE are equivalent in high SNR regime. The optimal power allocation
scheme for the general MIMO Gaussian channel is also obtained in \cite{Perez-Cruz2010TIT}.
For the real-valued non-diagonal MIMO Gaussian channels,
M. Payar\'{o} \textit{et al.} decompose the design of the optimal precoder
into three components: the left singular vectors, the singular values,
and the right singular vectors \cite{Payaro2009}.
The optimal left singular vectors
are proved to be the right singular vectors of the channel matrix.
Necessary and sufficient conditions of the optimal singular values
are established, from which the optimal singular values can be found
based on the standard numerical algorithm such as Newton method. The optimal solution of the right
singular vectors is a difficult problem. Based on standard numerical algorithm, a local optimum
design can be found. Moreover, the optimal designs in low and high SNR regime are analyzed in \cite{Payaro2009}.
For the complex-valued non-diagonal MIMO Gaussian channels,
M. Lamarca  proves that the optimal left singular vectors of the precoder
is still the right singular vectors of the channel matrix \cite{Lamarca2009}. More importantly,
when the entries of the channel matrix and the input signals are all real-valued,
M. Lamarca proves that the mutual information is a concave function of a quadratic function
of the precoder matrix in \cite{Lamarca2009}. Then, an iterative algorithm is proposed to maximize
the mutual information by updating this  quadratic function along the gradient
decent direction. For the complex-valued  non-diagonal MIMO Gaussian channel with finite alphabet inputs,
C. Xiao \textit{et al.} prove that the mutual information is still
a concave function of a quadratic function of the precoder matrix \cite{Xiao2011TSP}. Then,
a parameterized iterative algorithm is designed to find the optimal precoder
which achieves the global maximum of the mutual information.
Moreover, the coded bit error rate (BER) performance of the obtained precoder is simulated.
The BER gains of the obtained precoder comparing to the conventional designs
coincides with that of the mutual information gains, which indicate
that maximizing the mutual information is also an effective criterion for optimizing
the coded BER in practical systems.  X. Yuan \textit{et al.} consider a joint linear precoding and linear MMSE detection scheme
for finite alphabet input signals with perfect CSIT (PCSIT) and perfect CSI at the receiver (PCSIR) \cite{Yuan2014TIT}. The linear precoder is set to be
a product of three matrices: the right singular vector matrix  of the channel, the diagonal power allocation matrix,
and the normalized discrete Fourier transform matrix.
When the detector and the decoder
satisfy the curve-matching principle \cite[Eq. (23)]{Yuan2014TIT}, an asymptotic  achievable rate of the linear precoding and linear MMSE detection scheme
as the dimension of the transmit signal goes towards infinity is derived. The derived achievable rate
is a concave function of the power allocation matrix. As a result, the optimal power allocation scheme
for maximizing the achievable rate can be obtained based on the standard convex optimization method.
Very recently, K. Cao \textit{et al.} propose a precoding design in high SNR regime
by transforming the maximization of the
minimum Euclidean distance into a semi-definite programming optimization problem \cite{Cao2017EL}.
Moreover, for a large number of antennas, the implementation
complexity of digital beamforming may be significantly high.
To reduce the complexity, the MIMO beamforming can be implemented
in both the analog and digital domains, which is referred as
hybrid beamforming \cite{Han2015CM}.  R. Rajashekar \textit{et al.} investigate the hybrid beamforming designs
to maximize the mutual information
for millimeter wave MIMO systems with finite alphabet inputs \cite{Rajashekar2016TCom}.

\begin{table}
\centering
  \renewcommand{\multirowsetup}{\centering}
 \captionstyle{center}
  {
\caption{MIMO transmissions with finite alphabet inputs based on mutual
information maximization}
\label{table:MI-non-Gaussian}
\begin{lrbox}{\tablebox}
\begin{tabular}{|c|c|c|}
\hline
  Paper   & Model &   Main Contribution \\ \hline
\multirow{2}{*} { A. Lozano  \textit{et al.} \cite{Lozano2006TIT}}   &   Diagonal MIMO   & Obtain the optimal power allocation  \\
 &  PCSIT, PCSIR &  Obtain the optimal power allocation in low and high SNR regime
\\ \hline
\multirow{2}{*} {D. P. Palomar  \textit{et al.} \cite{Palomar2006TIT}}   &    Non-diagonal MIMO   & \multirow{2}{*} {Propose a gradient descent algorithm to search for the optimal precoder}  \\
 &  PCSIT, PCSIR &
\\ \hline
\multirow{2}{*} {C. Xiao  \textit{et al.} \cite{Xiao2008GlobalCom}}   &    Non-diagonal MIMO   & Establish necessary conditions for the optimal precoder \\
 &  PCSIT, PCSIR & Propose a numerical algorithm to search for the optimal precoder
\\ \hline
\multirow{2}{*} {C. Xiao  \textit{et al.} \cite{Xiao2009ICC}}   &    Non-diagonal MIMO   & Establish necessary conditions for the optimal precoder \\
 &  SCSIT, PCSIR & Propose a numerical algorithm to search for the optimal precoder
\\ \hline
\multirow{2}{*} {F. P\'{e}rez-Cruz \textit{et al.}  \cite{Perez-Cruz2010TIT}}   &    Non-diagonal MIMO   & Establish necessary conditions for the optimal precoder \\
 &  PCSIT, PCSIR & Analyze  the optimal precoder in low and high SNR regime
\\ \hline
\multirow{2}{*} {M. Payar\'{o} \textit{et al.} \cite{Payaro2009}}   &   Real-valued non-diagonal MIMO   & Obtain the  left  singular vectors and the singular
values of the optimal precoder \\
 &  PCSIT, PCSIR & Analyze  the optimal precoder in low and high SNR regime
\\ \hline
\multirow{4}{*} {M. Lamarca \cite{Lamarca2009}}   &  \multirow{2}{*} {Non-diagonal MIMO}     &
 Obtain the  left  singular vectors of the optimal precoder in complex-valued channels \\
 &   &   Prove that the mutual information is a concave function of a quadratic function \\
 & \multirow{2}{*} { PCSIT, PCSIR} &   of the precoder matrix for
 real-valued non-diagonal MIMO channels  \\
 &  &   Propose an iterative algorithm to maximize
the mutual information \\ \hline
\multirow{4}{*} {C. Xiao \cite{Xiao2011TSP}}   &   \multirow{2}{*} {Non-diagonal MIMO}   &  Prove that the mutual information is still a concave function of a quadratic function \\
 &   &  of the precoder matrix for
complex-valued non-diagonal MIMO channels \\
 &   \multirow{2}{*} {PCSIT, PCSIR} &   Propose a global optimal iterative algorithm to maximize the mutual information
\\
 &   &  Examine the coded BER performance of the obtained precoder
\\ \hline
\multirow{2}{*} {X. Yuan \textit{et al.} \cite{Yuan2014TIT}}   &   Non-diagonal MIMO   & Design the linear precoder by a product
of three matrices
\\
 &  PCSIT, PCSIR & Obtain an asymptotic  achievable rate of this linear precoder and linear MMSE detection
\\ \hline
\multirow{2}{*} {C. Kao \textit{et al.} \cite{Cao2017EL}}   &   Non-diagonal MIMO   &
Transform the maximization of the
minimum Euclidean distance in high SNR regime
\\
 &  PCSIT, PCSIR & into a semi-definite programming optimization problem
\\ \hline
\multirow{2}{*} {R. Rajashekar  \textit{et al.} \cite{Rajashekar2016TCom}}   &   Non-diagonal MIMO   &  Propose hybrid beamforming designs
to maximize   \\
 &  PCSIT, PCSIR & mutual information for millimeter wave MIMO systems
\\ \hline
\end{tabular}
\end{lrbox}
\scalebox{0.85}{\usebox{\tablebox}}
}
\end{table}

\subsubsection{Low Complexity Design}
Although the proposed algorithm in \cite{Xiao2011TSP}
finds the global optimal precoder, capable of
achieving the maximum of the mutual information for the complex-valued
non-diagonal MIMO Gaussian channel with finite alphabet inputs,
its implementation complexity is usually difficult
to afford in practice systems.
There are two main reasons for the high computational complexity.
First, both the mutual information
and the MMSE function lack closed-form expression. Therefore,
 the Monte Carlo method has to be used to evaluation both terms,
which requires an average over enormous noise realizations in order to maintain
the accuracy. More importantly, the evaluation of the mutual
information \cite[Eq. (5)]{Xiao2008GlobalCom} and the MMSE function \cite[Eq. (8)]{Xiao2008GlobalCom}
involves additions over the modulation signal space
which goes exponentially with the number of transmit antennas.
This results in a prohibitive computational complexity  when the dimension
of the transmit antennas is large.

By employing the Jensen's inequality and the
intergral over the exponential function,
a lower bound of the single-input single-output (SISO) mutual information with finite alphabet inputs
is derived in \cite[Eq. (9)]{Abhinav2011}. The evaluation
of this lower bound does not require
to  calculate the expectation over the noise.
As a result, W. Zeng  \textit{et al.} extend
this lower bound to the case of MIMO systems
to solve the first factor of the high computational complexity  due to
the Monte Carlo average \cite{Zeng2012WCL}.
An algorithm is proposed by maximizing
this lower bound. Numerical results indicate that
the proposed design in \cite{Zeng2012WCL} reduces
the computational complexity without performance loss.
W. Zeng  \textit{et al.} further investigate
the linear precoder design with SCSI at the transmitter (SCSIT) over
the Kronecker fading model \cite{zeng2012linear}. A lower
bound of the ergodic mutual information which avoids the expectations
over both the channel and the noise is derived.
Simulations over various fading channels show that with
a constant shift, this lower bound provides
an accurate approximation of the ergodic mutual information.
Moreover, it is proved that maximizing this lower bound
is asymptotically optimal in low and high SNR regime. An
iterative algorithm is further developed by maximizing this low bound.
Numerical results indicate that the proposed design in \cite{zeng2012linear}
achieves a near-optimal mutual information and BER performance with
reduced complexity. The lower bound in \cite{zeng2012linear} is utilized
to design the linear precoder for practical MIMO-orthogonal frequency division multiplexing (OFDM) systems in \cite{Zheng2013IETC}.

When perfect instantaneous CSI is available at the transmitter,
the MIMO channel can be decomposed into a set of parallel subchannels.
For the second factor of the high computational complexity,  S. K. Mohammed \textit{et al.}
propose  X-Codes to pair two subchannels into a group \cite{Mohammed2011TIT}.
X-Codes are fully characterized by the paring and a $2 \times 2$
single angle parameterized rotation matrix for each pair. Based
on X-Codes structure, the mutual information can be expressed
as  a sum of the mutual information of all the pairs.
In this case, the calculation of the mutual information only
involves additions over the modulation signal space
which goes polynomially with the number of transmit antennas.
This significantly reduces the computational complexity of evaluating
the mutual information. Designs
of the rotation angle and power allocation within each pair
and the optimization pairing and power allocation among pairs are further
proposed to increase the mutual information performance.
T. Ketseoglou \textit{et al.} extend this idea to pair multiple subchannels based on a Per-Group Precoding (PGP) technique in
\cite{Ketseoglou2015TWC}. The PGP technique decomposes the design of the power allocation matrix and the right
singular vector matrix into the design of several decoupled matrices with a much
small dimension. An iterative algorithm is proposed to find
the optimal solution of these decoupled matrices. Numerical results indicate that
the proposed algorithm achieves almost the same performance as the design in \cite{Xiao2011TSP}
but with a significant reduction of the computational complexity.
Moreover, T. Ketseoglou \textit{et al.} propose a novel and efficient method to evaluate the mutual information of
the complex-valued non-diagonal MIMO Gaussian channel with finite alphabet inputs based on the Gauss-Hermite
quadrature rule \cite{Ketseoglou2016}. This approximation method can be combined with the PGP technique to further
reduce the design complexity.

When only SCSI is available at the transmitter,
a low-complexity iterative algorithm to find the optimal
design over the Kronecker fading model is proposed in \cite{Wu2016ICC}. The proposed algorithm relies on
an asymptotic approximation of the ergodic mutual information in large system limit,
which avoids the Monte Carlo average over the channel. Moreover,
based on this approximation, the design of the power allocation matrix and the right singular vector matrix of the precoder
can also be decoupled into the design of small dimension matrices within different groups.
Simulations indicate that the proposed algorithm in \cite{Wu2016ICC} radically reduces the computational complexity of the design in \cite{zeng2012linear}  by orders of magnitude
with only minimal losses in performance.
Y. Wu \textit{et al.} propose a unified low-complexity precoder design for
the complex-valued non-diagonal MIMO Gaussian channel with finite alphabet inputs \cite{Wu2016}.
An asymptotic expression of the ergodic mutual information over the jointly-correlated Rician fading model
is derived in large system limit.  Combined with the approximation given in \cite{zeng2012linear},
the obtained  asymptotic expression can provide an accurate approximation of the
ergodic mutual information without the  Monte Carlo average over the channel and the noise.
The left singular matrix of the optimal precoder for this general model is obtained.
Structures of the power allocation matrix and the right singular matrix of the precoder are further established.
 The proposed structures decouple the independent data streams over parallel equivalent subchannels, which
 avoid a complete-search of the entire signal space for the precoder design. Based on these,
  a novel low computational complexity iterative algorithm is proposed to search for the optimal precoder
  with the general CSI availability at the transmitter, which includes the cases of PCSIT,
  imperfect CSIT (IPCSIT), and  SCSIT.  Here we provide Example \ref{complexity-analytical}
  and Example \ref{complexity-simulation} to analytically and numerically compare the complexity of
  the low-complexity algorithm in \cite{Wu2016} and the complete-search design in \cite{zeng2012linear}, respectively.
  As observed from Table \ref{tab:mac_dim}-\ref{runing_time_3},
  the computational complexity of the complete-search design in \cite{zeng2012linear} grows exponential and quickly becomes unwieldy.
  Simulations indicate that the proposed
  algorithm in \cite{Wu2016} drastically reduces the implementation complexity for finite alphabet precoder design,
  but in such a way that the loss in performance---established based on the 3GPP spatial channel model \cite{Salo2005}---is minimal.
  Moreover, some novel insights for the precoder design with SCSIT are revealed in \cite{Wu2016}.
 A brief summation of the above-mentioned
low-complexity MIMO transmissions with finite alphabet inputs is given in Table \ref{table:LC-non-Gaussian}.

\begin{example} \label{complexity-analytical}
Assume $N_{\mathrm t}$ and $N_{\mathrm s} $ denote the number of transmit antenna
and the number of streams in each subgroup, respectively.
Assume $N_{\mathrm s} =2 $ and QPSK modulation.
The numbers of additions required for calculating the mutual information and the MSE matrix in
the complete-search design in \cite{zeng2012linear} and the algorithm in \cite{Wu2016} are listed in Table \ref{tab:mac_dim} for different
numbers of transmit antenna.
\begin{table}[!h]
\centering
\caption{Number of additions required for calculating the mutual information and the MSE matrix.} \label{tab:mac_dim}
\vspace*{1.5mm}
\begin{tabular}{|c|c|c|c|c|c|c|c|}
\hline
 $N_{\mathrm t} $   & 4 &  8 &  16 &   32     \\ \hline
  Complete-search  design in \cite{zeng2012linear}  & 65536 & 4.29  e+009  & 1.84 e+019  &   3.4 e+038      \\ \hline
 Algorithm in \cite{Wu2016} &   512   &   1024  & 2048 &  4096   \\ \hline
\end{tabular}
\end{table}
\end{example}

\begin{example} \label{complexity-simulation}
Let us evaluate the complexity of the algorithm in \cite{Wu2016} for different values of $N_{\mathrm s}$.
Matlab is used on an Intel Core i7-4510U 2.6GHz processor.
Tables \ref{runing_time_1}--\ref{runing_time_3} provide the running time per iteration, for various numbers of antennas and constellations, with $\times$ indicating that the time exceeds one hour.

\begin{table}[!h]

\centering
 \captionstyle{center}
  {
\caption{Running time (sec.) per iteration with BPSK. } \label{runing_time_1}
\begin{tabular}{|c|c|c|c|}
\hline
  $N_{\mathrm t}$  &     $ N_{\mathrm s} = 2$   &   $ N_{\mathrm s} = 4$  &  $N_{\mathrm s} = N_{\mathrm t}$        \\ \hline
  $4$  & 0.0051   &   0.0190   &  0.0190   \\ \hline
 $8$    &  0.0112   &  0.0473  &  11.6209       \\ \hline
   $16$    &  0.0210    &   0.1939 &   $\times$     \\ \hline
      $32$    &  0.0570    &   0.4111 &   $\times$      \\ \hline
\end{tabular}
}
\end{table}

\begin{table}[!h]

\centering
\captionstyle{center}
  {
\caption{Running time (sec.) per iteration with QPSK.} \label{runing_time_2}
\begin{tabular}{|c|c|c|c|}
\hline
  $N_{\mathrm t}$  &     $ N_{\mathrm s} = 2$   &   $ N_{\mathrm s} = 4$  &  $N_{\mathrm s} = N_{\mathrm t}$        \\ \hline
  $4$  &  0.1149 &   21.5350   &  21.5350  \\ \hline
 $8$    &  0.2029   &   23.3442  &  $\times$        \\ \hline
   $16$    & 0.3001   &   48.1725 &   $\times$     \\ \hline
      $32$    & 0.7094   &   98.7853&   $\times$      \\ \hline
\end{tabular}
}
\end{table}

\begin{table}[!h]

\centering
 \captionstyle{center}
  {
\caption{Running time (sec.) per iteration with 16-QAM.} \label{runing_time_3}
\begin{tabular}{|c|c|c|}
\hline
  $N_{\mathrm t}$  &     $ N_{\mathrm s} = 2$    &  $N_{\mathrm s} = N_{\mathrm t}$        \\ \hline
  $4$  &  28.0744    &   $\times$  \\ \hline
 $8$    &  58.3433 &     $\times$        \\ \hline
   $16$    & 106.6022      &   $\times$     \\ \hline
      $32$    & 233.2293 &   $\times$      \\ \hline
\end{tabular}
}
\end{table}

\end{example}

\begin{table}
\centering
  \renewcommand{\multirowsetup}{\centering}
 \captionstyle{center}
  {
\caption{Low-complexity MIMO transmissions with finite alphabet inputs}
\label{table:LC-non-Gaussian}
\begin{lrbox}{\tablebox}
\begin{tabular}{|c|c|c|}
\hline
  Paper   & Model &   Main Contribution \\ \hline
\multirow{2}{*} { W. Zeng  \textit{et al.} \cite{Zeng2012WCL}}   &   MIMO   & \multirow{2}{*} {Design based on a lower bound without Monte Carlo average over the noise} \\
 &  PCSIT, PCSIR &
\\ \hline
\multirow{2}{*} { W. Zeng  \textit{et al.} \cite{zeng2012linear}}   &   MIMO   & Design based on a lower bound without Monte Carlo average  \\
 &  SCSIT, PCSIR & over the channel and the noise
\\ \hline
\multirow{2}{*} {Y. R. Zheng \textit{et al.} \cite{Zheng2013IETC}}   &   MIMO-OFDM  & Evaluate MIMO transmissions
with finite alphabet inputs \\
 &  practical channel estimation & in practical MIMO test-bed
\\ \hline
\multirow{2}{*} {S. K. Mohammed \textit{et al.} \cite{Mohammed2011TIT}}   &   MIMO  & Decouple the mutual information into a
sum of  \\
 &  PCSIT, PCSIR & the mutual information of small dimension matrices
\\ \hline
\multirow{2}{*} {T. Ketseoglou \textit{et al.} \cite{Ketseoglou2015TWC}}   &   MIMO  & Decompose the design of  the power allocation matrix and the right
singular vector matrix  \\
 &  PCSIT, PCSIR & into the design of several decoupled matrices with a much
small dimension
\\ \hline
\multirow{2}{*} {T. Ketseoglou \textit{et al.} \cite{Ketseoglou2016}}   &   MIMO  & Propose a novel and efficient method to evaluate the mutual information  \\
 &  PCSIT, PCSIR & based on the Gauss-Hermite
quadrature rule
\\ \hline
\multirow{2}{*} {Y. Wu \textit{et al.} \cite{Wu2016ICC}}   &   MIMO  & Propose a design which reduces the computational complexity of the design in \cite{zeng2012linear}  \\
 &  SCSIT, PCSIR &     by orders of magnitude with only minimal losses in performance
\\ \hline
\multirow{2}{*} {Y. Wu \textit{et al.} \cite{Wu2016}}   &   MIMO  & Propose a unified low-complexity precoder design which applies   \\
 &  general CSIT, PCSIR &    for the general CSI availability at the transmitter
\\ \hline
\end{tabular}
\end{lrbox}
\scalebox{0.85}{\usebox{\tablebox}}
}
\end{table}

\subsection{Design Based on MSE}
\subsubsection{Linear Transmission Design}
\begin{figure*}[!ht]
\centering
\includegraphics[width=0.8\textwidth]{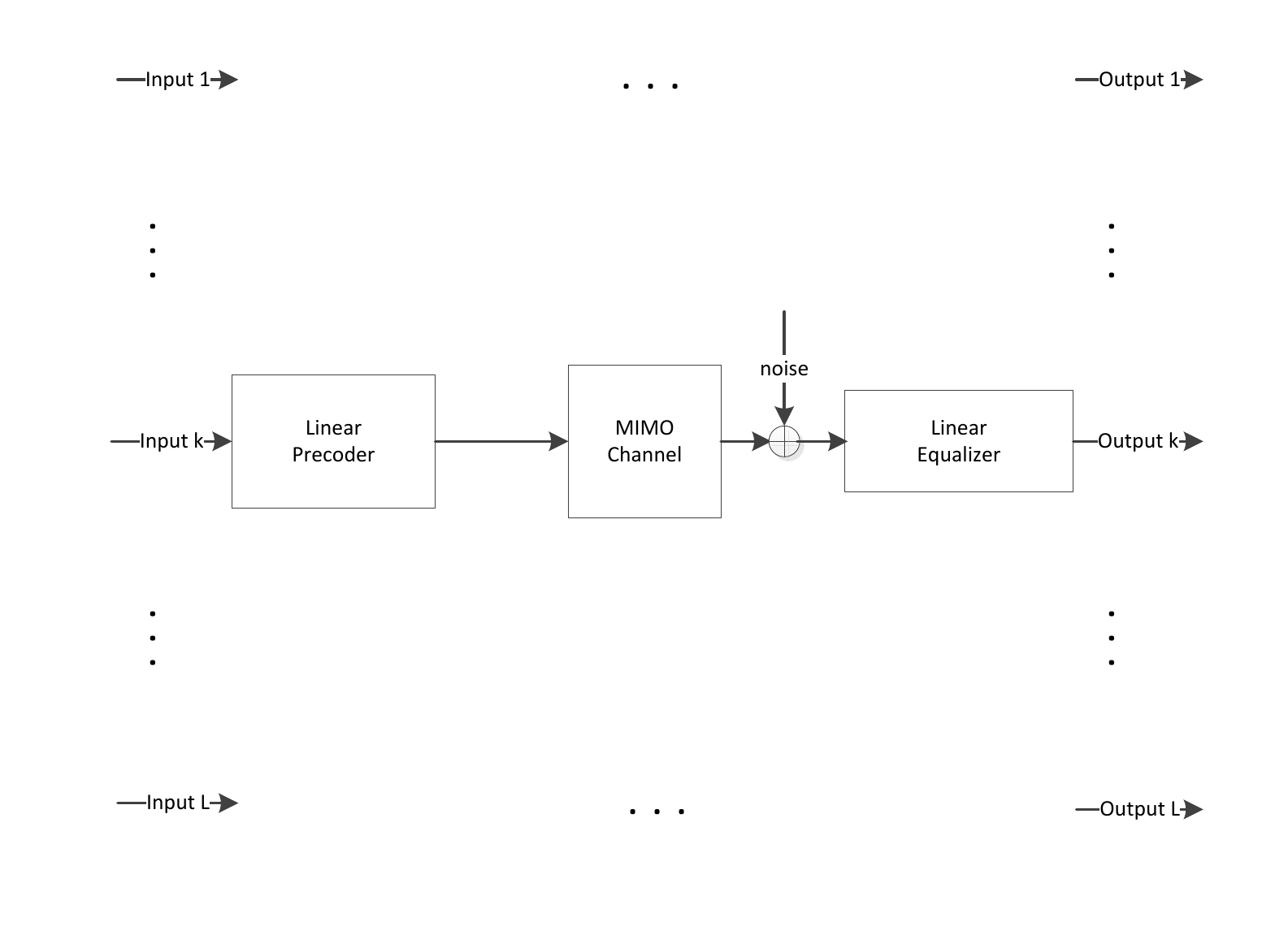}
\caption {\space\space Multi-carrier MIMO system with linear processing and carrier-noncooperative scheme.}
\label{LP-LR}
\end{figure*}

The MSE is an important criterion to measure the quality of the communication link with discrete input signals.
With PCSIT and PCSIR, an optimal
linear transceiver design to minimize the weighted sum MSE of all channel substreams under total average transmit power constraint
is proposed in \cite{Sampath2001TCOM}, developing upon early work in \cite{Lee1976TCOM,Salz1985,Yang1994,Scaglione1999TSP}.
The minimization of the sum MSE of all channel substreams under maximum eigenvalue constraint
is investigated in \cite{Scaglione2002TSP}. A unified framework to optimize the MSE function of all channel substreams under total average transmit power constraint for a multi-carrier MIMO system and linear processing at both ends of the link
is established in \cite{Palomar2003TSP}. It is assumed in \cite{Palomar2003TSP} that the signal constellation
and coding rate for all the subchannels have been determined before the transmission and
the carrier-noncooperative scheme is adopted.
The system model in \cite{Palomar2003TSP} is illustrated in Figure \ref{LP-LR}. It was proved
in \cite{Palomar2003TSP} that if the MSE function is Schur-concave, the optimal solution
is to diagonalize the channel in each subcarrier directly, and if the MSE function is Schur-convex,
the optimal solution is to pre-rotate the transmit symbol  and then
diagonalize the channel in each subcarrier. The above two families of objective functions embrace most of popular
criteria used in communication systems, including the MSE, the SINR, and
also the uncoded BER with identical constellations. The design in \cite{Palomar2003TSP} is extended to a general shaping precoder constraint in \cite{Palomar2004CL}.
Moreover, D. P. Palomar employs a primal decomposition approach to decompose the original optimization problem
over multiple (e.g.,multicarrier) MIMO channels
in \cite{Palomar2005TSP} into several subproblems, each of which can be independently solved over a single MIMO channel.
These subproblems are coordinated  by a simple master problem. As a result, the optimization problem in \cite{Palomar2003TSP}
can be solved in a simple and efficient way.
 In the mean time, the transmission design which
satisfies the specific MSE constraints with minimum transmit power is investigated in \cite{Palomar2004TSP}.
The optimal solution is to pre-rotate the transmit signal by a particular unitary matrix
and then diagonalize the channel. Similar problem is also considered in \cite{Palomar2005TSP_2}
 under the uncoded BER constraints instead of the MSE constraints.  It is assumed that the
 constellation used for each substream is different. Then, the uncoded BER is neither Schur-convex nor Schur-concave function
 of the MSE.  This optimization problem is solved  by the primal decomposition approach.
Moreover, L. G. Ord\'{o}\~{n}ez \textit{et al.} propose an adaptive number of substreams transmission scheme
to minimize the uncoded BER under fix transmission rate constraint and total average transmit power constraint \cite{Palomar2009TSP}.
 A widely linear MMSE
 transceiver which processes the in-phase and quadrature components of the input signals separately is designed
 in \cite{Sterle2007TSP}. The widely linear processing is implemented by using a real-value representation of the input
 signals.  Recently, a linear precoder design to minimize the sum MSE of all channel substreams under the joint
  total average transmit power constraint and maximum eigenvalue constraint is proposed in \cite{Dai2012TCOM}.

However, perfect CSI is only reasonable for the slow fading scenario where the channel
remains deterministic for a long period. Otherwise, the CSI estimated at the transmitter and the
receiver may have mismatches with the actual channel. Therefore, S. Serbetli  \textit{et al.}
begin to investigate the optimum transceiver structure in sense of minimizing
the sum MSE of transmit data streams via imperfect CSI at both ends \cite{Serbetli2006TWC}.
A MIMO system with correlated receive antennas is considered
and the effect of channel estimation is investigated in \cite{Serbetli2006TWC}.
X. Zhang  \textit{et al.} investigate the statistically robust design of linear MIMO transceivers based on minimizing the MSE function
under total average transmit power constraint \cite{Zhang2008TSP}. The statistically robust design exploits
the channel mean and channel covariance matrix (equivalently, the channel estimate and the estimation error)
to design linear MIMO transceivers. Two cases are considered in \cite{Zhang2008TSP}: i) IPCSIT and imperfect channel state information at the receiver
(IPCSIR) with no transmit antenna correlation;
 ii) IPCSIT and PCSIR. For the former case, the optimal solution  has the same
structure as the perfect CSI case by
replacing the instantaneous channel realization matrix and the noise covariance matrix with the channel mean
matrix and equivalent covariance matrix (e.g., channel covariance matrix plus noise covariance matrix), respectively.
For the latter case, a lower bound of the average MSE of all substreams
is derived and the linear precoder minimizing this lower bound is obtained.
The statistically robust design for the IPCSIT and IPCSIR case with transmit antenna correlation is investigated in \cite{Ding2009TSP}. By finding
the optimal solution which satisfies the KKT
conditions, the structures of the optimal precoder and equalizer minimizing
the sum MSE of transmit data streams under total average transmit power constraint are obtained in \cite{Ding2009TSP}.
In the mean time,
J. Wang \textit{et al.} investigate a deterministic robust design for MIMO system
under total average transmit power constraint \cite{Wang2010TSP}. The
deterministic robust design assumes the estimated imperfect CSI is within
a neighborhood of actual channel. The case of IPCSIT and IPCSIR is considered in \cite{Wang2010TSP}.  Given a pre-fixed linear receiver structure,
the optimal transmission direction and the optimal power allocation for the linear precoder matrix minimizing
the MSE of transmit data streams are found in \cite{Wang2010TSP}. For the same scenario, the optimal linear precoder structure given a pre-fixed linear receiver and
the optimal linear equalizer structure given a pre-fixed linear precoder are obtained in \cite{Wang2011SPL}. Therefore, the precoder
and the equalizer can be designed alternatively by an iterative algorithm.
For the scenario considered in \cite{Wang2010TSP}, the optimal linear MIMO transceiver design is found in \cite{Tang2015}.
On the other hand, to reduce the cost of analog-to-digital converter (ADC) chains at the receiver,
MIMO analog beamformers are designed in \cite{Venkateswaran2010TCOM} to minimize both the output MSE of the digital beamformer
and the power consumption of the ADC by exploiting the statistical knowledge of the channel.
A brief summation
of linear transmission designs based on MSE
is given in Table \ref{table:MMSE_Transmission}, where the notations
LP and LE denote abbreviations for
linear precoder and linear equalizer, respectively.

\begin{table}
  \renewcommand{\multirowsetup}{\centering}
 \captionstyle{center}
  {
\caption{Linear transmission designs with discrete input signals  based on the MSE}
\label{table:MMSE_Transmission}
\begin{lrbox}{\tablebox}
\begin{tabular}{|c|c|c|c|c|c|}
\hline
  Paper   & Systems &   Criterion  &  Transceiver  &  CSI & Constraint \\ \hline
\multirow{2}{*} {H. Sampath \textit{et al.} \cite{Sampath2001TCOM}}   & \multirow{2}{*}{MIMO} & Weighted  & LP  &  PCSIT  & \multirow{2}{*}{Sum power}   \\ &  & sum MSE & LE   &  PCSIR &
\\ \hline
\multirow{2}{*}{A. Scaglione \textit{et al.} \cite{Scaglione2002TSP}} & Multi-carrier   & \multirow{2}{*}{Sum MSE} &  LP    &  PCSIT  &  \multirow{2}{*}{Maximum eigenvalue}  \\
 & MIMO  &   &  LE    & PCSIR &  \\ \hline
\multirow{2}{*}{D. P. Palomar \textit{et al.} \cite{Palomar2003TSP}} &   Multi-carrier & \multirow{2}{*}{MSE function} &  LP    &  PCSIT  &  \multirow{2}{*}{Sum power}  \\
 & MIMO  &  &   LE    &  PCSIR &    \\ \hline
\multirow{2}{*}{D. P. Palomar  \cite{Palomar2004CL}}  & \multirow{2}{*}{MIMO} &  \multirow{2}{*}{MSE function} &  LP   &  PCSIT  &  Covariance matrix   \\
 &  &  &  LE    &   PCSIR & shaping  \\ \hline
\multirow{2}{*}{D. P. Palomar  \cite{Palomar2005TSP}} & Multi-carrier &  MSE function &  LP     &  PCSIT  & \multirow{2}{*}{Sum power}  \\
 & MIMO &  efficient solution  &  LE    &   PCSIR &   \\ \hline
\multirow{2}{*}{D. P. Palomar \textit{et al.} \cite{Palomar2004TSP}} & \multirow{2}{*}{MIMO} &  \multirow{2}{*}{Transmit power} &  LP    &  PCSIT & \multirow{2}{*}{MSE}  \\
 &  &   &  LE    &  PCSIR &   \\ \hline
\multirow{2}{*}{D. P. Palomar  \textit{et al.} \cite{Palomar2005TSP_2}} & Multi-carrier   &  Uncoded BER &  LP    &  PCSIT  & \multirow{2}{*}{Sum power} \\
 & MIMO &  different constellations &   LE    &  PCSIR &   \\ \hline
\multirow{2}{*}{L. G. Ord\'{o}\~{n}ez \textit{et al.} \cite{Palomar2009TSP}} &  \multirow{2}{*}{MIMO}  &  Uncoded BER & LP  &  PCSIT  & Sum power  \\
 &   &  adaptive streams &  LE    &  PCSIR &  fixed rate  \\ \hline
\multirow{2}{*}{F. Sterle \cite{Sterle2007TSP}}  & \multirow{2}{*}{MIMO} &  \multirow{2}{*}{Sum MSE} & Widely LP    &  PCSIT  & \multirow{2}{*}{Sum power}   \\
  & &   & Widely   LE   &  PCSIR &    \\ \hline
\multirow{2}{*}{J. Dai \textit{et al.} \cite{Dai2012TCOM}} & \multirow{2}{*}{MIMO} &  \multirow{2}{*}{Sum MSE} &   LP     &  PCSIT  & Sum power   \\
 &  &   &    LE    &   PCSIR &  maximum eigenvalue   \\ \hline
 \multirow{2}{*}{S. Serbetli \textit{et al.} \cite{Serbetli2006TWC}} &  \multirow{2}{*}{MIMO} &  Average sum MSE  &   LP    &  IPCSIT and IPCSIR &   \multirow{2}{*}{Sum power} \\
  &  & channel estimation effect &   LE    &  no transmit correlation &    \\ \hline
\multirow{2}{*}{X. Zhang \textit{et al.} \cite{Zhang2008TSP}} &  \multirow{2}{*}{MIMO} &  \multirow{2}{*}{ Average MSE function}  &   LP    &  IPCSIT and IPCSIR &   \multirow{2}{*}{Sum power} \\
  &  & &   LE    &  no transmit correlation &    \\ \hline
\multirow{2}{*}{X. Zhang \textit{et al.} \cite{Zhang2008TSP}} &  \multirow{2}{*}{MIMO} &  Average MSE function  &   LP    &  IPCSIT  &   \multirow{2}{*}{Sum power}  \\
 &   & lower bound &    LE    &   PCSIR &    \\   \hline
 \multirow{2}{*}{M. Ding \textit{et al.} \cite{Ding2009TSP}} & \multirow{2}{*}{MIMO}  &  \multirow{2}{*}{Average sum MSE} &   LP     &  IPCSIT and IPCSIR &   \multirow{2}{*}{Sum power} \\
 &  &   &    LE    & with transmit correlation &    \\ \hline
\multirow{2}{*}{J. Wang \textit{et al.} \cite{Wang2010TSP}} & \multirow{2}{*}{MIMO}  &  \multirow{2}{*}{Sum MSE} &   LP     &  IPCSIT  &   \multirow{2}{*}{Sum power} \\
 &  &   &   fixed LE    &   IPCSIR &    \\ \hline
 \multirow{2}{*}{J. Wang \textit{et al.} \cite{Wang2011SPL}} & \multirow{2}{*}{MIMO}  &  Sum MSE &  LP     &  IPCSIT  &   \multirow{2}{*}{Sum power}  \\
 &  & not optimal &   LE    &   IPCSIR & \\  \hline
 \multirow{2}{*}{H. Tang \textit{et al.} \cite{Tang2015}}  & \multirow{2}{*}{MIMO}  & Sum MSE &   LP    &  IPCSIT  &   \multirow{2}{*}{Sum power}  \\
 &  &  optimal &    LE    &  IPCSIR &   \\ \hline
 \multirow{2}{*}{V. Venkateswaran \textit{et al.} \cite{Venkateswaran2010TCOM}}  & \multirow{2}{*}{MIMO}  &  MSE, ADC power &   LP    &  SCSIT  &   \multirow{2}{*}{No}  \\
 &  &  optimal &    LE    &  SCSIR &   \\ \hline
\end{tabular}
\end{lrbox}
\scalebox{0.9}{\usebox{\tablebox}}
}
\end{table}

\subsubsection{Non-linear Transmission Design}

\begin{figure*}[!ht]
\centering
\includegraphics[width=0.8\textwidth]{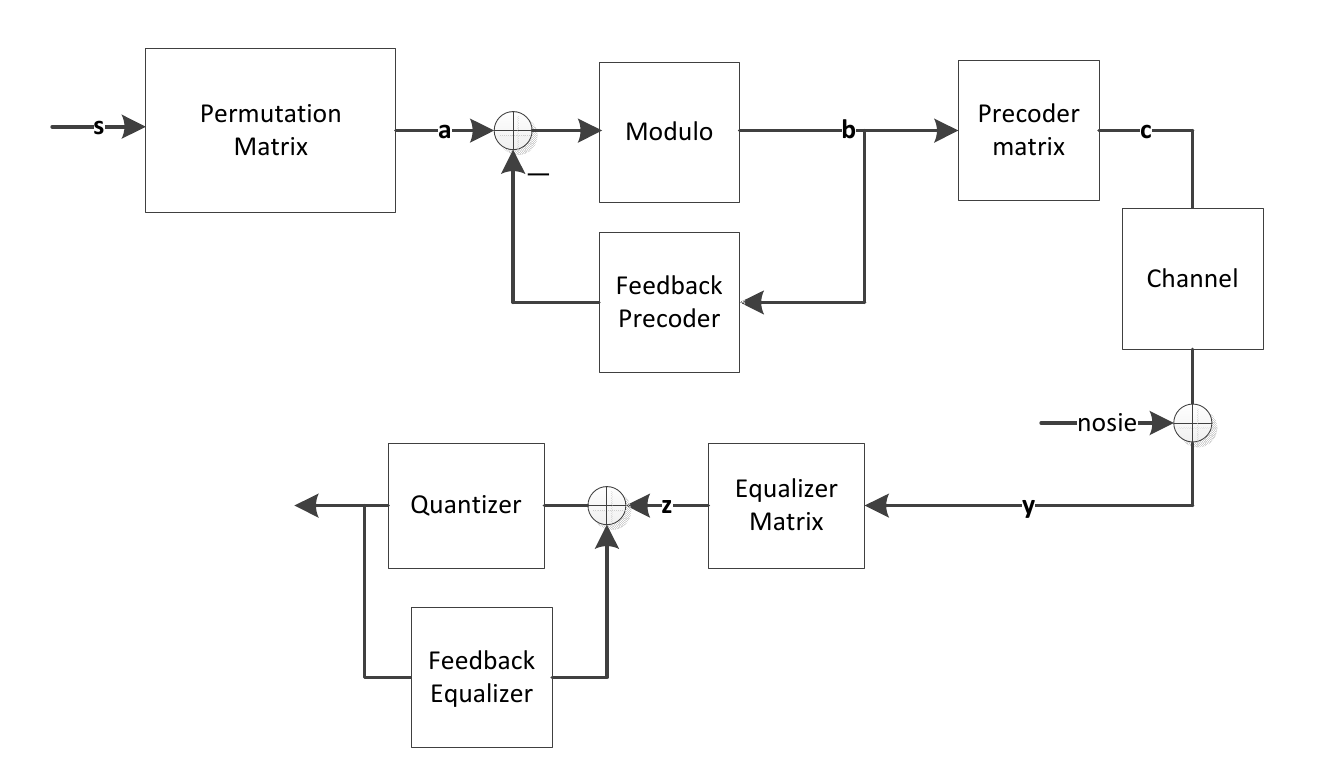}
\caption {\space\space MIMO transceiver with THP and DFE.}
\label{MIMO-THP-DFE}
\end{figure*}

Above-mentioned work are based on the linear processing at both ends of the link.
However, research have shown that non-linear processing
schemes can offer additional advantages over linear
processing schemes \cite{Fischer2002,Amico2008TWC,Xu2006TSP,Jiang2007,Shenouda2008JSTSP,Simeone2004TWC}. Two typical technologies are
Tomlinson-Harashima precoding (THP) at the transmitter
and decision-feedback equalization (DFE) at the receiver.
A unified transceiver structure of a MIMO system with THP and DFE
is illustrated in Figure \ref{MIMO-THP-DFE}.  Assuming PCSIT and PCSIR, a THP precoding with linear equalization which minimizes
the sum MSE of transmit data streams is proposed
in \cite{Fischer2002}. The work in \cite{Fischer2002} focuses on the design of the feedback precoder in Figure \ref{MIMO-THP-DFE}.
The permutation matrix and the precoder matrix in Figure \ref{MIMO-THP-DFE}  are set
to be identical matrix.
The THP design  with linear equalization is extended to MIMO-OFDM system in \cite{Amico2008TWC}, where the input signal is
pre-ordered by a permutation matrix  and an additional precoder matrix is designed
to improve the overall MSE performance. Two linear precoding designs with DFE at the receiver
are proposed in \cite{Xu2006TSP} and \cite{Jiang2007}, which optimize the sum MSE and the MSE function
of the transmit data streams, respectively. A unified framework
which is applicable for both THP with linear equalization and linear precoding with DFE
is established in \cite{Shenouda2008JSTSP}. The proposed designs in \cite{Shenouda2008JSTSP} optimize the MSE function
of the transmit data streams. It is proved that if the MSE function is
Schur-convex, the optimal design forces the MSE
of each transmit data stream to be equal, and if the MSE function is Schur-concave,
the optimal THP leads to linear precoding and the optimal DFE leads
to linear equalization. When only statistical CSI is available at the transmitter,
O. Simenone \textit{et al.} propose a THP with linear equalization design
based on minimizing a lower bound of the sum MSE of the transmit data streams \cite{Simeone2004TWC}.
 A brief summation
of non-linear transmission designs based on MSE
is given in Table \ref{table:MMSE_Transmission_non}, where
the notations
NLP and NLE denote abbreviations for
non-linear precoder and non-linear equalizer, respectively.

\begin{table}[!t]
\centering
  \renewcommand{\multirowsetup}{\centering}
 \captionstyle{center}
  {
\caption{Non-linear transmission designs with discrete input signals  based on the MSE}
\label{table:MMSE_Transmission_non}
\center
\begin{tabular}{|c|c|c|c|c|c|}
\hline
  Paper   & Systems &   Criterion  &  Transceiver  &  CSI & Constraint \\ \hline
\multirow{2}{*}{R. F. H. Fischer \textit{et al.} \cite{Fischer2002}} & \multirow{2}{*}{MIMO}  & \multirow{2}{*}{Sum MSE}  &   NLP   &  PCSIT  &   \multirow{2}{*}{Sum power}  \\
 &  &    &   LE    &   PCSIR &   \\ \hline
\multirow{2}{*}{A. A. D. Amico \textit{et al.} \cite{Amico2008TWC}} &  MIMO &  \multirow{2}{*}{Sum MSE}   &   NLP, LP    &  PCSIT  &   \multirow{2}{*}{Sum power}  \\
  & OFDM &   &    LE    &  PCSIR &   \\ \hline
\multirow{2}{*}{F. Xu \textit{et al.} \cite{Xu2006TSP}} & \multirow{2}{*}{MIMO} &  \multirow{2}{*}{Sum MSE}  &    LP    &  PCSIT & \multirow{2}{*}{Sum power}  \\
 &  &    &    LE, NLE    &  PCSIR &    \\ \hline
\multirow{2}{*}{Y. Jiang \textit{et al.} \cite{Jiang2007}} &  \multirow{2}{*}{MIMO} &  \multirow{2}{*}{MSE function}  &    LP   &  PCSIT  &  \multirow{2}{*}{Sum power} \\
 &  &    &    LE, NLE    &  PCSIR &    \\ \hline
\multirow{2}{*}{M. B. Shenouda  \textit{et al.} \cite{Shenouda2008JSTSP}} &  \multirow{2}{*}{MIMO}  &   \multirow{2}{*}{Sum MSE}    &   NLP, LP, LE &  PCSIT  &  \multirow{2}{*}{Sum power} \\
 &   &    &    LP, LE, NLE  &  PCSIR &   \\ \hline
\multirow{2}{*}{O. Simeone  \textit{et al.} \cite{Simeone2004TWC}}  &  \multirow{2}{*}{MIMO} &  Average  sum MSE   &   NLP, LP, LE   &  SCSIT  &  \multirow{2}{*}{Sum power}   \\
& &  lower bound  &    LP, LE, and NLE    &  PCSIR &  \\ \hline
\end{tabular}
}
\end{table}

\subsubsection{Performance Analysis}
Since the publication of \cite{Palomar2003TSP}, some work begin to analyze
the performance of the uncoded MIMO systems within the linear processing
framework (also refereed to MIMO multichannel-beamforming (MB) systems \cite{Jin_5}) in \cite{Palomar2003TSP}.
In \cite{Ordonez2007TSP}, the first-order polynomial expansions
for the marginal pdfs
of all of the individual ordered eigenvalues of
uncorrelated central Wishart-distributed matrix
are derived. Based on these results, the closed form expressions
for the outage probability and the average uncoded BER
performance for each subchannel of MIMO MB systems with fixed power allocation over an uncorrelated Rayleigh flat-fading channel
in high SNR regime are obtained \cite{Ordonez2007TSP}.
The obtained expressions reveal both the diversity gain and
the array gain in each subchannel.
Also, tight lower and upper bounds for the performance with nonfixed power allocation
such as waterfilling power allocation are obtained.
For the correlated central Wishart-distributed matrix,
the exact pdfs  expressions for the ordered eigenvalues
are derived in \cite{Ordonez2009TSP}. These expressions allow
us to analyze the MIMO MB systems
over a semicorrelated Rayleigh flat-fading channel under various performance
 criterions, including the outage probability and the average uncoded BER.
Simple closed-form first-order polynomial expansions for these marginal pdfs
are  derived in \cite{Palomar2009TSP}, from which the diversity gain and the array gain
of the MIMO MB systems are indicated.  In the mean time,
the exact pdfs  expressions for the uncorrelated noncentral
 Wishart-distributed matrix are independently derived in \cite{Jin_5}
 and \cite{Ordonez2009TSP}, respectively. The first order polynomial expansions
for these marginal pdfs are independently obtained in \cite{Palomar2009TSP}
 and \cite{Jin_5}, respectively.  Performance of MIMO MB systems
 over uncorrelated Rician fading channel can be analyzed  similarly based on these expressions.
 For MIMO MB over a more general Rayleigh-product fading  channel,
 H. Zhang \textit{et al.} derive the first order polynomial expansions of the ordered eigenvalues of the
channel Gram matrix \cite{Hzhang}. The diversity gain and the array gain for MIMO MB systems with fixed power allocation
over the Rayleigh-product fading  channel
are revealed in \cite{Hzhang}.  Later, the analysis for MIMO MB systems is extended to scenarios in the presence
of co-channel interference.  With fixed power allocation,
L. Sun \textit{et al.}
obtain  exact and asymptotic (low outage probability) expressions of outage probability of MIMO MB systems
in the presence of unequal power interferers \cite{Sun}, where
both the desired user and the interferes experience
semicorrelated Rayleigh flat-fading. Also,
S. Jin \textit{et al.} obtain  exact and asymptotic  expressions of outage probability
of MIMO MB systems in the interference-limited (without noise) scenarios
with equal power  interferers. It is assumed in \cite{Jin_3}
that the desired user experiences uncorrelated Rician fading and the interferers experience
uncorrelated Rayleigh fading. Y. Wu \textit{et al.} obtain exact and asymptotic  expressions of outage probability
of MIMO MB systems with arbitrary  power interferers, where the desired user experiences  Rayleigh-product
fading and the interferers experience uncorrelated Rayleigh fading \cite{Wu2012TWC}.
A brief summation
of performance analysis of MIMO MB systems is given in Table \ref{table:per_MIMO_MB}.

\begin{table}[!t]
\centering
  \renewcommand{\multirowsetup}{\centering}
 \captionstyle{center}
  {
\caption{Performance analysis of MIMO MB systems}
\label{table:per_MIMO_MB}
\center
\begin{lrbox}{\tablebox}
\begin{tabular}{|c|c|c|c|c|}
\hline
  Paper   &  user channel &   interferer channel &  interferer power  & outage probability   \\ \hline
  L. G. Ord\'{o}\~{n}ez \textit{et al.} \cite{Ordonez2007TSP}   &  Uncorrelated Rayleigh fading &  No &  No  & Asymptotic  \\ \hline
\multirow{2}{*}{L. G. Ord\'{o}\~{n}ez \textit{et al.} \cite{Ordonez2009TSP}}   &  Semicorrelated Rayleigh fading  & \multirow{2}{*}{No} &  \multirow{2}{*}{No}  & \multirow{2}{*}{Exact}  \\
    & Uncorrelated Rician fading  &   &    &   \\ \hline
\multirow{2}{*}{L. G. Ord\'{o}\~{n}ez \textit{et al.} \cite{Palomar2009TSP}}   &  Semicorrelated Rayleigh fading&  \multirow{2}{*}{No} &  \multirow{2}{*}{No}   & \multirow{2}{*}{Asymptotic} \\
  & Uncorrelated Rician fading &   &    &  \\ \hline
S. Jin \textit{et al.} \cite{Jin_5}   &  Uncorrelated Rician fading &  No &  No  & Exact, asymptotic \\ \hline
H. Zhang \textit{et al.} \cite{Hzhang}   &  Rayleigh-product fading &  No &  No  &  Asymptotic \\ \hline
L. Sun \textit{et al.} \cite{Sun}   &  Semicorrelated Rayleigh fading &  Semicorrelated Rayleigh fading &  Unequal  &  Exact \\ \hline
\multirow{2}{*}{S. Jin \textit{et al.} \cite{Jin_3}}   & \multirow{2}{*}{Uncorrelated Rician fading} &  \multirow{2}{*}{Uncorrelated Rayleigh fading} &
\multirow{2}{*}{Equal}   &   Exact, asymptotic \\
  &  &  &   &   Interference-limited \\ \hline
Y. Wu \textit{et al.} \cite{Wu2012TWC}   &  Rayleigh-product fading  & Uncorrelated Rayleigh fading &  Arbitrary  & Exact, asymptotic     \\ \hline
\end{tabular}
\end{lrbox}
\scalebox{0.95}{\usebox{\tablebox}}
}
\end{table}

\subsection{Diversity Driven Designs}
\subsubsection{Early research for Open-loop Systems}
\begin{figure*}[!ht]
\centering
\includegraphics[width=0.8\textwidth]{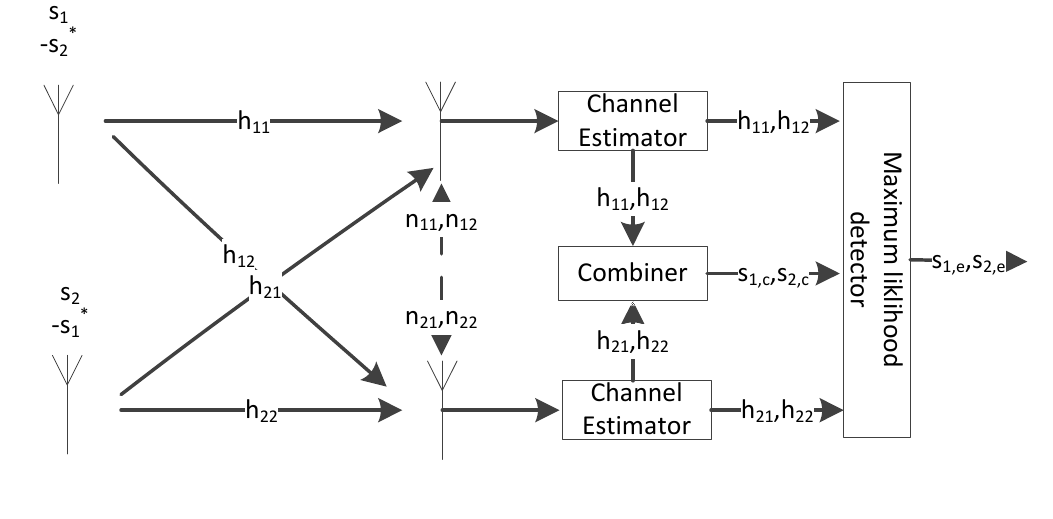}
\caption {\space\space The OSTBC for a $2 \times 2$ system.}
\label{Alamouti_OSTPC}
\end{figure*}

For the open-loop MIMO system, it is usually assumed that
the CSI is available at the receiver but not at the transmitter.
To acquire receive diversity, exploiting diversity combining
schemes at the receiver, such as equal gain
combining, selection combining, and maximum
ratio combining (MRC), can be tracked back to the 60 years ago
with references in the paper \cite{Brennan2003}.
Without CSIT, space-time techniques
are designed to combat the deleterious effects of small scale fading.
These space-time techniques can acquire additional
transmit diversity and thereby
improve the reliability of the link with discrete input signals.
A simple and
well known space-time technique is the Alamouti orthogonal space-time
block code (OSTBC) for a two transmit antennas and two receive antennas system \cite{Alamouti2006JASC},
as illustrated in Figure \ref{Alamouti_OSTPC}.
For OSTBC in Figure \ref{Alamouti_OSTPC}, in time slot $1$,
the symbols $s_{1}$ and $s_{2}$ are transmitted in
antenna $1$ and $2$, respectively.
In time slot $2$, the symbol $-s_{2}^{*}$ and $-s_{1}^{*}$
are transmitted in antenna $1$ and $2$, respectively.
After passing the channel and the combiner in Figure \ref{Alamouti_OSTPC},
the signal $s_{1,{\rm c}}$ and $s_{2,{\rm c}}$ sent to the
maximum likelihood (ML) detector are given by
\begin{align}
{s_{1,{\rm{c}}}} &= \left( {{{| {{h_{11}}} |}^2} + {{| {{h_{12}}} |}^2} + {{| {{h_{21}}} |}^2} + {{| {{h_{22}}} |}^2}} \right){s_0} + h_{11}^*{n_{11}} + {h_{12}}n_{12}^*
 + h_{22}^*{n_{22}} + {h_{21}}{n_{21}} \label{eq:s_1_c}\\
{s_{2,{\rm{c}}}}  & = \left( {{{| {{h_{11}}}|}^2} + {{| {{h_{12}}} |}^2} + {{| {{h_{21}}} |}^2} + {{| {{h_{22}}} |}^2}} \right){s_1} - {h_{11}}n_{11}^* + h_{12}^*{n_{12}}
  - {h_{22}}n_{22}^* + h_{21}^*{n_{21}}.  \label{eq:s_2_c}
\end{align}
(\ref{eq:s_1_c}) and (\ref{eq:s_2_c}) are equivalent to that of four-branch MRC.
Therefore, it achieves a diversity of four for the ML detector.
Additional transmit diversity is obtained  by utilizing OSTBC.
For both real and complex OSTBC, the construction
of OSTBC for any number of transmit antennas and receive antennas is
proposed in \cite{Tarokh1999TIT}.
The proposed design achieves maximum possible spatial
diversity order with a low complexity linear decoding.
The space-time block code (STBC) is extended to single-carrier MIMO system
in \cite{Zhou2003TIT} and MIMO-OFDM system in \cite{Liu2002TSP}.
The proposed designs achieve the maximum diversity but
with the increase of the decoding complexity
since the detector needs to jointly decode all subchannels.
To solve this problem, a subchannel grouping method
 with the maximum diversity performance and a reduced decoding complexity is proposed in \cite{Liu2002TSP}.
The transmission rate of above OSTBC design is less than
1 symbol per channel use (pcu) for more than two transmit antennas.
Therefore, a quasi-orthogonal
space-time block code (QOSTBC), which provides 1 symbol pcu
transmission rate and half of the maximum possible diversity for four transmit antennas,
is proposed in \cite{Jafarkhani2001TIT}. The QOSTBC divides the transmission matrix columns into different groups,
and makes the columns in different groups orthogonal to each other.
This quasi-orthogonal structure increases the transmission rate.
Also, based on this structure, the ML decoding
at the receiver can be done by searching symbols in each group individually instead of
searching all transmit signals jointly.
By rotating half of the transmit symbol constellations, QOSTBC can achieve 1 symbol pcu rate and maximum diversity
 for four transmit antennas \cite{Sharmai2003TCom} and  higher rate than
OSTBC and maximum diversity for MIMO systems \cite{Su2004TIT}.  Based on a different rotation method, C. Yuen \textit{et al.} construct
a class of QOSTBCs for MIMO systems, whose ML decoding
can be reduced to a joint detection of two real symbols \cite{Yuen2005TWC}.
On the other hand,
the space time trellis codes (STTC), which achieves both maximum spatial
diversity and large coding gains, is designed in \cite{Tarokh1998TIT}. However,
the decoding complexity of STTC grows exponentially with transmit rate
and becomes significantly high for high order modulation.

\begin{figure*}[!ht]
\centering
\includegraphics[width=0.8\textwidth]{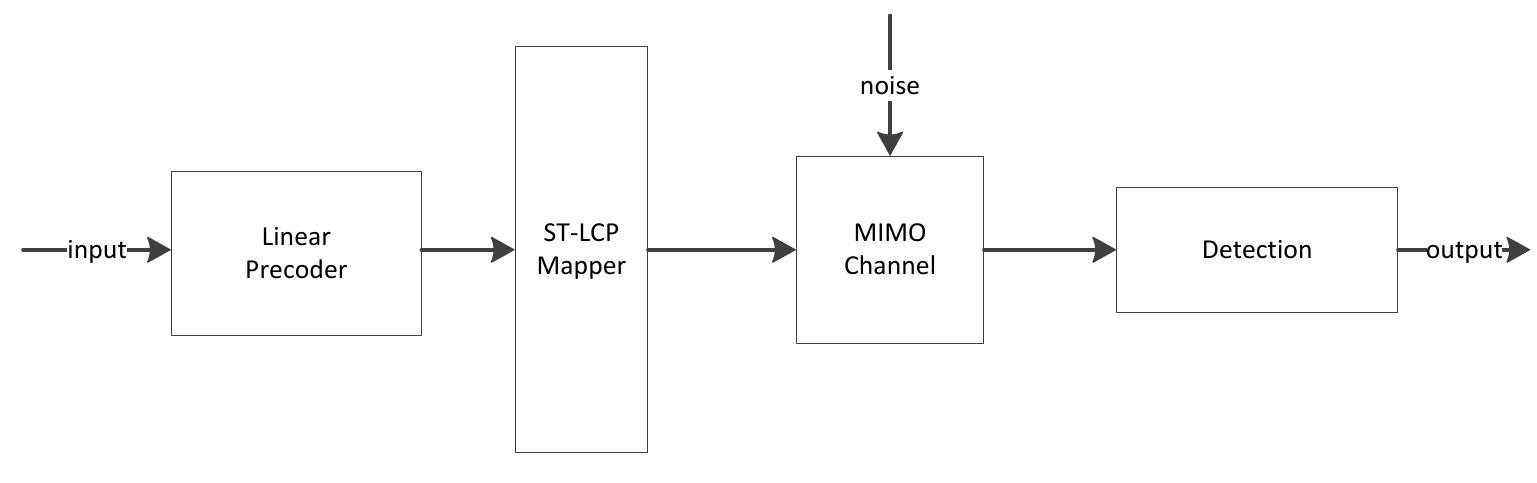}
\caption {\space\space The baseband equivalent framework of ST-LCP.}
\label{ST-LCP}
\end{figure*}

A space time linear constellation precoding (ST-LCP) is proposed in \cite{Xin2003TWC} as an
alternative scheme to obtain transmit diversity. The baseband equivalent framework
of ST-LCP is illustrated in Figure \ref{ST-LCP}. The key idea of ST-LCP is to perform
a pre-combination of the transmit signal by a linear precoder matrix as in Figure \ref{ST-LCP}.
Then, the pre-processed data is mapped to the transmit antenna by a diagonal matrix and
a unitary matrix. Based on the average PEP
expression, linear precoders which achieve the maximum diversity and perform
close to the upper bound of the coding gain are designed in closed-form. Near
optimal sphere decoding algorithms are employed to reduce the decoding complexity of ST-LCP.
Simulations indicate that ST-LCP achieve a better BER performance than
classic OSTBC. On the other hand, Liu \textit{et al.} study ST-LCP
for MIMO-OFDM system \cite{Liu2003TCOM}. The authors in \cite{Liu2003TCOM} combine ST-LCP with a subcarrier
grouping method to realize both the maximum diversity  and the low
decoding complexity.

Although the above-mentioned work achieves good BER performance, their
transmit rate can not exceed 1 symbol pcu. This limits
the transmission efficiency of MIMO system.
B. A. Sethuraman \textit{et al.} construct a high rate full diversity
STBC using division algebras with high decoding
complexity \cite{Sethuraman2003TIT}. Also, many
work try to construct full rate ($N_t$ symbol pcu) full diversity space time
codes. One approach is to extend the ST-LCP to the full rate full diversity
design \cite{Ma2003TSP}. The key idea is to replace the linear precoder in Figure \ref{ST-LCP}
with a linear complex-field  encoder (LCFE). The LCFE divides
the $N_t^2$ transmit symbols into $N_t$ layers, where $N_t$ is the number
of transmit antenna. In each layer, a closed form linear precoder matrix is used to combine the $N_t$
symbols. Then, the ST-LCP mapper in Figure \ref{ST-LCP}
is replaced with space time LCFE (ST-LCFE)
mapper, where the pre-precessed $N_t^2$ data is mapped into a $N_t \times N_t$ matrix.
Finally, the mapped data is transmitted by $N_t$ antennas
in $N_t$ time slot. In this case, $N_t$ symbol pcu transmission rate
is achieved. It is proved that the ST-LCFE design also obtains
the maximum spatial  diversity of MIMO system. The price of the ST-LCFE is the
exponential increase of the decoding complexity. Therefore,
 both diversity-complexity tradeoff and modulation-complexity
tradeoff schemes are proposed in \cite{Ma2003TSP}. Moreover, the ST-LCFE
design is extended to frequency-selective MIMO-OFDM and
time-selective fading channels in \cite{Ma2003TSP}. Another approach is to
employ threaded algebraic space time (TAST) code \cite{Gamal2003TIT}.
When using TAST code, the size of the
information symbols alphabet is increased
since the degree of the algebraic number field increases.
Therefore, full rate $N_t$ symbol pcu and
the maximum spatial diversity can be achieved simultaneously.
A golden STBC with
non-vanishing determinants  (NVD) property when the signal constellation
size approaches infinity is first constructed for $2\times2$,
$3\times3$, and $6\times6$
antenna systems in \cite{Belfiore2005TIT,Oggier2006TIT}, and extended to arbitrary number of antennas
MIMO systems in \cite{Elia2007TIT}. The NVD property guarantees that the STBC performs
well irrespectively of the transmit signal alphabet size.
The above-mentioned work are some important initial
space-time techniques with discrete input signals  and PCSIR.
These space-time techniques  focus on improving the
transmit diversity, coding gain, and transmission rate of the link.
A brief comparison of these work is given in Table \ref{table:SPT_Initial}.

\begin{table}[!t]
\centering
 \captionstyle{center}
  {
\caption{Some important initial space-time techniques with discrete input signals  and perfect CSIR }
\label{table:SPT_Initial}
\begin{lrbox}{\tablebox}
\begin{tabular}{|c|c|c|c|c|}
\hline
  Paper  &    System  &   Coding gain  & Rate  &  Decoding complexity  \\ \hline
S. M. Alamouti \cite{Alamouti2006JASC}  & $2 \times2 $  &  Not optimal &  1 symbol pcu  & Low \\ \hline
V. Tarokh \textit{et al.} \cite{Tarokh1999TIT} &  MIMO   & Not optimal  &  $< 1$ symbol pcu &  Low    \\ \hline
 S. Zhou \textit{et al.}  \cite{Zhou2003TIT}   &  Single-carrier  MIMO & Not optimal &  $< 1$ symbol pcu    &  High   \\ \hline
Z. Liu  \textit{et al.} \cite{Liu2002TSP}  &   MIMO-OFDM    &  Large  &  $< 1$ symbol pcu   &   Lower than \cite{Zhou2003TIT}   \\ \hline
 H. Jafarkhani   \cite{Jafarkhani2001TIT}  &    $4 \times 4$    &  Not optimal (half diversity)  &  $1$ symbol pcu   &   Reduced \\ \hline
      N. Sharma \textit{et al.} \cite{Sharmai2003TCom}  &  $4 \times 4$   &  Not optimal  &  $1$ symbol pcu   &   Reduced  \\ \hline
           W. Su  \textit{et al.} \cite{Su2004TIT}  &  MIMO  &  Not optimal  &  $\leq 1$ symbol pcu   &   Reduced  \\ \hline
          C. Yuen \textit{et al.} \cite{Yuen2005TWC}  &  MIMO   &  Not optimal  & $\leq 1$ symbol pcu    &   Lower than \cite{Su2004TIT}  \\ \hline
V. Tarokh \textit{et al.} \cite{Tarokh1998TIT}  &  MIMO   &   Large &   $< 1$ symbol pcu    & High    \\ \hline
Y. Xin \textit{et al.} \cite{Xin2003TWC}  &  MIMO   &   Near-optimal &   1 symbol pcu    & High    \\ \hline
Z. Liu \textit{et al.} \cite{Liu2003TCOM}  &  MIMO-OFDM  &   Not optimal &   1  symbol pcu    & Lower than \cite{Zhou2003TIT}      \\ \hline
B. A. Sethuraman \textit{et al.} \cite{Sethuraman2003TIT}  &  MIMO  &   Near-optimal &   $> 1$ symbol pcu    & High     \\ \hline
\multirow{2}{*}{X. Ma \textit{et al.} \cite{Ma2003TSP}}  &  MIMO, MIMO-OFDM &   \multirow{2}{*}{Not optimal} &   \multirow{2}{*}{$N_t$ symbol pcu}    &  \multirow{2}{*}{High}  \\
  &   MIMO time selective channels &  &     &    \\ \hline
H. E. Gammal \textit{et al.} \cite{Gamal2003TIT}  &  MIMO  &   Not optimal &   $N_t$ symbol pcu    &  High  \\ \hline
J.-C. Belfiore \textit{et al.} \cite{Belfiore2005TIT}  &  $2 \times2 $   &   Large (NVD) &   2 symbol pcu    &  High  \\ \hline
F. Oggier \textit{et al.} \cite{Oggier2006TIT}  &  $2 \times2 $, $3 \times 3 $, $6 \times 6 $   &    Large (NVD) &   $N_t$ symbol pcu    &  High  \\ \hline
P. Elia \textit{et al.} \cite{Elia2007TIT}  &  MIMO  &    Large (NVD) &   $N_t$ symbol pcu    &  High  \\ \hline
\end{tabular}
\end{lrbox}
\scalebox{0.95}{\usebox{\tablebox}}
}
\end{table}

For the noncoherent scenario where the CSI is neither available at the transmitter nor at the receiver,
V. Tarokh \textit{et al.} combine differential modulation  with OSTBC to formulate
 transmission schemes, which can still obtain transmit diversity with PSK signals \cite{Tarokh2000JSAC,Jafarkhani2001TIT_2}.
Based on unitary group codes, differential STBCs to achieve the maximum transmit diversity of MIMO systems
are proposed in \cite{Hochwald2000TCom,Hughes2000TIT}, where the optimal modulation scheme for two transmit antennas
is obtained in \cite{Hughes2000TIT}. A double differential STBC is designed for time selective fading channels
in \cite{Liu2001TCom}.  In addition, a differential STBC using QAM signals is constructed in \cite{Hwang2003TSP}.
For the partially coherent scenario where the CSI is not available at the transmitter and only imperfect CSI is available
at the receiver, robust STBCs are designed in \cite{Siwamogsatham2002TSP,Baccarelli2004TSP,Giese2007TIT}. The effects of mapping on
the error performance of the coded STBC have been studied in \cite{Li2005TCom,Tran2007TCom}.
A rate-diversity tradeoff for STBC under finite alphabet input constraints is obtained in \cite{Lu2003TIT} and
the corresponding optimal signal constructions for BPSK and QPSK constellation are provided.
The space-time techniques
have also been extended to MIMO CDMA systems \cite{Hochwald2001JSAC,Derryberry2002CM,Hanzo2003book}.

\subsubsection{Recent Advance for Open-loop Systems}
Recent space-time techniques with discrete input signals  focus on designing high rate full diversity STBCs with NVD and
low decoding complexity. For a multiple input single output (MISO) system equipped with a linear detector, a general criterion for designing
full diversity STBC is established in \cite{Liu2008TIT}. The STBC with Toeplitz structure
satisfies this criterion. Also, the rate of this Toeplitz STBC approaches 1 symbol pcu when the number
of channel use is sufficiently large. For MISO system equipped with ML detector, this Toeplitz STBC
achieves the maximum coding gain.
A new $2\times2$ full rate full diversity STBC is proposed by
reconstructing the generation matrix of OSTBC \cite{Paredes2008TSP}. The proposed STBC reduces the complexity of ML
decoding to the order of the standard sphere decoding.
Another $2\times2$ full rate full diversity STBC is
designed in \cite{Sezginer2009TCom}. During the ML decoding exhaustive search process for this STBC,
 the Euclidean distance can be divided into groups and each group is minimized independently.
Henceforth, this STBC leads to an obvious decoding complexity reduction comparing to
 the Golden code in \cite{Belfiore2005TIT} with a slight performance loss.
 A unified design framework for full rate full diversity $2\times2$
fast-decodable STBC is provided in \cite{Biglieri2009TIT}.
The ML decoding complexity of the designed code
is reduced to the complexity of standard sphere decoding.
For a $4\times2$ system, two QSTBCs are combined to formulate a new 2 pcu symbol rate and  half diversity
STBC with simplified ML decoding complexity \cite{Biglieri2009TIT}. K. P. Srinath \textit{et al.}
construct a $2\times2$ STBC which has the same performance as the Golden code in \cite{Belfiore2005TIT} and a lower
ML decoding complexity for non-rectangular QAM inputs \cite{Srinath2009JSTSP}. K. P. Srinath \textit{et al.} further
construct a  2 pcu symbol rate and full diversity $4\times2$ STBC which offers a large coding gain with a reduced
ML decoding complexity \cite{Srinath2009JSTSP}. H. Wang \textit{et al.} construct a QOSTBC which
experiences the same low decoding complexity as the design in \cite{Yuen2005TWC}, but has
a better coding gain for rectangular QAM inputs \cite{Wang2009TIT}.
K. R. Kumar \textit{et al.} design a STTC for MIMO systems based on lattice coset coding \cite{Kumar2009TIT}. The
constructed code has a good coding gain and
a reduced decoding complexity for large block length
with a DFE and lattice decoding.

In \cite{Karmakar2009TIT}, a general framework of multigroup ML decodable STBC is introduced to reduce the decoding
complexity of MIMO STBC. Multigroup ML decodable STBC is a full diversity STBC.
It allows ML decoding to be implemented for the symbols within each individual group.
As a result, multigroup ML decodable STBC provides a tradeoff between rate
and ML decoding complexity. A specific multigroup ML decodable STBC based on extended Clifford algebras,
named Clifford unitary weight, is constructed to meet the optimal tradeoff between the
rate and the ML decoding complexity in \cite{Rajan2010TIT}.
 Moreover, the structure of fast decodable code
is integrated into the multigroup code to formulate a fast-group-decodable
STBC \cite{Ren10}. A new class of full rate full diversity STBC which exploits the block-orthogonal property of
STBC (BOSTBC) is proposed in \cite{Ren2011JSTSP}. Simulation shows that this BOSTBC achieves a good BER performance
with a reduced decoding complexity when QR decomposition decoder with M paths (QRDM decoder)
is employed at the receiver. Also, this block-orthogonal property can be utilized \cite{Sinnokrot2010TWC} to reduce
the decoding complexity of the Golden code in \cite{Belfiore2005TIT} with ML decoder.
Rigorous theoretical analyses for the block-orthogonal structure of various
STBCs are given in \cite{Jithamithra2014TWC}. Some fast-decodable asymmetric STBCs are designed in \cite{Vehkalahti2012TIT,Markin2013TIT,Srinath2014TIT}.
In particular, R. Vehkalahti \textit{et al.} establish a family of fast-decodable
STBCs from division algebras for multiple-input double-output (MIDO) systems \cite{Vehkalahti2012TIT}. These codes
are full diversity with rate no more than $N_t/2$ symbol pcu.
By properly constructing the geometric structures of these codes, the ML decoding complexity can
by reduced by at least 37.5\%. In addition, explicit constructions
for $4\times 2$, $6\times 2$, and $6\times 3$ codes are given in \cite{Vehkalahti2012TIT}.
N. Markin \textit{et al.} propose an iterative STBC design method
for $N_t \times N_t/2$ MIMO systems \cite{Markin2013TIT}. The iterative design constructs
two $N_t/2 \times N_t/2$ original codewords with  algebraic codewords based on
cyclic algebra to create a new $N_t \times N_t$ codewords. The resulting code achieves $N_t/2$ symbol pcu rate
and full diversity with fast decoding complexity. This iterative
design is used to construct new $4\times2$ and $6\times3$ STBCs in \cite{Markin2013TIT} based on some well known STBCs such as Golden code in \cite{Belfiore2005TIT}.
Moreover, a generalization framework of this iterative design is used in \cite{Srinath2014TIT} to construct 2 symbol pcu rate and full diversity
STBC with large coding gain and fast decoding complexity for MIDO systems. Based on this framework, explicit
 $4\times 2$, $6\times 2$, $8\times 2$, and $12\times 2$ STBCs are constructed in \cite{Srinath2014TIT}.
A brief comparison of above low decoding complexity STBC designs is given in Table \ref{table:SPT_low}.

\begin{table}[!t]
\centering
 \captionstyle{center}
  {
\caption{Recent low decoding complexity space-time techniques with  discrete input signals}
\label{table:SPT_low}
\begin{lrbox}{\tablebox}
\begin{tabular}{|c|c|c|c|c|}
\hline
  Paper  &    System  &   Coding gain  & Rate  &  Decoding complexity  \\ \hline
J. Liu  \textit{et al.} \cite{Liu2008TIT}  &  MISO   &  NVD &  Approach 1 symbol pcu  & Linear \\ \hline
J. M. Paredes \textit{et al.} \cite{Paredes2008TSP}  & $ 2\times2 $  &  NVD &  2 symbol pcu  & Reduced ML \\ \hline
S. Sezginer \textit{et al.} \cite{Sezginer2009TCom}  & $ 2\times2 $  &  Not optimal &  2 symbol pcu  & Reduced ML \\ \hline
\multirow{2}{*}{E. Biglieri \textit{et al.} \cite{Biglieri2009TIT}}  & $ 2\times2 $  &  Not optimal &  2 symbol pcu  & Reduced ML \\
  &  $ 4\times2 $  &  Not optimal (half diversity) &  2 symbol pcu  & Reduced ML \\ \hline
\multirow{2}{*}{K. P. Srinath \textit{et al.} \cite{Srinath2009JSTSP}}  & $ 2\times2 $  &  NVD &  2 symbol pcu  & Lower than Golden code \cite{Belfiore2005TIT} \\
  &  $ 4\times2 $  &  NVD &  2 symbol pcu  & Reduced \\ \hline
H. Wang \textit{et al.} \cite{Wang2009TIT}  & MIMO  &  Larger than \cite{Yuen2005TWC} &  $1 \leq$ symbol pcu  &     Same as  \cite{Yuen2005TWC} \\ \hline
K. R. Kumar \textit{et al.} \cite{Kumar2009TIT}  & MIMO  &  Good &   High  & Reduced ML\\ \hline
S. Karmakar \textit{et al.} \cite{Karmakar2009TIT}  & MIMO  &  Not optimal &   Flexible   & Trade off with rate \\ \hline
G. S. Rajan \textit{et al.} \cite{Rajan2010TIT}  & MIMO  &  Not optimal &   Flexible   & Trade off with rate \\ \hline
T. P. Ren \textit{et al.} \cite{Ren10}   & MIMO  &  Not optimal &   Flexible   & Lower than \cite{Biglieri2009TIT} \\ \hline
T. P. Ren \textit{et al.} \cite{Ren2011JSTSP}   & MIMO  &  Not optimal &    $N_t$ symbol pcu    & Reduced QRDM \\ \hline
M. O. Sinnokrot \textit{et al.} \cite{Sinnokrot2010TWC}   & $2 \times 2$ Golden code \cite{Belfiore2005TIT}  &  NVD &    2 symbol pcu    & Reduced ML \\ \hline
G. R. Jithamithra  \textit{et al.} \cite{Jithamithra2014TWC} & MIMO  &  Classic STBCs &   Flexible    & Reduced SD \\ \hline
R. Vehkalahti  \textit{et al.} \cite{Vehkalahti2012TIT} & MIDO &  NVD &   $N_t/2$ symbol pcu   & Reduced ML \\ \hline
N. Markin  \textit{et al.} \cite{Markin2013TIT} & $N_t \times N_t/2$ &  NVD &   $N_t/2$ symbol pcu    & Reduced ML \\ \hline
\multirow{2}{*}{K. P. Srinath  \textit{et al.} \cite{Srinath2014TIT}} & $4\times2$, $6\times2$  & NVD, large  &   \multirow{2}{*}{2 symbol pcu}   & \multirow{2}{*}{Reduced ML} \\
 & $8\times2$, $12\times2$ & NVD, large &     &  \\ \hline
\end{tabular}
\end{lrbox}
\scalebox{0.9}{\usebox{\tablebox}}
}
\end{table}

Besides reducing the decoding complexity, there are some other space-time techniques recently, e.g.,
a full rate full diversity improved perfect STBC  is designed in \cite{Srinath2013TIT},
which has larger coding gain than the perfect STBC in \cite{Oggier2006TIT}.
A full rate STBC under the BICM structure is designed in \cite{Srinath2013TIT} to optimize the entire components of the
transmitter jointly, including an error-correcting code,
an interleaver, and a symbol mapper.  A new spatial modulation technology which only simultaneously activates
a few number of antennas among the entire antenna arrays is described in \cite{Renzo2011CM,Renzo2013TVT,Kadir2015TCST}.
Since the publication of \cite{Alamouti2006JASC}, space-time techniques
with discrete input signals have been studied extensively around the world \cite{Jafarkhani2005book,Giannakis2007book,Ahmed2015book}.

\subsubsection{Close-loop Systems}
\begin{figure*}[!ht]
\centering
\includegraphics[width=0.8\textwidth]{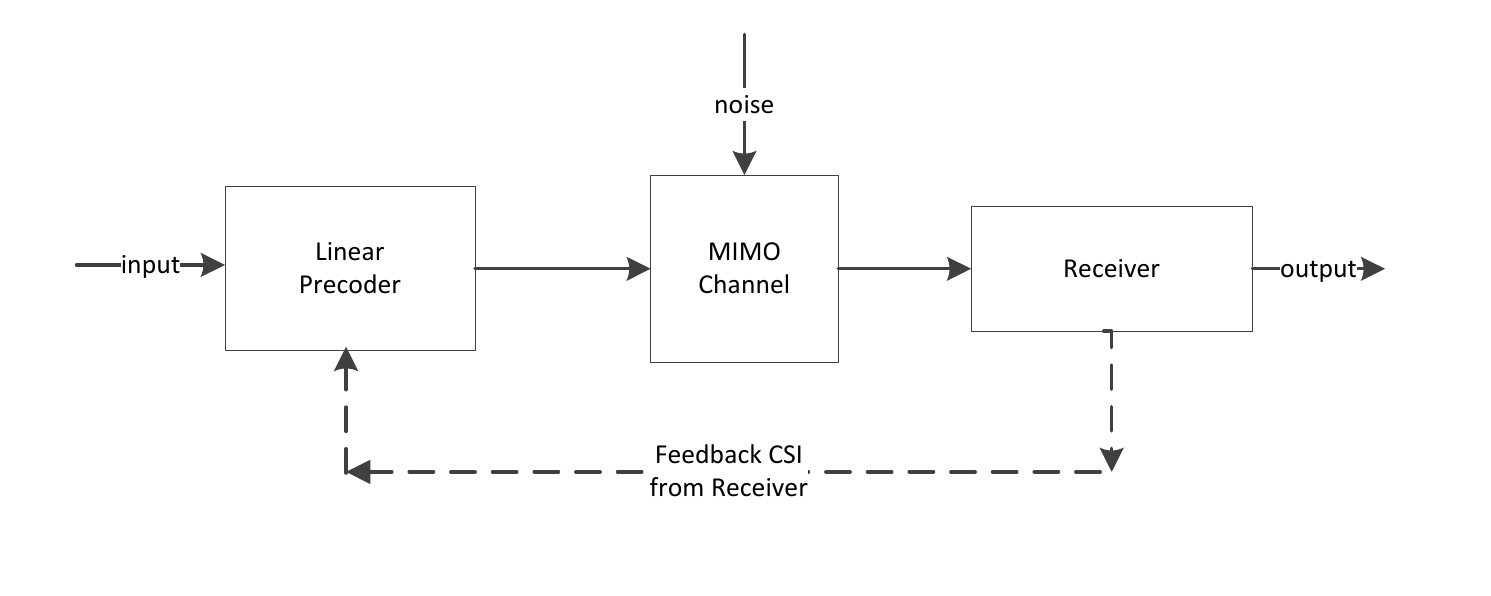}
\caption {\space\space The baseband equivalent framework of close-loop system.}
\label{LP-MIMO}
\end{figure*}

The space-time technologies achieve the transmit diversity without CSI at the transmitter,
which are applicable for open-loop systems. Alternatively, for close-loop MIMO systems in Figure \ref{LP-MIMO},
the CSI is available at the transmitter either by a feedback link from the receiver (frequency division duplex
systems) or a channel estimation
from the pilot signal of the receiver (time division duplex systems). In this case, proper transmission designs can be formulated
to exploit the close loop transmit diversity of MIMO systems with discrete input signals.
For example, a bit interleaved coded multiple beamforming technology is introduced in \cite{Akay2007TWC} to achieve the full diversity
and full spatial multiplexing for MIMO and MIMO-OFDM systems over i.i.d. Rayleigh fading channels.
The decoding complexity of the bit interleaved coded multiple beamforming design is further reduced
in \cite{Li2013TCOM}. An important criterion to enhance the transmit diversity for the close loop MIMO systems
is to maximize the minimum Euclidean distance
of the receive signal points. This criterion
is directly related to the SER of the MIMO system
if the ML detector is used and is proved to achieve the optimal
mutual information performance in high SNR regime  \cite{Perez-Cruz2010TIT}.
For a $2 \times 2$ system with BPSK and QPSK inputs,
an optimal solution for a non-diagonal precoder, which maximizes
the minimum Euclidean distance of the receive signal points (max-$d_{\rm min}$),
is proposed in \cite{Collin2004TSP}. It is shown in \cite{Collin2004TSP} that for BPSK inputs,
the optimal transmission is to focus the power on
the strongest subchannel, which is the beamforming design.
For QPSK inputs, the beamforming design is optimal when
the channel angle is less than a threshold. When the channel angle
is larger than the threshold, the optimal design can be found
by numerically searching the angle of the power allocation matrix.
In \cite{Vrigneau2008IJSTSP}, a suboptimal precoder design for MIMO systems with
an even number of transmit antenna and 4-QAM inputs is proposed.
The authors transform the MIMO channels into parallel subchannels
by the singular value decomposition (SVD). Each two subchannels are paired
into one group to formulate a set of $2\times2$
subsystems. Then, $2\times2$ sub-precoders
are designed similar as in \cite{Collin2004TSP}.
The design in \cite{Vrigneau2008IJSTSP} is denoted as the X-structure design.
It is noted this X-structure design enjoys both reduced
precoding and decoding complexities.
For $M$-QAM inputs, finding the optimal
 max-$d_{\rm min}$ MIMO precoder is rather complicated and the optimal solution
varies for different modulations. Therefore,
Ngo \textit{et al.} propose a suboptimal design by selecting $2\times2$ sub-precoders between
two deterministic structures \cite{Ngo2012TSP}. These sub-precoders
are then combined together to formulate a MIMO precoder based on the X-structure.
The suboptimal design in \cite{Ngo2012TSP} has good uncoded BER performance with a reduced implementation complexity.
An exact expression for the pdf of the minimum Euclidean distance under this suboptimal precoder design
is derived in \cite{Oyedapo2015TWC}. Also, a max-$d_{\rm min}$ MIMO precoder design for a three parallel
data-stream scheme and $M$-QAM inputs is proposed in \cite{Ngo2012TWC}.
This work allows us to decompose arbitrary MIMO channels into
$2\times2$ and $3\times3$ subchannels and then design the max-$d_{\rm min}$ MIMO precoder
efficiently. On the other hand, a suboptimal real-value max-$d_{\rm min}$ MIMO precoder for $M$-QAM inputs
is proposed in \cite{Srinath2011TSP}. The proposed design has a order of magnitude lower ML decoding complexity than
the precoders in \cite{Vrigneau2008IJSTSP,Ngo2012TSP}. Moreover, it is proved in \cite{Srinath2011TSP} that the max-$d_{\rm min}$ precoder achieves full
diversity of MIMO systems. In the mean time, two low decoding complexity precoders derived from
a rotation matrix and the X-structure, called X-precoder and Y-Precoder, are proposed in \cite{Mohammed2011TIT}.
X-precoder has a closed-form expression for $4$-QAM inputs.
Y-precoder has an explicit expression for arbitrary $M$-QAM inputs
but with a worse error performance than X-precoder. Overall,
both X-precoder and Y-Precoder are suboptimal in error performance and loses out
in word error probability when comparing to max-$d_{\rm min}$ precoder \cite{Vrigneau2008IJSTSP,Ngo2012TSP}.
Two suboptimal max-$d_{\rm min}$ MIMO precoder designs for $M$-QAM inputs
by optimizing the diagonal elements of the ``SNR-like" matrix and
the lower bound of the minimum Euclidean distance are proposed in \cite{Ngo2013EURASIP} and \cite{Lin2015TVT}, respectively.

An alternative approach of exploiting the close loop transmit diversity is to design the precoder
based on the lattice structure.
A lattice-based linear MIMO precoder design to minimize the
transmit power under fixed block error rate and fixed total data rate
is proposed in \cite{Bergman2008TSP}. The precoder design is decomposed into
the design of four matrices. The optimal rotation matrix is given in a closed form
structure, the bit load matrix and the basis reduction matrix are optimized alternatingly
via an iterative algorithm, and the lattice base matrix is selected based on a lower bound
of the transmit power. Linear precoders which achieve full
diversity of $N_t \times N_t$ MIMO systems and apply for
integer-forcing receiver are designed in \cite{Sakzad2015TWC} based on full-diversity
lattice generator matrices. Some work also use lattice theory
to design max-$d_{\rm min}$ precoder.
For a $2 \times 2$ system with a square QAM constellation inputs,
the optimal real-value max-$d_{\rm min}$  precoder is found by expanding an ellipse within
a 2-dimensional lattice to maximum extent \cite{Xu2012CLetter}. For MIMO systems with $4$-QAM inputs,
a real-value max-$d_{\rm min}$ precoder is directly constructed based on well-known dense
packing lattices \cite{Xu2014TCOM}. The obtained precoder offers better minimum Euclidean distance and BER performance
than the precoder based on the X-structure  \cite{Vrigneau2008IJSTSP}.
For $N_t \times 2$ systems with infinite signal constellation, an explicit
structure of the max-$d_{\rm min}$ precoder is obtained from
the design of the lattice generator matrix \cite{Kapetanovic2013TWC}. Numerical results
indicate that the obtained precoder achieves good mutual information performance
with large QAM inputs. This precoder design is extended to MIMO systems in \cite{Kapetanovic2015TIT}.
A brief summation of transmission designs with discrete input signals to optimize the transmit diversity for
the close loop MIMO systems is given in Table \ref{table:close-diversity}.

\begin{table}[!t]
\centering
 \captionstyle{center}
  {
\caption{Techniques for the transmit diversity of close loop MIMO systems with discrete input signals}
\label{table:close-diversity}
\begin{lrbox}{\tablebox}
\begin{tabular}{|c|c|c|c|c|c|}
\hline
  Paper  &   System  &   Modulation  &  minimum Euclidean distance  & Design complexity  & Decoding complexity  \\ \hline
 \multirow{2}{*}{E. Akay \textit{et al.}  \cite{Akay2007TWC}}  &  MIMO & \multirow{2}{*}{$M$-QAM}   &  \multirow{2}{*}{Not optimal}  & \multirow{2}{*}{Low} & \multirow{2}{*}{High} \\
     &  MIMO-OFDM  &   &    &  &  \\  \hline
     B. Li  \textit{et al.}  \cite{Li2013TCOM}  &  MIMO-OFDM &  $M$-QAM  &  Not optimal  & Low &  Reduced \\  \hline
L. Collin \textit{et al.}  \cite{Collin2004TSP}  &  $2\times2$ & BPSK,  QPSK  &  Optimal  & Low & Low \\ \hline
B. Vrigneau \textit{et al.} \cite{Vrigneau2008IJSTSP} &  MIMO   & QPSK  &  Near-optimal &  Reduced & Reduced   \\ \hline
Q.-T. Ngo \textit{et al.}  \cite{Ngo2012TSP}   &   MIMO &  $M$-QAM  &  Suboptimal   &  Reduced  & Reduced \\ \hline
Q.-T. Ngo \textit{et al.} \cite{Ngo2012TWC}  &   MIMO    &  $M$-QAM  &  Suboptimal   &  Reduced &  Reduced  \\ \hline
K. P. Srinath \textit{et al.} \cite{Srinath2011TSP}  &  MIMO   &   $M$-QAM &   Smaller than \cite{Vrigneau2008IJSTSP}  & Lower than \cite{Ngo2012TSP} & Lower than \cite{Ngo2012TSP}  \\ \hline
S. K. Mohammed \textit{et al.} \cite{Mohammed2011TIT}  &  MIMO   &  $M$-QAM    &  Not optimal &  Lower than \cite{Ngo2012TSP}   &  Lower than  \cite{Ngo2012TSP}  \\ \hline
Q.-T. Ngo \textit{et al.} \cite{Ngo2013EURASIP}  &  MIMO  & $M$-QAM  &   Larger than \cite{Ngo2012TSP}    &  Reduced  & High \\ \hline
C.-T. Lin \textit{et al.} \cite{Lin2015TVT}  &  MIMO &   $M$-QAM &  Close to \cite{Mohammed2011TIT} &  Lower than \cite{Mohammed2011TIT} & High \\ \hline
S. Bergman \textit{et al.} \cite{Bergman2008TSP}  & MIMO &   $M$-QAM &    Not optimal & Reduced & High \\ \hline
A. Sakzad \textit{et al.} \cite{Sakzad2015TWC}  &  $N_t \times N_t$ &   $M$-QAM &   Not optimal & Reduced & Reduced \\ \hline
X. Xu \textit{et al.} \cite{Xu2012CLetter}  &  $2\times2$ &   Square $M$-QAM &    Optimal &  Low & Low \\ \hline
X. Xu \textit{et al.} \cite{Xu2014TCOM}  & MIMO   &   QPSK  &  Larger than \cite{Ngo2012TSP}   &  Reduced & High \\ \hline
D. Kapetanovi\'c \textit{et al.} \cite{Kapetanovic2013TWC}  & $N_t \times 2$   &   Infinite Constellation   &  Asymptotic optimal  &  Low & Low \\ \hline
D. Kapetanovi\'c \textit{et al.} \cite{Kapetanovic2015TIT}  & MIMO   &   Infinite Constellation   &  Asymptotic optimal  &  Reduced & High \\ \hline
\end{tabular}
\end{lrbox}
\scalebox{0.9}{\usebox{\tablebox}}
}
\end{table}

\begin{figure*}[!ht]
\centering
\includegraphics[width=0.8\textwidth]{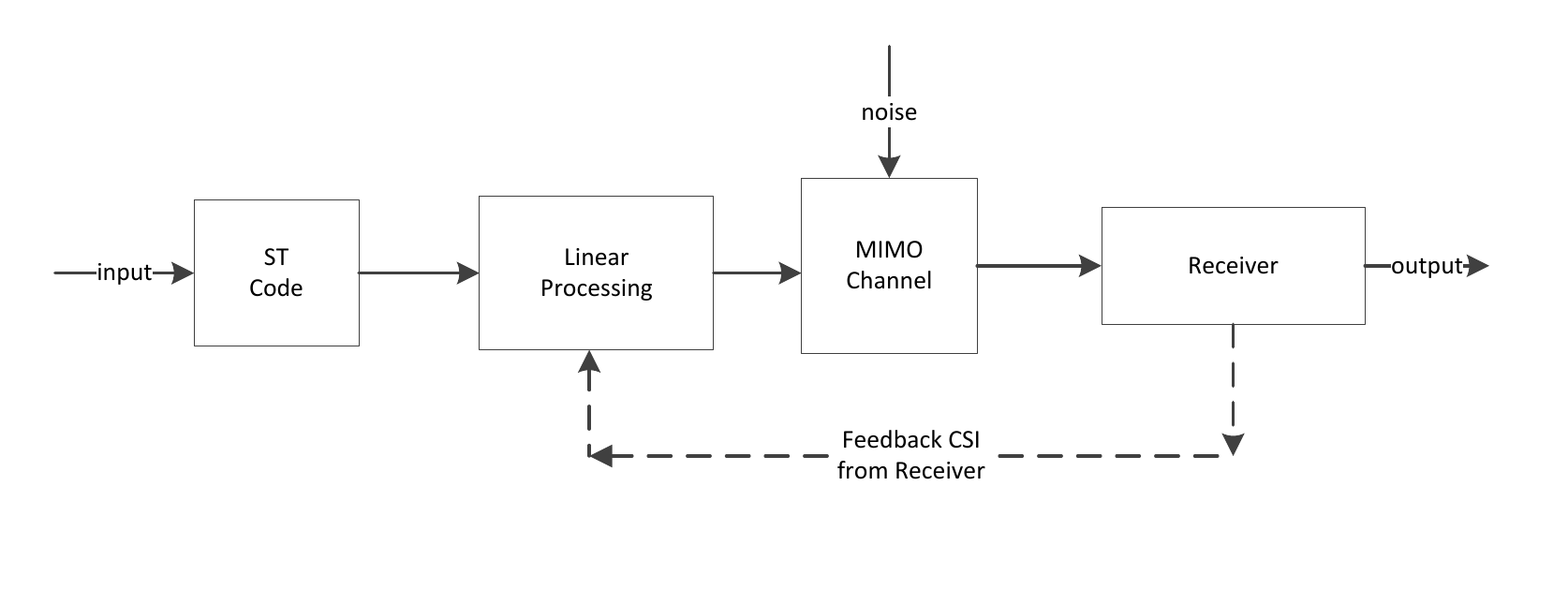}
\caption {\space\space A close loop  STBC system}
\label{LP-STC-MIMO}
\end{figure*}

Some work combine the STBC with the linear processing step in close loop system to further improve the
performance of MIMO systems. The structure of a close loop STBC systems is illustrated in Figure \ref{LP-STC-MIMO}.
It is important to note that the main difference between the linear processing in Figure \ref{ST-LCP} and the linear precoder in Figure
\ref{LP-STC-MIMO} is the usage of CSI. The linear precoder in Figure \ref{ST-LCP} does not need any knowledge of CSI
and the linear processing in Figure \ref{LP-STC-MIMO} requires some forms of the CSI.
With PCSIT, H. J. Park \textit{et al.} exploit the constellation precoding scheme in \cite{Xin2003TWC}
to design a full-diversity multiple beamforming scheme for uncoded systems \cite{Park2011TCOM}.
For the cases of both PCSTT and SCSIT, D. A. Gore \textit{et al.} investigate antenna selection techniques
to minimize the SER for MIMO systems employing OSTBC \cite{Gore2002TSP}.
Jongren \textit{et al.} combine
the beamforming design and the OSTBC with partial
knowledge of the channel over frequency-nonselective fading channel \cite{Jongren2002TIT}.
The linear precoder design for a STBC system over a transmit correlated Rayleigh fading channel
is proposed in \cite{Sampath2002Cletter}. The transmitter utilizes the transmit correlation matrix to design  a linear precoding matrix,
which minimizes an upper bound of the average PEP of the STBC system.
This design is extended to transmit correlated Rician fading channel and arbitrary correlated Rician fading channel
in \cite{Vu2006TSP} and in \cite{Bhatnagar2010TWC}, respectively. Linear precoding for a limited feedback STBC systems is
proposed in \cite{Wang2013TVT}. For the noncoherent scenario where the CSI is not available at the receiver,
linear precoder designs for differential STBC systems over Kronecker fading channel and arbitrary correlated Rayleigh fading channels
are provided in \cite{Cai2006TWC} and \cite{Bhatnagar2009TWC}, respectively. For the partially coherent
scenario where the imperfect channel estimation is performed at the receiver and the estimation error covariance matrix is perfectly known at
the transmitter via feedback, STBC designs are proposed in \cite{Borran2009TWC} and \cite{Yadav2012TCOM} for the i.i.d. Rayleigh fading channels based on
criterions of Kullback-Leibler distance and the cutoff rate, respectively.
The design in \cite{Yadav2013TCOM} is extended to the correlated Rayleigh fading channels where the transmit and receive correlation matrices
are perfectly known at the transmitter. Moreover, a novel and efficient bit mapping scheme is proposed in \cite{Yadav2013TCOM}.
A linear precoder is further designed in \cite{Yadav2014TWC} to adapt to the degradation caused by the imperfect CSIR.

\subsection{Adaptive MIMO Transmission}
The above-mentioned multiplexing, hybrid, and
diversity  transmission schemes can be employed
adaptively according to the channel condition.
Adaptive MIMO transmission is an effective
approach to increase the data rate and decrease
the error rate for wireless communication systems \cite{Catreux2002TWC,Catreux2002TCM,Chae2010TCM}.
Adaptive MIMO transmission techniques are investigated
for both uncoded and coded systems.

\subsubsection{Uncoded Systems}
\begin{figure*}[!ht]
\centering
\includegraphics[width=0.8\textwidth]{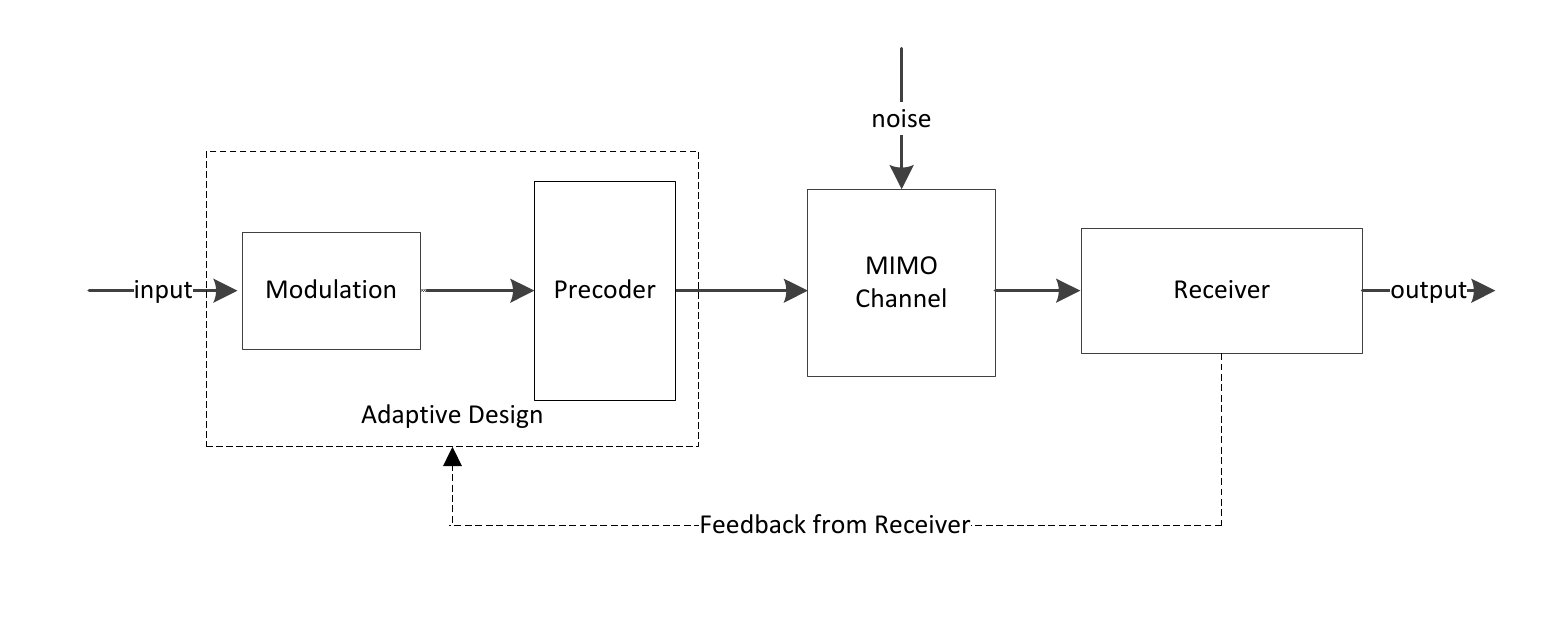}
\caption {\space\space Adaptive MIMO transmission for uncoded systems}
\label{AT-MIMO-UNCODE}
\end{figure*}

The capacity-achieving coding usually has long (infinitely) codewords
and brings delay to the system. Therefore, for the delay-sensitive applications,
the modulation part of physical layer is usually designed separately before
the outer error-correcting coding is applied, as shown in Figure \ref{AT-MIMO-UNCODE}.
For the uncoded SISO system subject to
the average power and the BER (instantaneous or average) constraints, the maximization
of the spectral efficiency is obtained by jointly optimizing the transmit rate
and power over Rayleigh fading channels in \cite{Goldsmith1997TCOM,Chung2001TCOM} and Nakagmi fading channels in \cite{Alouini2000KJWC},
developing upon the previous adaptive transmission techniques \cite{Hayes1968,Cavers1972,Otsuki1995EL,Webb1995TCOM,Kamio1995,Vucetic1991TCOM,Alamouti1994TCOM,Ue1995,Goeckel1999TCOM,Hole2000JSAC,Hu2001}.
Z. Zhou \textit{et al.} investigate adaptive transmission schemes
for  i.i.d.  Rayleigh flat fading MIMO channels with perfect or imperfect CSIT and CSIR \cite{Zhou2005TVT}.
With perfect CSIT and CSIR, the
MIMO channels are decomposed into a set of parallel SISO subchannels by
SVD.  Then, the transmit rate and power allocation of each subchannel
 need to be optimally designed to maximize the average spectral efficiency
 under the average transmit power and instantaneous subchannel BER
 constraints. It is proved that this optimization problem can be
 transformed into several subproblems, each of which only involves one
 subchannel. Then, the adaptive transmission designs for SISO channels
 in \cite{Goldsmith1997TCOM,Chung2001TCOM} can be exploited to find the optimal solution. Moreover, the effect
 of the discrete rate constraint and the imperfect CSIT on system performance
 are evaluated in \cite{Zhou2005TVT}. When only partial CSI, i.e., the channel mean feeback,
 is available at the transmitter, S. Zhou \textit{et al.} provide an adaptive two-dimensional beamformer
  to exploit the transmit diversity by incorporating the Alamouti coded \cite{Zhou2004TWC}.
Based on this beamformer, the beamvectors, the power allocation policy,
and the modulation schemes are designed  for MIMO and MIMO-OFDM systems in \cite{Zhou2004TWC} and \cite{Xia2004TSP}, respectively,
which maximize the average discrete spectral efficiency under the constant power and the average
BER constraints. To further study  the impact of the CSI error  on adaptive MIMO transmission,
S. Zhou \textit{et al.} consider the case where the transmitter performs
 a pilot signal assisted channel prediction and examine the effect of
 the prediction error on the BER performance for a one-dimensional adaptive beamformer \cite{Zhou2004TWC_2}.
Numerical results show that when the normalized estimation error is below a
critical threshold, the channel prediction error has a minor impact
on the BER performance. For correlated MISO  Rayleigh fading channels with a limited
number of feedback bits at the transmitter, P. Xia \textit{et al.}
develop a strategy to adaptively select the transmission mode to maximize
the average spectral efficiency under the average power and the average BER constraints \cite{Xia2005TCOM}.
For correlated MIMO  fading channels where the correlation matrices are available at the transmitter
and equal power is allocated to each transmit antenna, a strategy to determine the number
of active transmit antenna and the corresponding transmit constellations is developed
by maximizing a low bound of the minimum SNR margin with a fixed data rate \cite{Narasimhan2003TSP}.
R. W. Heath Jr \textit{et al.} propose a simple MIMO adaptive transmission
scheme by switching between spatial multiplexing and transmit diversity \cite{Heath2005TCOM}.
With perfect CSIR, both the minimum Euclidean distances of spatial multiplexing
scheme and STBC scheme which achieve a fixed transmit rate are computed at the receiver.
The scheme with larger minimum Euclidean distances
is chosen and the decision is feedback to the transmitter via a low-rate feedback channel.
For double-input multiple-output (DIMO) systems with both PCSIT and SCSIT, an adaptive scheme switching between OSTBC and
spatial multiplexing with zero-forcing (ZF) receiver based on the average spectral efficiency
is proposed in \cite{Huang2009TWC}. For an adaptive MIMO-OFDM system employing maximum ratio transmission (MRT) and MRC transceiver structure, the statistic properties of the number of bits transmitted per
OFDM block and the number of outage per OFDM block over i.i.d. Rayleigh fading channels
are derived in \cite{Kongara2008}.  A brief comparison of above adaptive MIMO transmitter
designs is given in Table \ref{table:Adaptive-Transmitter}.

\begin{table}[!t]
\centering
 \captionstyle{center}
  {
\caption{Adaptive MIMO transmitter designs for uncoded systems}
\label{table:Adaptive-Transmitter}
\begin{lrbox}{\tablebox}
\begin{tabular}{|c|c|c|c|c|}
\hline
  Paper  &    System  &  Criterion  & CSI &  Constraint \\ \hline
\multirow{2}{*}{Z. Zhou \textit{et al.} \cite{Zhou2005TVT}}  &  MIMO   &  Maximize average &  PCSIT, IPCSIT & Instantaneous BER \\
 &  i.i.d. Rayleigh fading   &  spectral efficiency  &  PCSIR & Average transmit power \\ \hline
 \multirow{2}{*}{S. Zhou \textit{et al.} \cite{Zhou2004TWC}}  &  MIMO   &  Maximize average &  Channel mean feedback & Average BER \\
 &  i.i.d. Rician fading   &  spectral efficiency  &  PCSIR &  Constant transmit power \\ \hline
  \multirow{2}{*}{P. Xia \textit{et al.} \cite{Xia2004TSP}}  &  MIMO-OFDM   &  Maximize average &  Channel mean feedback & Average BER \\
 &  i.i.d. Rician fading   &  spectral efficiency  &  PCSIR &  Constant transmit power \\ \hline
   \multirow{2}{*}{S. Zhou \textit{et al.} \cite{Zhou2004TWC_2}}  &  MIMO   &  Maximize average &   \multirow{2}{*}{Channel prediction error} & Instantaneous BER \\
 &  i.i.d. Rayleigh fading   &  spectral efficiency   &   &  Constant transmit power \\ \hline
    \multirow{2}{*}{P. Xia \textit{et al.} \cite{Xia2005TCOM}}  &  MISO   &  Maximize average &  Limited feedback & Average BER \\
 &  i.i.d. Rayleigh fading   &  spectral efficiency   &  PCSIR &  Average transmit power \\ \hline
     \multirow{2}{*}{R. Narasimhan \cite{Narasimhan2003TSP}}  &  MIMO   &  Maximize minimal  &  SCSIT & Fixed data rate \\
 &  correlated Rayleigh fading   &  SNR margin  &  PCSIR &  Equal power allocation \\ \hline
      \multirow{2}{*}{R. W. Heath Jr \cite{Heath2005TCOM} } &   \multirow{2}{*}{MIMO}   &  Spatial Multiplexing  &  Feedback decision &  \multirow{2}{*}{Fixed data rate} \\
 &  & OSTBC switch    &  PCSIR &  \\ \hline
       \multirow{2}{*}{J. Huang \cite{Huang2009TWC} } &   DIMO   &  Spatial Multiplexing  &  PCSIT, SCSIT &  \multirow{2}{*}{ZF receiver} \\
 & correlated Rayleigh fading  & OSTBC switch    &  PCSIR &  \\ \hline
        \multirow{2}{*}{K. P. Kongara \cite{Kongara2008} } &   MIMO-OFDM   &   \multirow{2}{*}{Performance Analysis}   &  PCSIT & MRT \\
 & i.i.d. Rayleigh fading  &    &  PCSIR &  MRC \\ \hline
\end{tabular}
\end{lrbox}
\scalebox{1}{\usebox{\tablebox}}
}
\end{table}

Some work study the adaptive MIMO transceiver designs. For perfect CSIT
and CSIR, D. P. Palomar \textit{et al.} investigate the transmit constellation
selection and the linear transceiver design to minimize the transmit power
with fixed transmit rate, where each subchannel is under the same
BER constraint \cite{Palomar2005TSP_Ada}.  Assuming the symbol error probability of each subchannel
is the same, the gap approximation method is used to select the transmit
constellations.  Then, necessary and sufficient conditions are
established to examine the optimality of  the subchannel diagonalization design.
Moreover,  an upper bound of the transmit power loss incurred by using the subchannel diagonalization design
is presented. On the other hand, an optimal bit loading scheme among subchannels
is developed in \cite{Yasotharan2005TCOM}. The proposed scheme minimizes a lower
bound of the average BER of the system subject to the fixed transmit data, constant transmit power, and
ZF transceiver structure constraints.
S. Bergman \textit{et al.} further obtain the optimal transceiver design and
the bit loading scheme among subchannels with DFE receiver \cite{Bergman2009TSP}.
The p-norms of weighed MSEs is minimized under the
fixed transmit data and the constant transmit power constraints.
The subchannel diagonalization design is proved to be always
optimal for the system model considered in \cite{Bergman2009TSP}. C.-C. Li \textit{et al.}
establish a duality relationship between
the joint linear transceiver and bit loading design of maximizing the transmit rate
and that of minimizing the transmit power in \cite{Li2011TSP}. Then, an
algorithm can be developed to find the optimal design for the transmit
rate maximization by using the solution of transmit power minimization problem.
A brief comparison of above adaptive MIMO transceiver
designs is given in Table \ref{table:Adaptive-Tranceiver}.

\begin{table}[!t]
\centering
 \captionstyle{center}
  {
\caption{Adaptive MIMO transceiver designs for uncoded systems}
\label{table:Adaptive-Tranceiver}
\begin{lrbox}{\tablebox}
\begin{tabular}{|c|c|c|c|c|}
\hline
  Paper  &    System  &  Criterion  & CSI &  Constraint \\ \hline
\multirow{2}{*}{D. P. Palomar \textit{et al.} \cite{Palomar2005TSP_Ada}}  &  \multirow{2}{*}{MIMO}   &  Minimize  &  PCSIT  & Instantaneous BER \\
 &  &  transmit power &  PCSIR & Fixed transmit rate \\ \hline
 \multirow{2}{*}{A. Yasotharan \textit{et al.} \cite{Yasotharan2005TCOM}}  &  \multirow{2}{*}{MIMO}   &  Minimize   &  PCSIT  & ZF transceiver, constant transmit power \\
 &  &  average BER &  PCSIR & Fixed transmit rate \\ \hline
  \multirow{2}{*}{S. Bergman \textit{et al.} \cite{Bergman2009TSP}}  &  \multirow{2}{*}{MIMO}   &  Minimize   &  PCSIT  & DFE receiver, constant transmit power  \\
 &  &  weighted MSE &  PCSIR & Fixed transmit rate \\ \hline
   \multirow{2}{*}{C.-C. Li \textit{et al.} \cite{Li2011TSP}}  &  \multirow{2}{*}{MIMO}   &  Maximize  &  PCSIT  & Instantaneous BER    \\
 &  &  transmit rate &  PCSIR &  Constant transmit power \\ \hline
\end{tabular}
\end{lrbox}
\scalebox{1}{\usebox{\tablebox}}
}
\end{table}

With perfect CSIR, Y. Ko \textit{et al.} propose a rate adaptive modulation scheme
combined with OSTBC for i.i.d. Rayleigh fading MIMO channels \cite{Ko2006TWC}.
Close-form expressions for the average spectral efficiency
and a tight upper bound of the average BER in presence of feedback delay are derived.
Then, optimal SNR switching thresholds for adaptive modulation,
which maximize the average spectral efficiency under the average BER, the outage
BER, and the constant transmit power constraints, are designed at the receiver.
The constellation chosen at the receiver is feeded back to the transmitter and
the impact of feedback delay on the average BER, throughout, and outage probability
are analyzed.  The design in \cite{Ko2006TWC} is extended to spatially correlated
Rayleigh fading channels in \cite{Huang2009TVT}, where a low complexity method to determine a
set of near optimal SNR switching thresholds is provided. On the other
hand, the design in \cite{Ko2006TWC} is also extended to the case with the channel
estimation error and the average transmit power constraint in \cite{Yu2009TVT}.
For MIMO-OSTBC systems, the channel estimation process
of the rate adaptive modulation scheme is investigated \cite{Duong2007TWC}.
The transmit constellation and the allocation of the time and power between
the pilot and data symbols are optimally designed to maximize the average spectral
efficiency under the instantaneous BER constraint.
The adaptive MIMO transmission
scheme for the energy efficiency maximization of MIMO-OSTBC systems with the channel estimation error  subject
to an instantaneous BER constraint is proposed in \cite{Chen2013TVT}.
By utilizing the imperfect CSIT,  Q. Kuang \textit{et al.} incorporate the design of spatial power allocation
between different transmit antennas into the adaptive modulation scheme in \cite{Yu2009TVT}
to further increase the average spectral efficiency \cite{Kuang2012TCOM}. A brief comparison
of above adaptive  transmission designs for MIMO-OSTBC systems is given in Table \ref{table:Adaptive-Tranceiver}.

\begin{table}[!t]
\centering
 \captionstyle{center}
  {
\caption{Adaptive transmission designs for MIMO-OSTBC uncoded systems}
\label{table:Adaptive-Tranceiver}
\begin{lrbox}{\tablebox}
\begin{tabular}{|c|c|c|c|c|}
\hline
  Paper  &    System  &  Criterion  & CSI &  Constraint \\ \hline
\multirow{2}{*}{Y. Ko \textit{et al.} \cite{Ko2006TWC}}  &  MIMO   &  Maximize average &  Feedback decision with delay  & Average BER, outage BER \\
 & i.i.d. Rayleigh fading  &  spectral efficiency  &  PCSIR &  Constant transmit power \\ \hline
 \multirow{2}{*}{J. Huang \textit{et al.} \cite{Huang2009TVT}}  &  MIMO   &  Maximize average &  Perfect channel estimation & Instantaneous BER, average BER  \\
 & correlated Rayleigh fading  & spectral efficiency  & Imperfect channel estimation   &  Constant transmit power \\ \hline
  \multirow{2}{*}{X. Yu \textit{et al.} \cite{Yu2009TVT}}  &  MIMO   &  Maximize average &  \multirow{2}{*}{Imperfect channel estimation}  & Average BER  \\
 & i.i.d. Rayleigh fading  & spectral efficiency  &   &  Average  transmit power \\ \hline
   \multirow{2}{*}{D. V. Duong \textit{et al.} \cite{Duong2007TWC}}  &  MIMO   &  Maximize average &  \multirow{2}{*}{Channel estimation optimization}  & Instantaneous  BER  \\
 & i.i.d. Rayleigh fading  & spectral efficiency  &   &  Equal power allocation \\ \hline
    \multirow{2}{*}{L. Chen \textit{et al.} \cite{Chen2013TVT}}  &  MIMO   &  Maximize  &  IPCSIT  & Instantaneous  BER  \\
 & i.i.d. Rayleigh fading  & energy  efficiency  &  PCSIR &  Equal power allocation \\ \hline
     \multirow{2}{*}{Q. Kuang \textit{et al.} \cite{Kuang2012TCOM}}  &  MIMO   &  Maximize average  &  IPCSIT  & Average  BER  \\
 & i.i.d. Rayleigh fading  &  spectral   efficiency  &  PCSIR &  Equal power allocation \\ \hline
\end{tabular}
\end{lrbox}
\scalebox{0.9}{\usebox{\tablebox}}
}
\end{table}

Other adaptive MIMO transmission schemes with outdated CSIT and imperfect CSIT are investigated in \cite{Zhou2005}
and \cite{Paris2006,Wang2006CLetter}, respectively. X. Zhang \textit{et al.} and T. Delamotte  \textit{et al.} consider the adaptive power allocation for each subchannel with the instantaneous BER constraint based on the modified
water-filling and mercury water-filling algorithms in \cite{Zhang2003} and \cite{Delamotte2013}, respectively.
T. Haustein \textit{et al.} develop an adaptive power allocation strategy for each subchannel
by minimizing the MSEs at the receiver under the maximum average BER constraint \cite{Rahman2007}.
Some adaptive MIMO transmission schemes by providing the reconfigurable operation modes of the
antenna arrays are discussed in \cite{Cetiner2004CM,Cetiner2006ALetter}.

\subsubsection{Coded Systems}
The capacity expression in \cite{Telatar1999} has provided a fundamental transmission limit for MIMO system.
Since then, adaptive modulation and coding (AMC) schemes have been extensively investigated to approach
this performance limit. M. R. McKay \textit{et al.} consider the AMC for the correlated Rayleigh fading MIMO channels \cite{McKay2007TVT}.
The BICM is used for the coding. With SCSIT,
two simple transmission schemes are switched: statistical beamforming
or spatial multiplexing with a zero-forcing receiver. Then,
the transmitter selects the best combination of the code rate,
the modulation formate, and the transmission scheme to maximize
the data rate subject to a general BER
union bound. Adaptive DIMO switching  between the spatial multiplexing
and STBC based on the spectral efficiency maximization
for a non-selective channel with practical constraints, impairments, and
receiver designs is investigated
in \cite{Choi2008JSAC}. For spatial multiplexing or STBC,
the spectral efficiency thresholds to select the best coding rate and the modulation format
is determined based on the packet error rate (PER) vs. instantaneous capacity
simulation curves (convolutional code).
The selection is performed at the receiver with perfect CSIR and then
feeds back to the transmitter. Also, for a time- and frequency-selective channel, a new physical layer  abstraction
and switching algorithm is proposed in \cite{Choi2008JSAC}, where a weighed sum
of channel qualities is minimized to reduce the variance of the channel qualities.
P. H. Tan \textit{et al.} investigate the AMC
for MIMO-OFDM systems in presence of channel estimation errors
in slow fading channels \cite{Tan2008JSAC}. The convolutional or low density parity check codes (LDPC) coded BER
at the output of the detector is evaluated via curving
fitting method. The packet error rate (PER) for each modulation and coding scheme
is accurately approximated based on the obtained coded BER expression. Then, the code rate, the modulation format,
and the number of transmit streams are selected by maximizing the system
throughout under a fixed PER constraint. A supervised learning algorithm, called $k$-nearest neighbor, is
used to select the AMC parameters by maximizing the data rate under frame error rate (FER)
constraints for MIMO-OFDM systems with convolutional codes in  frequency-
and spatial-selective channels \cite{Daniels2010TVT}. Assuming perfect CSIT and CSIR, M. D. Dorrance \textit{et al.}
consider the AMC for a vertical bell laboratories layer space-time (V-BLAST) MIMO system with rate-compatible LDPC codes and an incremental
redundancy hybrid automatic repeat request (H-ARQ) scheme \cite{Dorrance2010TCOM}. A constrained water-filling algorithm for finite alphabet
input signals is provided to allocate power for each subchannel. Then, the modulation format used for each subchannel is
determined  based on a look-up table. Z. Zhou \textit{et al.} incorporate the BICM into the adaptive transmission scheme
in \cite{Zhou2005TVT} to improve the BER robustness against the CSI feedback delay in \cite{Zhou2011TWC}. Moreover, BICM technology is
modified by using a multilevel puncturing and interleaving technique and combined with rate-compatible
punctured code/turbo code to achieve a near full multiplexing gain.
An AMC strategy for BICM MIMO-OFDM systems with soft Viterbi decoding is
proposed in \cite{Stupia2012TSP}. A system performance metric called the expect goodput is defined in \cite{Stupia2012TSP},
which is a function of the code rate, the modulation format,
the subchannel power allocation, and the space-time-frequency precoder.
The transmitter determines these AMC parameters by maximizing the expect goodput
via the feedback of the perfect CSI and the effective SNR from the receiver.
Spectral efficiencies of the adaptive MIMO transmission schemes in real-time practical
test-bed are shown in \cite{Haustein2006EURASIP}. A brief comparison
of above AMC schemes for MIMO systems is given in Table \ref{table:AMC-MIMO}.

\begin{table}[!t]
\centering
 \captionstyle{center}
  {
\caption{AMC schemes for MIMO systems}
\label{table:AMC-MIMO}
\begin{lrbox}{\tablebox}
\begin{tabular}{|c|c|c|c|c|}
\hline
  Paper  &    System  &  Criterion  & CSI &  Constraint \\ \hline
\multirow{2}{*}{M. R. McKay \textit{et al.} \cite{McKay2007TVT}}  &  MIMO   &  Statistical beamforming &  SCSIT & \multirow{2}{*}{BER union bound} \\
 & correlated  Rayleigh fading  & spatial multiplexing switch   &  PCSIR &   \\ \hline
 \multirow{2}{*}{Y.-S. Choi \textit{et al.} \cite{Choi2008JSAC}}  &  DIMO, non-selective   &  Spatial multiplexing &  Feedback decision & \multirow{2}{*}{Practical PER} \\
 & time- and frequency-selective  & STBC switch   &  PCSIR &   \\ \hline
  \multirow{2}{*}{P. H. Tan \textit{et al.} \cite{Tan2008JSAC}}  &  MIMO-OFDM   &  Maximize  & \multirow{2}{*}{Channel estimation error}  & \multirow{2}{*}{Fixed PER} \\
 & slow fading & system throughout   &   &   \\ \hline
   \multirow{2}{*}{R. C. Daniels \textit{et al.} \cite{Daniels2010TVT}}  &  MIMO-OFDM   & Maximize average  & PCSIT & \multirow{2}{*}{Fixed FER} \\
 & frequency-
and spatial-selective &  spectral efficiency     & PCSIR  &   \\ \hline
   \multirow{2}{*}{M. D. Dorrance\textit{et al.} \cite{Dorrance2010TCOM}}  & V-BLAST MIMO  &  Maximize average  & PCSIT & Rate-constrained  \\
 & H-ARQ &  spectral efficiency   &  PCSIR &  power, bit allocation  \\ \hline
 \multirow{2}{*}{Z. Zhou \textit{et al.} \cite{Zhou2011TWC}}  &  MIMO   &  Maximize average &   IPCSIT & Instantaneous BER \\
 &  i.i.d. Rayleigh fading   &  spectral efficiency  &  PCSIR & Average transmit power \\ \hline
  \multirow{2}{*}{I. Stupia \textit{et al.} \cite{Stupia2012TSP}}  &    \multirow{2}{*}{MIMO-OFDM}   &  Maximize  &   PCSIT & \multirow{2}{*}{Soft Viterbi decoding} \\
 &   &  expect goodput  &  PCSIR & \\ \hline
   \multirow{2}{*}{T. Haustein \textit{et al.} \cite{Haustein2006EURASIP}}  &   Practical MIMO   &  Maximize average  & Channel estimation   & \multirow{2}{*}{Average BER} \\
 &  test bed &  spectral efficiency &  in practical systems & \\ \hline
 \end{tabular}
\end{lrbox}
\scalebox{0.9}{\usebox{\tablebox}}
}
\end{table}

The current wireless standards such as 3GPP LTE, IEEE 802.16e,
and IEEE 802.11n define concrete implementation procedures
for adaptive MIMO transmission in practice.
We take the downlink transmission in LTE standard as an example.
The code rate and modulation format are selected based on the received Channel Quality
Indicator (CQI) user feedback. For example, the $4$-bit CQI indices and their
 interpretations in LTE \cite{3GPPV880} are given in Table \ref{table:AM-LTE}.
 Then, the modulated transmit symbols are mapped into one or several layers.
This antenna mapping depends on the Rank Indicator user feedback, which includes a
single antenna port, transmit diversity, and spital multiplexing \cite[Table 7.2.3-0]{3GPPV880}. The precoding matrices
for these transmission modes depend on the Precoding Matrix Indicator user feedback, which are
defined in details in \cite[Sec. 6.3.4]{3GPPV900}.  The adaptive MIMO transmission schemes
 in IEEE 802.16e and IEEE 802.11n are similar.
 In IEEE 802.16e,  52 combinations
 of codes, code rate, and modulation format are defined as in \cite[Table 8.4]{Andrews2007book}.
 The transmission mapping matrices includes \cite[Eq. (8.7)]{Andrews2007book}: Matrix A, exploiting only diversity;
 Matrix B, combining diversity and spatial multiplexing; Matrix C, employing spatial multiplexing.
 Then, a transmission mode is selected based on the link quality evaluation.
In IEEE 802.11n,  16 mandatory modulation and coding schemes are defined \cite[Table 1]{Nee2006WPC}.
A spatial multiplexing matrix is applied to convert the transmitted data
streams into the transmit antennas and an additional cyclic deay can be applied
per transmitter to provide transmit cyclic delay diversity \cite[Figure 1]{Nee2006WPC}.
Some practical AMC simulation results in these standards are shown in \cite{Daniels2010TVT,Nee2006WPC,Chae2010TCM,Peng2007,Mogensen2007,Marabiss2008,Mehlfuhrer2008EURASIP,Mehlfuhrer2009,Guo2013}.

\begin{table}[!t]
\centering
 \captionstyle{center}
  {
\caption{$4$ bit CQI table in LTE standard}
\label{table:AM-LTE}
\begin{lrbox}{\tablebox}
\begin{tabular}{|c|c|c|c|}
\hline
  CQI index  &    Modulation   &  Code rate $\times$ 1024  & Efficiency  \\ \hline
  0  &       \multicolumn{3}{|c|}{out of range}   \\ \hline
  1 &    QPSK   &  78  & 0.1523  \\ \hline
 2  &    QPSK   &  120  & 0.2344  \\ \hline
3   &    QPSK   &  193 & 0.3770  \\ \hline
 4  &    QPSK   &  308 & 0.6016 \\ \hline
  5  &    QPSK   &  449 & 0.8770  \\ \hline
 6 &    QPSK   &  602 & 1.1758  \\ \hline
 7  &    16QAM   &  378  & 1.4766  \\ \hline
  8  &    16QAM   & 490 & 1.9141  \\ \hline
9  &    16QAM   &  616 & 2.4063  \\ \hline
10  &    64QAM   &  466  & 2.7305  \\ \hline
11  &    64QAM   &   567 & 3.3223  \\ \hline
12  &    64QAM   &  666 & 3.9023  \\ \hline
13  &    64QAM   & 772  & 4.5234  \\ \hline
14  &    64QAM   & 873  & 5.1152  \\ \hline
15  &    64QAM   &  948  & 5.5547  \\ \hline
\end{tabular}
\end{lrbox}
\scalebox{1}{\usebox{\tablebox}}
}
\end{table}

\begin{figure*}[!ht]
\centering
\includegraphics[width=0.8\textwidth]{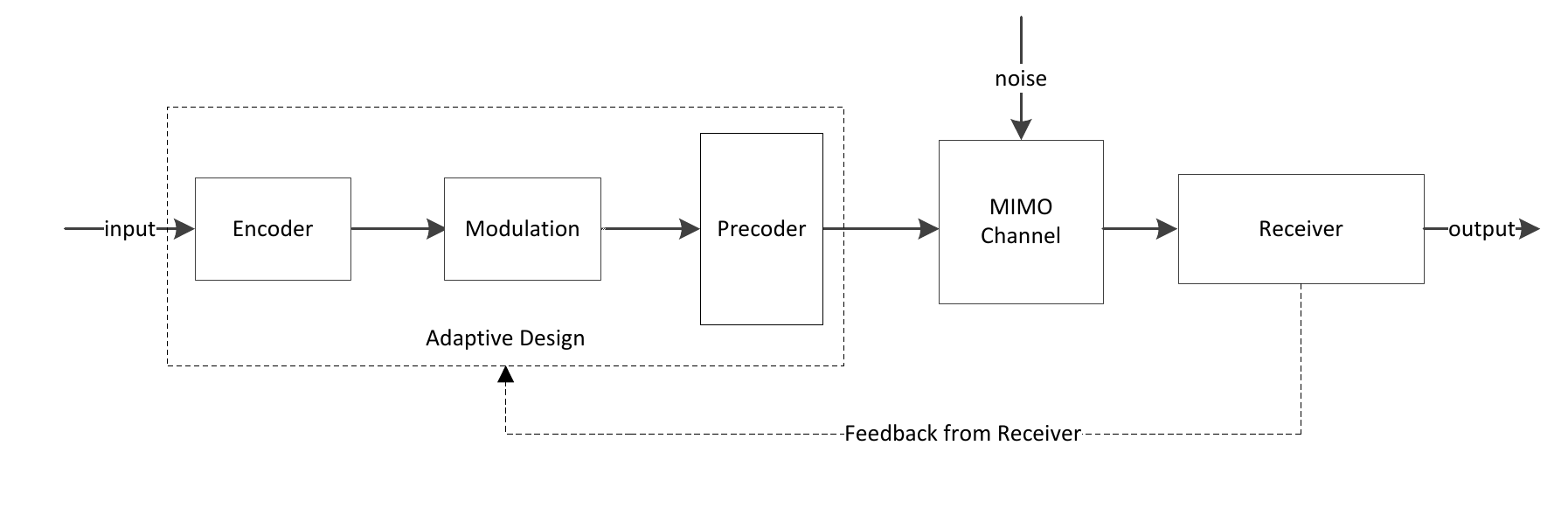}
\caption {\space\space Adaptive MIMO transmission for coded systems}
\label{AT-MIMO-CODE}
\end{figure*}

\section{Multi-user MIMO Systems}
In this section, we introduce the transmission designs with discrete input signals
for multiuser MIMO systems. We discuss the following four  scenarios: i) MIMO uplink transmission;
ii) MIMO downlink transmission; iii) MIMO interference channel; iv) MIMO wiretap channel.

\subsection{MIMO Uplink Transmission}
\subsubsection{Mutual Information Design}
For the multiple access channel (MAC), the maximum mutual information
region with uniformly distributed finite alphabet inputs is defined
as the constellation-constrained capacity region \cite{Harshan2011TIT,Harshan2013TWC}.
For the MIMO MAC with PCSIT,
M. Wang \textit{et al.} derive the constellation-constrained
capacity region for an arbitrary number of
users and arbitrary antennas configurations \cite{Wang2011TWC}.
The boundary of this region can be achieved by solving the weighted sum-rate (WSR)
maximization problem with finite alphabet and individual power constraints.
Necessary conditions of the optimal precoders of all users are established
and an iterative algorithm is proposed to find these optimal precoders.
Moreover, a LDPC coded system with iterative detection and decoding
is provided  to examine the BER performance of the obtained precoders.
Numerical results indicate the obtained precoders achieve considerably
better performance than the non-precoding design and the Gaussian input
design in terms of both mutual information and coded BER.

For the MIMO MAC with SCSIT, asymptotic expressions
for the mutual information of the MIMO MAC with arbitrary inputs over Kronecker fading channels
are derived in large system limits \cite{Wen2007TIT}. The linear precoder design
for the MIMO MAC is further developed  based on the sum-rate maximization \cite{Girnyk2014TWC}.
The proposed design embraces the case where non-Gaussian interferers are present.
For a more general Weichelberger's fading model in \cite{Weichselberger2006TWC},
W. Yu \textit{et al.} investigate the linear precoder design based on the WSR
maximization \cite{Wu2015TWC}. The asymptotic (in large system limits) optimal left singular matrix of each user's optimal
precoder is proved to be the eigenmatrix of the transmit correlation matrix of the user.
This results facilitates the derivation of an efficient iterative algorithm for computing
the optimal precoder for each user.  A brief summation
of above  mutual information based precoder designs for the MIMO uplink transmission with discrete input signals is given in Table \ref{table:MIMO-MAC-MI}.

\begin{table}
\centering
  \renewcommand{\multirowsetup}{\centering}
 \captionstyle{center}
  {
\caption{Mutual information based precoder designs for the MIMO uplink transmission  with discrete input signals}
\label{table:MIMO-MAC-MI}
\begin{lrbox}{\tablebox}
\begin{tabular}{|c|c|c|}
\hline
  Paper   & Model &   Main Contribution \\ \hline
\multirow{2}{*} { M. Wang \textit{et al.}  \cite{Wang2011TWC}}   & MIMO MAC   & Propose an iterative algorithm to find  \\
 &PCSIT, PCSIR &   the optimal precoders which maximize the WSR
\\ \hline
\multirow{2}{*} { C.-K. Wen \textit{et al.}  \cite{Wen2007TIT}}   & MIMO MAC,  Kronecker fading    & Derive asymptotic expressions for the mutual information  \\
 & SCSIT, PCSIR  &   of the MIMO MAC with arbitrary inputs
\\ \hline
\multirow{3}{*} { M. A. Girnyk \textit{et al.}  \cite{Girnyk2014TWC}}   & MIMO MAC,  Kronecker fading  & \multirow{2}{*} {Propose an iterative algorithm to find}  \\
 & SCSIT, PCSIR  &  \multirow{2}{*}{the optimal precoders which maximize the sum-rate}  \\
  &  non-Gaussian interferers &
\\ \hline
\multirow{2}{*} {Y. Wu \textit{et al.}  \cite{Wu2015TWC}}   & MIMO MAC, Weichelberger's fading    & Find the asymptotic
optimal left singular matrix of each user's optimal precoder  \\
 & SCSIT, PCSIR  &  Propose an efficient iterative algorithm to maximize WSR
\\ \hline
\end{tabular}
\end{lrbox}
\scalebox{0.9}{\usebox{\tablebox}}
}
\end{table}

\subsubsection{MSE Design}
For the MIMO MAC with PCSIT and PCSIR,  E. A. Jorswieck \textit{et al.} first design the
transmit covariance matrices of the users to minimize the sum MSE with the multiuser MMSE receiver
and individual user power constraints \cite{Jorswieck2003TSP}. A power allocation scheme among users is developed
under a sum power constraint to further reduce the sum MSE. Based on the KKT conditions,
the optimal transmit covariance matrices and the power allocation scheme can be found by
an iterative algorithm. In addition, the achievable MSE region is studied in \cite{Jorswieck2003TSP}.
S. Serbetli \textit{et al.} investigate how to determine
the number of transmit symbols for each user \cite{Serbetli2004TSP}. Z.-Q. Luo \textit{et al.}
extend the MMSE transceiver design to MIMO MAC OFDM systems \cite{Luo2004TSP}.
I.-T. Lu \textit{et al.} further study the impact of the practical
per-antenna power and peak power constraints \cite{Lu2009SS}.
The MMSE transceiver design for MIMO MAC with channel mean or covariance SCSIT feedback
and PCSIR subject to individual user power constraints is investigated in \cite{Jorswieck2006,Zhang2006}.
For both sum and individual user power constraints, X. Zhang \textit{et al.} study the average sum MSE
minimization problem for two user MIMO MAC with correlated estimation error
IPCSIT and IPCSIR \cite{Zhang2007}.  P. Layec \textit{et al.} further consider the $K$
user MIMO MAC case with additional quantization noise and individual user power constraints \cite{Layec2007}.
A near optimal closed-form transceiver structure, which minimizes the average sum MSE
for MIMO MAC with i.i.d estimation error IPCSIT and IPCSIR
subject to a sum power constraint,
is provided in \cite{Huang2007WCNC}.
A brief summation
of above MMSE precoder designs for the MIMO uplink transmission with discrete input signals is given in Table \ref{table:MIMO-MAC-MMSE}.

\begin{table}
\centering
  \renewcommand{\multirowsetup}{\centering}
 \captionstyle{center}
  {
\caption{MMSE precoder designs for the MIMO uplink transmission  with discrete input signals}
\label{table:MIMO-MAC-MMSE}
\begin{lrbox}{\tablebox}
\begin{tabular}{|c|c|c|}
\hline
  Paper   & Model &   Main Contribution \\ \hline
\multirow{3}{*} {E. A. Jorswieck \textit{et al.} \cite{Jorswieck2003TSP}}   & MIMO MAC   &   \\
 & PCSIT, PCSIR & Propose an iterative algorithm to minimize the sum MSE \\
  & Individual/sum power constraints &
\\ \hline
\multirow{3}{*} {S. Serbetli \textit{et al.} \cite{Serbetli2004TSP}}   & MIMO MAC   &   \\
 & PCSIT, PCSIR & Investigate how to determine
the number of transmit symbols  \\
  & Individual power constraints &
\\ \hline
\multirow{3}{*} {Z.-Q. Luo \textit{et al.} \cite{Luo2004TSP}}   & MIMO MAC OFDM  &   \\
 & PCSIT, PCSIR &  Extend the MMSE transceiver design to OFDM systems  \\
  & Individual power constraints &
\\ \hline
\multirow{3}{*} {I.-T. Lu \textit{et al.}\cite{Lu2009SS} } & MIMO MAC   &   \\
 & PCSIT, PCSIR &  Propose iterative algorithms to minimize the sum MSE\\
  & Per-antenna/peak power constraints &
\\ \hline
\multirow{3}{*} {{ \cite{Jorswieck2006,Zhang2006}} } & MIMO MAC   &   \\
 & SCSIT, PCSIR &  Propose iterative algorithms to minimize the average sum MSE\\
  & Individual power constraints &
\\ \hline
\multirow{3}{*} {{ X. Zhang \textit{et al.} \cite{Zhang2007} } } & Two-user MIMO MAC   &   \\
 & IPCSIT, IPCSIR &  Propose iterative algorithms to minimize the average sum MSE\\
  & Individual/sum power constraints &
\\ \hline
\multirow{3}{*} {{ P. Layec \textit{et al.}  \cite{Layec2007} } } & MIMO MAC   &   \\
 & IPCSIT, IPCSIR, Quantization noise&  Propose iterative algorithms to minimize the average sum MSE\\
  & Individual power constraints &
\\ \hline
\multirow{3}{*} {{ J. W. Huang \textit{et al.}  \cite{Huang2007WCNC} } } & MIMO MAC   &   \\
 & IPCSIT, IPCSIR &  Provide closed-form near optimal MMSE precoder designs\\
  & A sum power constraint &
\\ \hline
\end{tabular}
\end{lrbox}
\scalebox{0.9}{\usebox{\tablebox}}
}
\end{table}

\subsubsection{Diversity Design}
For open-loop systems, based on the framework for analyzing
the dominant error event regions in \cite{Gallager1985TIT},
M. E. G\"{a}rtner \textit{et al.} derive
space-time/frequency code design criterions
for two-user fading MIMO MAC \cite{Gatner2006}.
Moreover, an explicit two-user  $2 \times 2$
 coding scheme is constructed  by
concatenating two Alamouti schemes with a column
swapping for one user's codeword to achieve
a minimum rank of three.
A algebraic construction of STBC based on diversity and multiplexing tradeoff
for MIMO MAC is proposed in \cite{Badr2008}. Similar STBC designs are
also provided for the asynchronous single-input multiple-output (SIMO) MAC in \cite{Badr2009}, non-uniform user transmit power MIMO MAC in \cite{Badr2009_2},
and the two-user MIMO multiple-access AF relay channel in \cite{Badr2008_2}.
A space-frequency code for MIMO-OFDM MAC is designed in \cite{Zhang2010TSP},
which achieves full diversity for every user.
However, the codes in \cite{Zhang2010TSP} suffer from high large peak-to-average power ratio
since some elements in the codeword matrices are zero.
Another group of STBC for MIMO MAC
is proposed in \cite{Hong2009TVT} by minimizing a truncated union-bound approximation.
Simulations show that the codes in \cite{Hong2009TVT} achieve
better error probability than the codes in \cite{Gatner2006,Zhang2010TSP}.
H.-F. Lu \textit{et al.} propose  sphere-decodable STBCs for two-user
MIMO MAC and provide an explicit construction for $2\times 2$ case \cite{Lu2009JSTSP}.
STBCs for  two-user MIMO MAC with reduced average sphere decoding complexity
are further constructed in \cite{Harshan2009}.  Moreover,
A differential STBC for two-user MIMO MAC is proposed in \cite{Bhatnagar2012TWC}.
Another  differential STBC  for two-user MIMO MAC with low complexity non-coherent
decoders is further designed in \cite{Poorkasmaei2013TCom}.
For close-loop systems, by exploiting the outdated CSI,
J. W. Huang \textit{et al.} investigate
the linear precoding design for MIMO MAC with OSTBC
to minimize PEP subject to individual transmit power constraint of each user \cite{Huang2007}.
  Also, Y.-J. Kim \textit{et al.} propose a STBC
for two-user DIMO MAC with a linear detection by
utilizing the phase feedback transmitted to one user \cite{Kim2011ICC}.
By exploiting PCSIT, F. Li \textit{et al.} propose full-diversity precoder designs for two-user
and $K$-user MIMO MAC in \cite{Li2009JSTSP} and \cite{Li2011TCom}, respectively.
A brief comparison of above work is given in Table \ref{table:SPT-MAC}.

\begin{table}
\centering
  \renewcommand{\multirowsetup}{\centering}
 \captionstyle{center}
  {
\caption{Space-time techniques for the MIMO uplink transmission with discrete input signals}
\label{table:SPT-MAC}
\begin{lrbox}{\tablebox}
\begin{tabular}{|c|c|c|}
\hline
  Paper   & Model &   Main Contribution \\ \hline
\multirow{2}{*} {M. E. G\"{a}rtner \textit{et al.}  \cite{Gatner2006}}   &  MIMO MAC  & Derive
space-time/frequency code design criterions  \\
 & Two-user &  Construct an explicit coding scheme for two-user  $2 \times 2$
\\ \hline
\multirow{2}{*} {M. Badr \textit{et al.}  \cite{Badr2008,Badr2008_2,Badr2009,Badr2009_2}}   &  MIMO MAC, asynchronous SIMO MAC, unbalanced MIMO MAC  & Construct STBCs based on
 \\
 & Two-user MIMO multiple-access AF relay &  diversity and multiplexing tradeoff
\\ \hline
\multirow{2}{*} {W. Zhang \textit{et al.}  \cite{Zhang2010TSP}}   & \multirow{2}{*} { MIMO-OFDM MAC} & Design a space-frequency code \\
 & &  which achieves full diversity for every user
\\ \hline
\multirow{2}{*} {Y. Hong \textit{et al.}  \cite{Hong2009TVT}}   & \multirow{2}{*} { MIMO MAC} & Design STBCs based on \\
 & &  truncated union-bound approximation
\\ \hline
\multirow{2}{*}  {\cite{Lu2009JSTSP,Harshan2009}}   &  MIMO MAC & \multirow{2}{*} {Design  sphere-decodable STBCs} \\
 &Two-user &
\\ \hline
\multirow{2}{*} {\cite{Bhatnagar2012TWC,Poorkasmaei2013TCom}  }  &  MIMO MAC & \multirow{2}{*} {Design differential STBCs} \\
 & Two-user &
 \\ \hline
\multirow{2}{*} {J. W. Huang \textit{et al.} \cite{Huang2007}}  &  MIMO MAC  &  {Investigate
the linear precoding design } \\
 & Outdated CSIT & subject to individual user's transmit power constraint
\\ \hline
\multirow{2}{*} { Y.-J. Kim \textit{et al.} \cite{Kim2011ICC}}  &  DIMO MAC  & \multirow{2}{*} {Design a STBC with a linear detection} \\
 & Two-user, feedback &
\\ \hline
\multirow{2}{*} { F. Li \textit{et al.} \cite{Li2009JSTSP,Li2011TCom}}  & Two-user MIMO MAC, $K$-user MIMO MAC & \multirow{2}{*} {Propose full-diversity precoder designs} \\
 & PCSIT &
\\ \hline
\end{tabular}
\end{lrbox}
\scalebox{0.85}{\usebox{\tablebox}}
}
\end{table}

\subsubsection{Adaptive Transmission}
For MIMO-OFDM MAC with PCSIT and PCSIR, Y. J. Zhang \textit{et al.} propose
an adaptive resource allocation scheme when each user's signal
is transmitted along the maximum singular of its own channel \cite{Zhang2005TCOM}.
The proposed scheme selects the subcarrier allocation,
the power allocation, and modulation modes to minimize the total transmit power
subject to each user's uncoded BER and data rate constraints. It is assumed that
these uplink transmission parameters are determined at the base station and then sent
to the users via the error-free channel. Simulation results
indicate that within a reasonable region of Doppler spread,
the proposed scheme is also robust to the channel time variation.
A low-complexity design with a
neighborhood search for the optimal resource allocation solution is further proposed in \cite{Zhang2005TCOM_2}.
For MIMO-OFDM MAC where each user employs the transmission design in \cite{Xia2005TCOM}
with its own SCSI, R. Chemaly \textit{et al.} propose an adaptive resource allocation scheme to maximize
the sum data rate subject to each user's uncoded BER and the total power constraints \cite{Chemaly2005}.
By taking an additional fairness consideration among different users, another adaptive resource allocation scheme is given
in \cite{Maw2011ISAS} to maximize the sum data rate subject to each user's uncoded BER and the total power constraints.
A brief comparison of above work is given in Table \ref{table:AT-MAC}.

\begin{table}
\centering
  \renewcommand{\multirowsetup}{\centering}
 \captionstyle{center}
  {
\caption{Adaptive transmissions for the MIMO uplink transmission with discrete input signals}
\label{table:AT-MAC}
\begin{lrbox}{\tablebox}
\begin{tabular}{|c|c|c|}
\hline
  Paper   & Model &   Main Contribution \\ \hline
\multirow{2}{*} {Y. J. Zhang \textit{et al.} \cite{Zhang2005TCOM}}   &  MIMO-OFDM MAC, perfect CSI  & Propose an adaptive resource allocation \\
 & Each user's uncoded BER and data rate constraints &  to minimize the total transmit power
\\ \hline
\multirow{2}{*} {Y. J. Zhang \textit{et al.} \cite{Zhang2005TCOM_2}}   &  MIMO-OFDM MAC, perfect CSI  & Propose a low complexity adaptive resource  \\
 & Each user's uncoded BER and data rate constraints &  allocation to minimize the total transmit power
\\ \hline
\multirow{2}{*} { R. Chemaly \textit{et al.}  \cite{Chemaly2005}}   &  MIMO-OFDM  MAC, SCSIT, PCSIR  & Propose an adaptive resource allocation \\
 & Each user's uncoded BER and total power constraints &   to maximize  the sum data rate
\\ \hline
\multirow{2}{*} {M. S. Maw \textit{et al.} \cite{Maw2011ISAS}}   &  MIMO-OFDM  MAC, perfect CSI & Propose an adaptive resource allocation
considering the    \\
 & Each user's uncoded BER and total power constraints &  users' fairness and maximizing  the sum data rate
\\ \hline
\end{tabular}
\end{lrbox}
\scalebox{0.9}{\usebox{\tablebox}}
}
\end{table}

\subsection{MIMO Downlink Transmission}

\subsubsection{Mutual Information Design}
For the 2-user MISO BC with PCSIT, PCSIR, and finite alphabet inputs,
the precoder designs to maximize the sum rate and the corresponding simplified receiver structure
are studied in \cite{Ghaffar2009PIMRC,Ghaffar2010ICC,Ghaffar2009Asilomar}.
Y. Wu \textit{et al.} further investigate the linear precoder design
to  maximize the WSR of the  general MIMO BC \cite{Wu2012TWC_2}.
Explicit expressions for the achievable rate region
are derived, which is applicable to an arbitrary number of users
with generic antenna configurations. Then, it is revealed that the
sum rate loss will occur in high SNR regime for the improper precoder design
because of the non-uniquely decodable transmit signals.
An iterative algorithm is proposed to optimize precoding matrices for all
users. A downlink multiuser system with  LDPC code and
iterative detection and decoding is further developed. Simulations illustrate
that the proposed precoding design provides substantial WSR and
coded BER gains over the best rotation design for SISO BC \cite{Deshpande2009}, the non-precoding design,
and the Gaussian input design. W. Wu \textit{et al.} investigate the linear precoder design
for cooperative multi-cell MIMO downlink systems with PCSIT and finite alphabet inputs \cite{WWu2015TWC}.
To reduce the complexity of evaluating the non-Gaussian interferers,
W. Wu \textit{et al.} use a Gaussian interferer to approximate the sum of
non-Gaussian interferers at each user's receive side and propose two iterative algorithms
to maximize the approximated sum-rate
under per base station power constraints.  Comparing to the algorithm in \cite{Wu2012TWC_2}, the proposed algorithms in \cite{WWu2015TWC}
reduce the implementation complexity but with negligible performance losses.
S. X. Wu \textit{et al.} propose two transmit schemes to maximize the multicast rate
of MISO downlink multicasting systems with PCSIT and finite alphabet inputs \cite{Wu2015SPL}.
 A brief summation
of above  mutual information based precoder designs for the MIMO downlink transmission with discrete input signals is given in Table \ref{table:MIMO-BC-MI}.

\begin{table}
\centering
  \renewcommand{\multirowsetup}{\centering}
 \captionstyle{center}
  {
\caption{Mutual information based precoder designs for the MIMO downlink transmission with discrete input signals}
\label{table:MIMO-BC-MI}
\begin{lrbox}{\tablebox}
\begin{tabular}{|c|c|c|}
\hline
  Paper   & Model &   Main Contribution \\ \hline
\multirow{2}{*} { R. Ghaffar  \textit{et al.}  \cite{Ghaffar2009PIMRC,Ghaffar2010ICC,Ghaffar2009Asilomar}}   & MISO BC, 2-user   & Propose the precoder designs and the corresponding  \\
 &PCSIT, PCSIR &   simplified receiver structure to maximize the sum-rate
\\ \hline
\multirow{2}{*} { Y. Wu  \textit{et al.}  \cite{Wu2012TWC_2}}   & MIMO BC& Propose an iterative algorithm to find \\
 &PCSIT, PCSIR &   the optimal precoders which maximize the WSR
\\ \hline
\multirow{2}{*} { W. Wu \textit{et al.}  \cite{WWu2015TWC}}   &  Cooperative multi-cell MIMO downlink  & Propose two low-complexity algorithms to find \\
 &PCSIT, PCSIR &   the optimal precoders which maximize the sum-rate
\\ \hline
\multirow{2}{*} { S. X. Wu \textit{et al.}  \cite{Wu2015SPL}}   & MISO downlink multicasting & Propose two transmit schemes  \\
 &PCSIT, PCSIR &    to maximize the multicast rate
\\ \hline
\end{tabular}
\end{lrbox}
\scalebox{0.9}{\usebox{\tablebox}}
}
\end{table}

\subsubsection{MSE Design}
For the linear processing, early research in \cite{Tenenbaum2004,Zhang2005Cletter} consider to extend the MMSE transceiver design in \cite{Palomar2003TSP}
to MIMO BC.  For MIMO BC with PCSIT, PCSIR, and a sum power constraint,
S. Shi \textit{et al.} establish a uplink-downlink  duality
framework between the downlink and uplink MSE feasible regions \cite{Shi2007TSP}.
Based on this framework,  the downlink sum-MSE minimization can be transformed into
an equivalent uplink problem. Then, two globally
optimum algorithms are proposed to find the optimal MMSE transceiver design.
S. Shi \textit{et al.} further propose near-optimal low complexity iterative
algorithms to minimize the maximum ratio of MSE and
given requirement under a sum power constraint
and to minimize the transmit power under various MSE requirements \cite{Shi2008TSP}.
Moreover, R. Hunger \textit{et al.} propose a low complexity approach
to evaluate the MSE duality \cite{Hunger2009TSP}, which is able to support
the switched off data streams and the passive users correctly.
The MSE duality in \cite{Shi2007TSP} is extended to the imperfect CSI case for MISO BC \cite{Ding2007}.
D. S.-Murga \textit{et al.} provide a transceiver design framework for
MIMO-OFDM BC  based on the channel Gram matrices feedback,
where the CSI quantization error is also taken into consideration \cite{Murga2012TWC}.
As an example, a robust precoder design with fixed decoder which minimizes
the sum MSE subject to a sum power constraint is given in \cite{Murga2012TWC}.
M. Ding \textit{et al.}
propose a robust transceiver design for network MIMO systems with imperfect
backhaul links \cite{Ding2012TVT}, which minimizes the maximum substream MSE subject to
per base station power constraints.

For the robust design where the actual channel
is considered to be within an uncertain region around the channel
estimate, a robust power allocation scheme among users is proposed
for MISO BC to minimize the transmit power subject to each user's MSE
requirement \cite{Payaro2007JSAC}.
N. Vi\v{c}i\'{c} \textit{et al.} extend this to
robust transceiver design for MISO BC
and MIMO BC in \cite{Vuclc2009TSP} and \cite{Vuclc2009TSP_2}, respectively.
X. He \textit{et al.} further investigate this robust
transceiver design subject to the probabilistic MSE
requirement of each user \cite{He2013TWC}.
A robust resource allocation design to optimize the function
of worst case MSE subject to quadratic power constraints
for multicell MISO downlink transmission is proposed in \cite{Bjornson2012TSP}.

Some work consider the MMSE precoder design with the fixed decoder.
A. D. Dabbagh \textit{et al.} obtain an MMSE based precoding technique
for MISO BC with IPCSIT and a sum power constraint \cite{Dabbagh2008TCom}.
The limited feedback system model is also investigated in \cite{Dabbagh2008TCom}.
H. Sung \textit{et al.} develop a generalized MMSE precoder
for MIMO BC with perfect and imperfect CSI \cite{Sung2009TCom}.
The designs for the minimization of the sum MSE subject to a sum power constraint
and the minimization of the interference-plus-noise power at the receiver subject
to individual user's power constraints are obtained.
P. Xiao \textit{et al.} propose an improved MIMO BC MMSE precoder with  perfect and imperfect CSI
subject to a sum power constraint  for improper signal constellations such as amplitude shift-keying, offset
quadrature phase shift keying, etc \cite{Xiao2010TSPMar}.  A brief summation
of above linear MMSE precoder designs for the MIMO downlink transmission with discrete input signals is given in Table \ref{table:MIMO-BC-MMSE}.
There are also some work studying the non-linear MMSE precoder designs
for the MIMO downlink transmission with discrete input signals  \cite{Shi2007TSP,Fischer2004,Hardjawana2009JSTSP,Masouros2012TSP,Liu2013TSP,Zu2014TCom,Sun2015TCom}.

\begin{table}
\centering
  \renewcommand{\multirowsetup}{\centering}
 \captionstyle{center}
  {
\caption{Linear MMSE precoder designs for the MIMO downlink transmission  with discrete input signals}
\label{table:MIMO-BC-MMSE}
\begin{lrbox}{\tablebox}
\begin{tabular}{|c|c|c|}
\hline
  Paper   & Model &   Main Contribution \\ \hline
\multirow{3}{*} {S. Shi \textit{et al.} \cite{Shi2007TSP}}   & MIMO BC   &   \\
 & PCSIT, PCSIR & Establish a uplink-downlink  MSE duality framework \\
  & A sum power constraint &
\\ \hline
\multirow{2}{*} {S. Shi \textit{et al.} \cite{Shi2008TSP}}   & MIMO BC   &  {Minimize the maximum ratio of MSE and
given requirement} \\
 & PCSIT, PCSIR &  {Minimize the transmit power under various MSE requirements } \\
 \hline
\multirow{3}{*} {R. Hunger \textit{et al.} \cite{Hunger2009TSP}}   & MIMO BC   &  \multirow{2}{*} {Establish a MSE duality, which is able to support} \\
 & PCSIT, PCSIR & \multirow{2}{*} {the switched off data streams and the passive users correctly} \\
  & A sum power constraint &
\\ \hline
\multirow{3}{*} {M. Ding \textit{et al.} \cite{Ding2007}}   & MISO BC   & \\
 & IPCSIT, IPCSIR &  Extend the MSE duality in \cite{Shi2007TSP} to the imperfect CSI case  \\
  & A sum power constraint &
\\ \hline
\multirow{3}{*} {D. S.-Murga \textit{et al.} \cite{Murga2012TWC}}   & MIMO-OFDM BC   &   \multirow{2}{*} { Provide a transceiver design framework} \\
 & Channel Gram matrix &  \multirow{2}{*} { based on the channel Gram matrices feedback} \\
  & A sum power constraint &
\\ \hline
   \multirow{3}{*} {M. Ding \textit{et al.} \cite{Ding2012TVT}}   &  Network MIMO systems   &   \\
 & Imperfect backhaul links  &  Minimize the maximum substream MSE   \\
  & Per base station power constraints &   \\  \hline
\multirow{3}{*} {M. Payar\'{o} \textit{et al.} \cite{Payaro2007JSAC}}   & MISO BC   &  \multirow{2}{*} {Propose a robust power allocation scheme among users} \\
 & IPCSIT, IPCSIR &   \multirow{2}{*} {to minimize the transmit power} \\
  & Each user's MSE requirement &
\\ \hline
 \multirow{3}{*}{N. Vi\v{c}i\'{c} \textit{et al.} \cite{Vuclc2009TSP}}   & MISO BC   &  \multirow{2}{*} {Propose a robust transceiver design} \\
 & IPCSIT, IPCSIR &  \multirow{2}{*} {to minimize the transmit power} \\
   & Each user's MSE requirement &  \\ \hline
 \multirow{2}{*} {N. Vi\v{c}i\'{c} \textit{et al.} \cite{Vuclc2009TSP_2}}   & MIMO BC   &  Propose robust transceiver designs \\
 & IPCSIT, IPCSIR & for several MSE-optimization problem   \\ \hline
 \multirow{3}{*} {X. He \textit{et al.} \cite{He2013TWC}}   & MIMO BC   &  \multirow{2}{*} {Propose a robust transceiver design} \\
 & IPCSIT, IPCSIR &  \multirow{2}{*} {to minimize the transmit power} \\
   & Each user's probabilistic MSE requirement &  \\ \hline
    \multirow{3}{*} {E. Bj\"{o}rnson \textit{et al.} \cite{Bjornson2012TSP}}   & MISO BC   &  \multirow{2}{*} {Propose a robust resource allocation design} \\
 & IPCSIT, IPCSIR &  \multirow{2}{*} {to optimize the function
of worst case MSE} \\
   & Quadratic power constraints &  \\ \hline
       \multirow{3}{*} {A. D. Dabbagh \textit{et al.} \cite{Dabbagh2008TCom}}   & MISO BC, fixed decoder  &   \\
 & IPCSIT, PCSIR &  Obtain an MMSE based precoding technique \\
   &  A sum power constraint &  \\ \hline
\multirow{2}{*} {H. Sung \textit{et al.}  \cite{Sung2009TCom}}   & MIMO BC, fixed decoder  & Minimize  the sum MSE subject to a sum power constraint  \\
 & Perfect and imperfect CSI &   Minimize the interference-plus-noise power  subject
to individual user's power constraints \\  \hline
 \multirow{3}{*} {P. Xiao \textit{et al.} \cite{Xiao2010TSPMar}}   & MIMO BC, fixed decoder  &   \multirow{2}{*} {Propose an improved MIMO BC MMSE precoder} \\
 & Perfect and imperfect CSI &  \multirow{2}{*} {for improper signal constellations }  \\
  & A sum power constraint  &   \\  \hline
\end{tabular}
\end{lrbox}
\scalebox{0.8}{\usebox{\tablebox}}
}
\end{table}

\subsubsection{Diversity Design}
In the downlink transmission, mitigating the multiuser interference is essential for the reliable
communication  and this normally needs the availability of some forms of CSI.
Therefore, there are not much work for the open loop downlink systems.
Q. Ma \textit{et al.} propose a trace-orthogonal space-time coding, which is suitable
for MIMO BC \cite{Ma2007}. The proposed space-time coding reduce the complexity of ML  with little performance loss.
Some differential STBC designs for the data encoded using layered source coding are proposed
for MIMO BC \cite{Larsson2003CLetter,Kuo2007TB}. For the close loop systems, the space-time techniques
are usually combined with the linear precoding design to eliminate the
co-channel interference and improve the diversity performance. R. Chen \textit{et al.}
propose a unitary downlink precoder design for multiuser MIMO STBC systems
with PCSIT \cite{Chen2004ICC}. Using enough antenna dimension, the proposed precoder can effectively cancel
the co-channel interference at each user's side. Then, the additional antenna dimension
can be used to select the precoder matrix for the diversity gain. R. Chen \textit{et al.}
further provide a more general transmission design for MIMO BC \cite{Chen2007TSP}, which
combines the unitary downlink precoder for co-channel interference mitigation
with the space-time precoder design and antenna selection technique for SER minimization.
Simulations indicate that the proposed design achieves significant diversity gains in terms of SER.
D. Wang \textit{et al.} combine nonlinear THP with two diversity techniques: dominant eigenmode
transmission and OSTBC to improve the BER performance of MIMO BC \cite{Wang2005VTC}.
B. Clerckx \textit{et al.} investigate the transmission
strategy for two-user
MISO BC with outdated CSIT and obtain the analytical expressions for
the corresponding average PEP over correlated Rayleigh fading channel \cite{Clerckx2015TCom}. By exploiting the further
SCSIT, signal constellations are optimized to improve the average PEP performance.
D. Lee \textit{et al.} analyze the outage probability of the average effective SNR
for MIMO BC employing OSTBC \cite{Lee2010TWC}. The analytical results indicate that
the opportunistic scheduling scheme provide the effective SNR and the diversity gains
compared to the block diagonalization-precoding scheme.
With PCSIT and no CSIR, L. Li \textit{et al.} propose a interference cancellation scheme
for MIDO BC, where the transmitter sends the precoded Alamouti code to every user
simultaneously \cite{Li2012TWC}.  Numerical results indicate that the proposed design in \cite{Li2012TWC}
has better diversity gains than the block diagonalization designs in \cite{Chen2004ICC,Lee2010TWC}.
By using a regularized block diagonalization precoder to suppress
the multiuser interference in the first step,
C.-T. Lin  \textit{et al.} extend the X-Structure precoder in \cite{Vrigneau2008IJSTSP} to MIMO BC in \cite{Lin2013}.
A brief comparison of above work is given in Table \ref{table:SPT-BC}.

\begin{table}
\centering
  \renewcommand{\multirowsetup}{\centering}
 \captionstyle{center}
  {
\caption{Space-time techniques for the MIMO downlink transmission with discrete input signals}
\label{table:SPT-BC}
\begin{lrbox}{\tablebox}
\begin{tabular}{|c|c|c|}
\hline
  Paper   & Model &   Main Contribution \\ \hline
Q. Ma \textit{et al.}  \cite{Ma2007}   & MIMO  BC  & Propose a trace-orthogonal space-time coding \\
\hline
\multirow{2}{*} {\cite{Larsson2003CLetter,Kuo2007TB} }  & \multirow{2}{*} {MIMO  BC}  & Propose differential STBC designs  \\
& &  for the data encoded using layered source coding
\\ \hline
\multirow{2}{*} {R. Chen \textit{et al.}\cite{Chen2004ICC} }  & MIMO  BC  & Combine a unitary downlink precoder   \\
& PCSIT, PCSIR &  with STBC
\\ \hline
\multirow{2}{*} {R. Chen \textit{et al.}\cite{Chen2007TSP} }  & MIMO  BC  & Combine a unitary downlink precoder   \\
& PCSIT, PCSIR &  with STBC and antenna selection
\\ \hline
\multirow{2}{*} {D. Wang \textit{et al.} \cite{Wang2005VTC} }  & MIMO  BC  & Combine THP with dominant eigenmode    \\
& PCSIT, PCSIR &  transmission and OSTBC
\\ \hline
\multirow{2}{*} {B. Clerckx \textit{et al.}  \cite{Clerckx2015TCom} }  & MISO  BC  & Optimize signal constellation    \\
& two-user, outdated CSIT, PCSIR &  to improve the average PEP
\\ \hline
\multirow{2}{*} {D. Lee \textit{et al.} \cite{Lee2010TWC} }  & MIMO  BC  & Analyze opportunistic scheduling and diagonalization-precoding   \\
& PCSIT, PCSIR &  for MIMO BC with OSTBC
\\ \hline
\multirow{2}{*} {L. Li \textit{et al.} \cite{Li2012TWC} }  & MIDO  BC  & Combine a interference cancellation scheme \\
& PCSIT, no CSIR &  with the Alamouti code
\\ \hline
\multirow{2}{*} {C.-T. Lin  \textit{et al.} \cite{Lin2013} }  & MIMO  BC  & Extend X-Structure precoder  \\
& PCSIT, PCSIR &  to MIMO BC
\\ \hline
\end{tabular}
\end{lrbox}
\scalebox{0.9}{\usebox{\tablebox}}
}
\end{table}

\subsubsection{Adaptive Transmission}
For uncoded systems, most adaptive transmission literatures focus on
the adaptive resource allocation for multiuser MIMO-OFDM systems. Typically,
with perfect CSIT and CSIR, C.-F. Tsai \textit{et al.} devise an adaptive resource allocation algorithm
for multiuser downlink MISO orthogonal frequency division multiplexing access (OFDMA)/spatial division multiple access (SDMA) systems with multimedia traffic \cite{Tsai2008TWC}.
A dynamical priority scheduling scheme is developed
to provide service to urgent users based on time to expiration.
Then, the resource for power, subchannel, and bit allocation is iteratively allocated  to maximize
the spectrum  efficiency subject to various quality of service constraints.
C.-M. Yen \textit{et al.} extend this adaptive resource allocation design  to multiuser downlink MIMO OFDMA systems \cite{Yen2010TVT},
where a utility function is defined to distinguish  the real time service and
the nonreal time service more clearly. Other resource allocation schemes for practical multiuser MIMO-OFDM systems can
be found in \cite{Hu2004WCNC,Ho2009TVT,Sadr2009TCST}.

For coded systems, Y. Hara \textit{et al.} design
an efficient downlink scheduling algorithm for multiuser downlink MIMO systems by optimizing an equivalent
uplink scheduling problem \cite{Hara2008TVT}.  With perfect knowledge of the channel, the proposed algorithm selects
the user for each transmit beam, the transmit weight for the user, and the corresponding modulation and coding scheme
to maximize the overall system throughout subject to PER constraints. Moreover, M. Fsslaoui \textit{et al.} propose a two-step adaptive
transmission scheme for multiuser downlink MISO-OFDM systems \cite{Esslaoui2011ISWCS}. In the first step, the users that should be simultaneously
transmitted are selected based on  the degree of orthogonality criterion. In the second step,
the modulation and coding scheme for each user is determined adaptively by maximizing the throughput
while satisfying the effective SNR constraint for each user. The proposed design is simulated for
the IEEE 802.11ac.  It is worth mentioning
 that as shown in \cite[Fig. 4]{Esslaoui2011ISWCS}, even the adaptive transmission technology is employed, there are still obvious gaps between
 the practical schemes and the theoretical capacity achieved by Gaussian input. The design in \cite{Esslaoui2011ISWCS} is extended
 to multiuser downlink MIMO-OFDM systems and the fairness among users is taken into consideration additionally \cite{Esslaoui2012ISWCS}.
J. Wang \textit{et al.} propose another transmission scheme for multiuser downlink MIMO-OFDMA systems,
where  the power allocation, the resource allocation policy, and the modulation and coding scheme for each user
are adaptively determined to maximize the system throughout \cite{Wang2009PIMRC}.    The proposed design is simulated for IEEE 802.16e and achieves
good performance. The adaptive transmission for multiuser downlink MIMO-OFDM systems with limited feedback is provided  in \cite{Chen2010Wicon},
where only a single spatial stream is transmitted for each user. A. R.-Alvari\={n}o \textit{et al.} propose
an adaptive transmission scheme for multiuser downlink MIMO-OFDM systems with limited feedback,
which allows multiple spatial streams for each user \cite{Alvarino2014TWC}. A greedy algorithm is used to select the
users and the spatial modes.  A machine learning classifier is used to select
the modulation and coding schemes for users by maximizing the overall system throughout subject
to FER constraints. A brief comparison of above work is given in Table \ref{table:AT-BC}.

\begin{table}
\centering
  \renewcommand{\multirowsetup}{\centering}
 \captionstyle{center}
  {
\caption{Adaptive transmissions for the MIMO downlink transmission with discrete input signals}
\label{table:AT-BC}
\begin{lrbox}{\tablebox}
\begin{tabular}{|c|c|c|}
\hline
  Paper   & Model &   Main Contribution \\ \hline
\multirow{2}{*} {C.-F. Tsai \textit{et al.} \cite{Tsai2008TWC}}   &  Uncoded MISO-OFDMA/SDMA, perfect CSI  & Devise an adaptive resource allocation algorithm \\
 & Various quality of service constraints  &    to maximize
the spectrum  efficiency
\\ \hline
\multirow{2}{*} {C.-M. Yen \textit{et al.} \cite{Yen2010TVT}}   &  Uncoded MIMO-OFDMA, perfect CSI  & Define a utility function to distinguish the \\
 & Various quality of service constraints  &    real time service and the nonreal time service
\\ \hline
\multirow{2}{*} {Y. Hara \textit{et al.} \cite{Hara2008TVT}}   &  Coded MIMO, perfect CSI & Design
an efficient downlink scheduling algorithm \\
 &  PER constraints  &   to maximize the overall system throughout
\\ \hline
\multirow{2}{*} {M. Fsslaoui \textit{et al.} \cite{Esslaoui2011ISWCS}}   & Coded MISO-OFDM,  perfect CSI  &  Propose a two-step adaptive
transmission scheme  \\
 &   Effective SNR constraints &  to maximize the spectral efficiency
\\ \hline
\multirow{2}{*} {M. Fsslaoui \textit{et al.} \cite{Esslaoui2012ISWCS}}   & Coded MIMO-OFDM,  perfect CSI  &  Fairness among users is taken \\
 &   Effective SNR constraints &   into consideration of system designs
\\ \hline
\multirow{2}{*} {J. Wang \textit{et al.} \cite{Wang2009PIMRC}}   & Coded MIMO-OFDMA,  perfect CSI  &  Propose an adaptive scheme for the power allocation, \\
 &   Effective SNR constraints &   the resource allocation policy, and the modulation and coding scheme
\\ \hline
\multirow{2}{*} {Z. Chen \textit{et al.} \cite{Chen2010Wicon}}   & Coded MIMO-OFDM,  feedback CSI  &  \multirow{2}{*} {Propose an adaptive scheme to maximize the spectrum efficiency} \\
 &   Effective SNR constraints &
\\ \hline
\multirow{2}{*} {A. R.-Alvari\={n}o \textit{et al.} \cite{Alvarino2014TWC}}   & Coded MIMO-OFDM,  feedback CSI  &  Propose an adaptive scheme to maximize the overall system \\
 &  FER constraints  &   throughout, which allows  multiple spatial streams for each user
\\ \hline
\end{tabular}
\end{lrbox}
\scalebox{0.9}{\usebox{\tablebox}}
}
\end{table}

\subsection{MIMO Interference Channel}

\subsubsection{Mutual Information Design}
By directly extending the classical interference alignment scheme used for Gaussian input signal,
B. Hari Ram \textit{et al.} and Y. Fadlallah \textit{et al.} investigate
the precoder designs for $K$-user MIMO interference channel by maximizing the mutual information of the finite alphabet signal sets in
\cite{Ram2013CL} and \cite{Fadlallahr2013PIMRC}, respectively. Following the interference alignment scheme,
the precoders in \cite{Ram2013CL,Fadlallahr2013PIMRC} only use half of the transmit antenna
dimension to send the desired signals. H. Huang \textit{et al.} consider to relax this dimension constraint
by allowing the interfering signals to overlap with the desired signal and propose a partial interference
alignment and interference detection scheme for $K$-user MIMO interference channel with finite alphabet inputs \cite{Huang2011TSP}.
For the interfering signals which can not be aligned at the each receiver, the receiver detects and cancels the residual
interference based on the constellation map. By treating the interference as noise, Y. Wu \textit{et al.} and
A. Ganesan  \textit{et al.} independently derive the mutual information expressions for  $K$-user MIMO interference channel with finite alphabet inputs
in \cite{Wu2013TCom} and \cite{Ganesan2014TWC}, respectively. Then, gradient decent based iterative algorithms are proposed in \cite{Wu2013TCom,Ganesan2014TWC}
to find the optimal precoders of all transmitters.  Instead of degree of freedom used
for Gaussian input signal, B. Hari Ram \textit{et al.} define a new performance metric to analyze the
transmission efficiency for $K$-user MIMO interference channel with finite alphabet inputs: number of
symbols transmitted per transmit antenna per channel user (SpAC) \cite{Ramr2013Asilomar}.  It is revealed in \cite{Wu2013TCom,Ganesan2014TWC}
that simple transmission schemes can achieve the maximum 1 SpAC in high SNR regime. However, the joint detection
of all the transmitter signals including the desired signals and the interfering signals at each receiver is
needed in \cite{Wu2013TCom,Ganesan2014TWC} to achieve 1 SpAC. B. H. Ram \textit{et al.} further propose a fractional
interference alignment scheme, which does not decode any of the interfering signals \cite{Ram2014}. The fractional
interference alignment can achieve any value of SpAC in range $[0,1]$ and a good
error rate performance for both uncoded and coded BER.   A brief summation
of above  mutual information based precoder designs for the MIMO interference channel with discrete input signals is given in Table \ref{table:MIMO-IC-MI}.

\begin{table}
\centering
  \renewcommand{\multirowsetup}{\centering}
 \captionstyle{center}
  {
\caption{Mutual information based precoder designs for the MIMO interference channel with discrete input signals}
\label{table:MIMO-IC-MI}
\begin{lrbox}{\tablebox}
\begin{tabular}{|c|c|c|}
\hline
  Paper   & Model &   Main Contribution \\ \hline
\multirow{2}{*} { \cite{Ram2013CL,Fadlallahr2013PIMRC} }  & MIMO interference channel  & Extend interference alignment scheme \\
 &Global PCSIT, PCSIR &   to precoder design with finite alphabet inputs
\\ \hline
\multirow{2}{*} { H. Huang \textit{et al.} \cite{Huang2011TSP} }  & MIMO interference channel  & Propose a partial interference
alignment   \\
 &Global PCSIT, PCSIR & and interference detection scheme
\\ \hline
\multirow{2}{*} { \cite{Wu2013TCom,Ganesan2014TWC} }  & MIMO interference channel  & Propose gradient decent based iterative algorithms
   \\
 &Global PCSIT, PCSIR & to find optimal precoders of all transmitters
\\ \hline
\multirow{2}{*} { \cite{Ramr2013Asilomar,Ram2014} }  & MIMO interference channel  & Propose a novel low-complexity
\\
 &Global PCSIT, PCSIR & fractional interference alignment scheme
\\ \hline
\end{tabular}
\end{lrbox}
\scalebox{0.9}{\usebox{\tablebox}}
}
\end{table}

\subsubsection{MSE Design}
For MIMO interference channel with global perfect CSI, H. Shen \textit{et al.} investigate
linear transceiver designs under a given and feasible degree of freedom \cite{He2010TWC}.
Two iterative algorithms are proposed to minimize both
the sum MSE and the maximum per-user MSE. The transceiver design
for MIMO interference channel with channel estimation error is also provided in \cite{He2010TWC}.
With perfect CSI, C.-E. Chen \textit{et al.} extend the design in \cite{Chen2010WCL} to minimize the maximum
per-stream MSE of all users. Also, F. Sun \textit{et al.} propose a low complexity
 transceiver design based on minimizing the MSE of the signal and interference leakage
 at each transmitter \cite{Sun2012SPL}. The robust transceiver design for MISO interference channel with random vector quantization feedback
 mechanism is proposed  in \cite{Tseng2015TVT}, where the impact of quantization error at the transmitter is considered.
The optimal transmit beamforming vectors and the receiver decoding scalars
for both the sum MSE and the maximum per-user MSE minimization are found by iterative algorithms.
A brief summation of above work is given in Table \ref{table:MIMO-IC-MMSE}.

\begin{table}
\centering
  \renewcommand{\multirowsetup}{\centering}
 \captionstyle{center}
  {
\caption{MMSE transceiver designs for  MIMO  interference channel  with discrete input signals}
\label{table:MIMO-IC-MMSE}
\begin{lrbox}{\tablebox}
\begin{tabular}{|c|c|c|}
\hline
  Paper   & Model &   Main Contribution \\ \hline
\multirow{2}{*} {H. Shen \textit{et al.}  \cite{He2010TWC}}   & MIMO interference channel   & \multirow{2}{*} {Minimize both
the sum MSE and the maximum per-user MSE} \\
 & Global perfect/Imperfect CSI &  \\ \hline
 \multirow{2}{*} {C.-E. Chen \textit{et al.} \cite{Chen2010WCL}}   & MIMO interference channel   & \multirow{2}{*} {Minimize the maximum
per-stream MSE of all users} \\
 & Global perfect CSI &  \\ \hline
  \multirow{2}{*} {F. Sun \textit{et al.} \cite{Sun2012SPL}}   & MIMO interference channel   & \multirow{2}{*} {Minimize the MSE of the signal and interference leakage} \\
 & Global perfect CSI &  \\ \hline
   \multirow{2}{*} {F.-S. Tseng \textit{et al.} \cite{Tseng2015TVT}}   & MIMO interference channel   & \multirow{2}{*} {Minimize both the sum MSE and the maximum per-user MSE} \\
 &  Random vector quantization feedback &  \\ \hline
\end{tabular}
\end{lrbox}
\scalebox{0.9}{\usebox{\tablebox}}
}
\end{table}

\subsubsection{Diversity Design}
For the open loop system, L. Shi \textit{et al.} design
full diversity STBCs for two-user MIMO interference channel
with the group ZF receiver \cite{Shi2012TWC}.
Y. Lu \textit{et al.} combine the blind interference alignment
scheme with three STBCs for $K$-user MISO interference channel employing
reconfigurable antenna \cite{Lu2014TCom}.
These combinations provide tradeoff between diversity,
 rate, and decoding complexity, where  threaded algebraic space-time codes
have full diversity and high rate with high decoding complexity,
OSTBCs have full diversity and low rate with the linear decoding complexity,
and Alamouti codes have high rate and low linear decoding complexity
but with full diversity loss. For the close loop system,
when the global CSI is available at all transmitters and receivers,
A. Sezgin \textit{et al.} combine the STBC with interference alignment
to achieve high diversity gains for MIMO interference channel \cite{Sezgin2009}.
However, the feasibility condition investigation for MIMO interference channel
reveals that the separation of space-time coding
and precoding design may not be optimal in general \cite{Ning2011TIT}. Moreover,
the diversity gains of various interference alignment schemes
for MIMO interference channels are analyzed in \cite{Ning2011TIT}.
L. Li \textit{et al.} combine the antenna selection technique
with the interference alignment design to improve the diversity gain of
the worst case user for 3-user $2\times2$ interference channel \cite{Li2013}.
A brief comparison of above work is given in Table \ref{table:SPT-IC}.

\begin{table}
\centering
  \renewcommand{\multirowsetup}{\centering}
 \captionstyle{center}
  {
\caption{Space-time techniques with discrete input signals for MIMO interference channel}
\label{table:SPT-IC}
\begin{lrbox}{\tablebox}
\begin{tabular}{|c|c|c|}
\hline
  Paper   & Model &   Main Contribution \\ \hline
\multirow{2}{*} {L. Shi \textit{et al.} \cite{Shi2012TWC} }  & Two-user MIMO interference channel  & \multirow{2}{*} {Design
full diversity STBCs} \\
& No PCSIT, PCSIR &
\\ \hline
\multirow{2}{*} {Y. Lu  \textit{et al.} \cite{Lu2014TCom} }  &  MISO interference, reconfigurable antenna &  Combine the blind interference alignment
scheme  \\
& No PCSIT, PCSIR & with three STBCs
\\ \hline
\multirow{2}{*} {A. Sezgin \textit{et al.} \cite{Sezgin2009} }  &  MIMO interference channel &  Combine STBC
  \\
& Global PCSIT, PCSIR & with the interference alignment design
\\ \hline
\multirow{2}{*} {H. Ning \textit{et al.} \cite{Ning2011TIT} }  &  MIMO interference channel &  Reveals that the separation of space-time coding
\\
& Global PCSIT, PCSIR & and precoding design may not be optimal
\\ \hline
\multirow{2}{*} {L. Li \textit{et al.} \cite{Li2013} }  &  3-user $2 \times2$ interference channel &  Combine the antenna selection technique
\\
& Global PCSIT, PCSIR & with the interference alignment design
\\ \hline
\end{tabular}
\end{lrbox}
\scalebox{0.9}{\usebox{\tablebox}}
}
\end{table}

\subsubsection{Adaptive Transmission}
B. Xie \textit{et al.} investigate the adaptive transmission for uncoded MIMO interference channels with imperfect CSIT \cite{Xie2013TWC}.
Interference alignment based bit-loading algorithms are proposed to minimize the average BER subject to a fixed sum-rate
constraint. Based on this, an adaptive transmission scheme switches among two interference alignment algorithms and the basic time-division
multiple access scheme to combat the CSI uncertainty is further provided.
M. Taki \textit{et al.} investigate the adaptive transmission for coded MIMO interference channels with imperfect CSIT \cite{Taki2014TWC}.
An interference alignment based adaptive transmission scheme is proposed, which adaptively selects the coding, modulation, and power allocation across users
to maximize the weighted sum rate subject to a sum power constraint and BER constraints for every transmit stream.

\subsection{MIMO Wiretap Channel}

\subsubsection{Mutual Information Design}
For the wiretap channel, the finite-alphabet inputs may allow both the receiver and the
eavesdropper to decode a transmitted message accurately if
the transmit power is high-enough. Therefore, the secrecy
rate may not be high at a high SNR. This phenomenon has been observed in
SISO single-antenna eavesdropper wiretap channels via simulations \cite{Rodrigues2010}.
Therefore,  S. Bashar \textit{et al.}  suggest to find an
optimum power control policy at the transmitter for maximizing
the secrecy rate \cite{Bashar2011CL}. For MISO single-antenna
eavesdropper (MISOSE) wiretap channel, S. Bashar \textit{et al.} investigate
the power control optimization issue at the transmitter
based on the conventional beamforming transmission for the
Gaussian input case and develop a numerical algorithm to
find the optimal transmission power \cite{Bashar2011CL}. This idea is extended to
MIMO multiple-antenna eavesdropper (MIMOME) wiretap channel  by exploiting a generalized SVD (GSVD) precoding structure to decompose
the MIMOME channel into a bank of parallel subchannels \cite{Bashar2012TCom}. Then, numerical algorithms are proposed to obtain the
adequate power allocated to each subchannel. Although the
precoding design in \cite{Bashar2012TCom} achieves significant performance gains
over the conventional design that relies on the Gaussian input
assumption, it is still suboptimal. The reasons are twofold. First,
part of the transmission symbols is lost by both the desired user
and the eavesdropper according to the GSVD structure. This
condition may result in a constant performance loss. Second,
the power control policy compels the transmitter to utilize only
a fraction of power to transmit in the high SNR regime, which
may impede the further improvement of the performance in
some scenarios (see \cite[Fig. 3]{Bashar2012TCom} as an example).
As a result, Y. Wu \textit{et al.} further investigate the linear precoder design
for MIMOME wiretap channel when
the instantaneous CSI of the eavesdropper is available at the transmitter \cite{Wu2012TVT}.
An iterative algorithm for secrecy rate maximization is developed
through a gradient method. For systems where the number of transmit
antennas is less than or equal to the number of the eavesdropper
antennas, only partial transmission power is necessary
for maximizing the secrecy rate at a high SNR. Accordingly,
 the excess  transmission power is used to construct
an artificial jamming signal to further improve the secrecy
rate.  By invoking the lower bound of the average
mutual information in MIMO Kronecker fading channels with the finite alphabet
inputs \cite{zeng2012linear}, Y. Wu \textit{et al.} also study the scenarios where the
transmitter only has the statistical CSI of the eavesdropper \cite{Wu2012TVT}.
W. Zeng \textit{et al.} study the precoder design for MIMO secure cognitive ratio systems
with finite alphabet inputs \cite{Zeng2016TWC}. With the SCSIT of the eavesdropper, an iterative
algorithm is proposed to maximize the secure rate of the secondary user under
the control of power leakage to the primary users.  The transmission
design for MIMO decode-and-forward (DF) relay systems with the IPCST and finite alphabet
inputs in presence of an user and $J$ non-colluding eavesdroppers
is investigated in \cite{Vishwakarma2015}.
Very recently, s secure transmission scheme
with a help of source node is proposed in \cite{Cao2017Access}.
A brief summation
of above  mutual information based precoder designs for the MIMO wiretap channel with discrete input signals is given in Table \ref{table:MIMO-wiretap-MI}.

\begin{table}
\centering
  \renewcommand{\multirowsetup}{\centering}
 \captionstyle{center}
  {
\caption{Mutual information based precoder designs for the MIMO wiretap channel with discrete input signals}
\label{table:MIMO-wiretap-MI}
\begin{lrbox}{\tablebox}
\begin{tabular}{|c|c|c|}
\hline
  Paper   & Model &   Main Contribution \\ \hline
\multirow{2}{*} {S. Bashar  \textit{et al.} \cite{Bashar2011CL} }  & MISOME  &  Find an
optimum power control policy  \\
 & SCSIT of the eavesdropper  &   at the transmitter for maximizing
the  secrecy rate
\\ \hline
\multirow{2}{*} {S. Bashar  \textit{et al.} \cite{Bashar2012TCom} }  & MIMOME  & Propose a GSVD precoder structure to  \\
 & PCSIT of the eavesdropper  &   increase the secure rate performance
\\ \hline
\multirow{2}{*} {Y. Wu  \textit{et al.} \cite{Wu2012TVT} }  & MIMOME  &  Propose iterative algorithms for secrecy rate maximization \\
 & PCSIT and SCSIT of the eavesdropper  &   Construct the jamming signal with the excess power
\\ \hline
\multirow{2}{*} {W. Zeng \textit{et al.} \cite{Zeng2016TWC} }  & MIMO cognitive radio secure &  Propose an iterative algorithm to maximize  \\
 &  SCSIT of the eavesdropper  &  the secure rate of the secondary user
\\ \hline
\multirow{2}{*} {S. Vishwakarma \textit{et al.} \cite{Vishwakarma2015} }  & MIMO DF relay secure &  Design the optimal source power, signal beamforming
 \\
 &  IPCSIT of all links &  and jamming signal matrix for worst case secrecy rate maximization
\\ \hline
\multirow{2}{*} {K. Cao \textit{et al.} \cite{Cao2017Access} }  & MIMOME with a helper & Design a secure transmission scheme
 \\
 &  SCSIT of the eavesdropper & with a help of source node
\\ \hline
\end{tabular}
\end{lrbox}
\scalebox{0.9}{\usebox{\tablebox}}
}
\end{table}

\subsubsection{MSE Design}
For MIMOME where the perfect instantaneous CSI is available at all parties, H. Reboredo \textit{et al.} investigate the transceiver design
minimizing the receiver's MSE  while guaranteeing the eavesdropper's MSE above a fixed threshold \cite{Reboredo2013TSP}.  A linear
zero-forcing precoder is used at the transmitter and either a zero-forcing decoder or an optimal linear Wiener decoder is designed
at the receiver. Moreover, the design is generalized to the scenarios where
only the statistical CSI is available. Simulations show that  maintaining
the eavesdropper's MSE above a certain level also limits eavesdropper's error probability.
This design is further extended to MIMO interference channel with a multiple antenna
eavesdropper \cite{Kong2016TIS}. M. Pei \textit{et al.} investigate the masked beamforming design for MIMO BC in presence of
 a multiple antenna eavesdropper  \cite{Pei2012TWC}. It is assumed that the transmitter obtains the perfect/imperfect CSI of the desired receivers
 but no CSI of the  eavesdropper.  The transmit power used to generate the artificial noise
 is maximized subject to the MSE requirements of the desired receivers.
Based on the same criterion, L. Zhang \textit{et al.} investigate the nonlinear THP design for MIMOME \cite{Zhang2015ICC}.
A brief summation of above work is given in Table \ref{table:MIMO-wiretap-MMSE}.

\begin{table}
\centering
  \renewcommand{\multirowsetup}{\centering}
 \captionstyle{center}
  {
\caption{MMSE transceiver designs for  MIMO  wiretap  channel with discrete input signals}
\label{table:MIMO-wiretap-MMSE}
\begin{lrbox}{\tablebox}
\begin{tabular}{|c|c|c|}
\hline
  Paper   & Model &   Main Contribution \\ \hline
\multirow{2}{*} {H. Reboredo \textit{et al.}  \cite{Reboredo2013TSP}}   & MIMOME   & Minimizing the receiver's MSE  while guaranteeing \\
 & Instantaneous/statistical CSI & the eavesdropper's MSE above a fixed threshold  \\ \hline
 \multirow{2}{*} {Z. Kong \textit{et al.}  \cite{Kong2016TIS}}   & MIMO interference channel secure   & Minimizing the receivers' sum MSE  while guaranteeing \\
 & Perfect CSI & the eavesdropper's MSE above a fixed threshold  \\ \hline
  \multirow{2}{*} {M. Pei \textit{et al.}  \cite{Pei2012TWC}}   & MIMO BC secure, Perfect/imperfect CSI of receivers   & Maximize the transmit power to  generate the artificial noise \\
 & No CSI of the eavesdropper & subject to the MSE requirements of the desired receivers \\ \hline
   \multirow{2}{*} {L. Zhang \textit{et al.} \cite{Zhang2015ICC}}   & MIMOME,  nonlinear THP design   & Maximize the transmit power to  generate the artificial noise \\
 & Perfect CSI of the receiver, no CSI of the eavesdropper & subject to the MSE requirements of the receiver\\ \hline
\end{tabular}
\end{lrbox}
\scalebox{0.9}{\usebox{\tablebox}}
}
\end{table}

\subsubsection{Diversity Design}
For two transmit antennas, two receive antennas, multiple antennas eavesdropper
wiretap channel, S. A. A. Fakoorian \textit{et al.} design  a secure STBC by exploiting
the perfect CSI of the receiver but no CSI of the eavesdropper \cite{Fakoorian2011}.
By combining QOSTBC in \cite{Jafarkhani2001TIT} with the artificial noise generation,
the proposed design in \cite{Fakoorian2011} effectively deteriorates the uncoded BER performance of the eavesdropper
 while maintains the good uncoded BER performance of the receiver. For the same STBC system, H. Wen \textit{et al.}
 propose a cross-layer scheme to further ensure the secure communication \cite{Wen2013TS}.
 T. Allen \textit{et al.} investigate  two/four transmit antennas, one receive antenna,
 and one antenna eavesdropper wiretap channels employing STBCs,  where no CSI of the receiver and the eavesdropper
 are available at transmitter \cite{Allen2014WCL}. By randomly rotating
 the transmit symbols at each transmit antenna based on the receive signal strength indicator,
 the proposed scheme provides full diversity of the receiver and enables the
 eavesdropper's diversity to be zero. For MISOSE wiretap channel with the same CSI assumption,
S. L. Perreau designs the transmit antennas weights based on STBC,
which formulates beamforming along the receiver's direction while disables
the eavesdropper's ability to discriminate the transmit signal \cite{Perreau2014}.
With an assumption that the transmitter has the perfect CSI of the receiver based on the
channel reciprocity while both the receiver and the eavesdropper have no channel knowledge,
X. Li \textit{et al.} investigate the secure communication for MISOME wiretap channel by using
M-PSK modulation and OSTBC \cite{Li2016TWC}. A phase shifting precoder is designed to align the transmit signals at
the receiver side so that the receiver can perform a non-coherent detection for the transmit signals.
At the same time, the information rate of the eavesdropper can be reduced to zero by the proposed design.
For the same channel model, another transmit antennas weights design method to deliberately randomize the eavesdropper's signal
but not the receiver's signal is proposed in \cite{Li2007JC}.
For MIMOME wiretap channel without the eavesdropper's CSI,
J.-C. Belfiore \textit{et al.} establish  a lattice wiretap code design criterion,
which minimizes the probability
of correctly decoding at the eavesdropper's side \cite{Belfiore2013TCom}.
T. V. Nguyen \textit{et al.} analyze the performance of the MISOME wiretap channel with perfect CSIT
of the receiver and SCSIT of the eavesdropper \cite{Nguyen2012}. The SER of confidential information
is derived and the corresponding secure diversity is obtained.
A brief comparison of above work is given in Table \ref{table:SPT-Wiretap}.

\begin{table}
\centering
  \renewcommand{\multirowsetup}{\centering}
 \captionstyle{center}
  {
\caption{Space-time techniques with discrete input signals for MIMO wiretap channel}
\label{table:SPT-Wiretap}
\begin{lrbox}{\tablebox}
\begin{tabular}{|c|c|c|}
\hline
  Paper   & Model &   Main Contribution \\ \hline
\multirow{3}{*} {S. A. A. Fakoorian \textit{et al.} \cite{Fakoorian2011} }  & Two transmit antennas and receive antennas
& Design a secure STBC \\
&  Multiple antennas eavesdropper  & with low uncoded BER of the receiver \\
&  Perfect CSI of the receiver, no CSI of the eavesdropper & and high uncoded BER of the eavesdropper
\\ \hline
\multirow{2}{*} {H. Wen \textit{et al.} \cite{Wen2013TS} }  & \multirow{2}{*} {Same as \cite{Fakoorian2011}}
& Propose a cross-layer scheme to \\
&    & further ensure the secure communication
\\ \hline
\multirow{3}{*} { T. Allen  \textit{et al.} \cite{Allen2014WCL} }  & Two/four transmit antennas, one receive antennas
& Design a secure STBC   \\
&  one antennas eavesdropper  & providing full diversity of the receiver \\
&   No CSI of the receiver and eavesdropper & and zero diversity of
 the eavesdropper
\\ \hline
\multirow{2}{*} {S. L. Perreau  \cite{Perreau2014} }  & MISOSE
& Design the transmit antennas weights based on STBC \\
&  No CSI of the receiver and eavesdropper  & to make the eavesdropper unable to decode the signal
\\ \hline
\multirow{2}{*} {X. Li \textit{et al.} \cite{Li2016TWC} }  & MISOME
& Design the secure communication scheme \\
&  No CSI of the receiver and eavesdropper  & by using
M-PSK modulation and OSTBC
\\ \hline
\multirow{2}{*} {X. Li \textit{et al.} \cite{Li2007JC}}  & MISOME
& Propose transmit antennas weights design method \\
&  No CSI of the receiver and eavesdropper  & to deliberately randomize the eavesdropper's signal
\\ \hline
\multirow{2}{*} {J.-C. Belfiore \textit{et al.} \cite{Belfiore2013TCom}}  & MIMOME
&   Establish  a lattice wiretap code design criterion minimizing \\
&   Perfect CSI of the receiver, no CSI of the eavesdropper &   the eavesdropper's correctly decoding probability
\\ \hline
\multirow{2}{*} {T. V. Nguyen \textit{et al.} \cite{Nguyen2012} }  & MISOME
&  \multirow{2}{*} { Analyze the SER of confidential information}\\
&  Perfect CSIT
of the receiver, SCSIT of the eavesdropper &
\\ \hline
\end{tabular}
\end{lrbox}
\scalebox{0.85}{\usebox{\tablebox}}
}
\end{table}

\section{Other Systems Employing MIMO Technology}
In this section, we overview the transmission design with discrete input signals for
other systems employing MIMO technology.  We consider the following three typical scenarios:
i) MIMO Cognitive Radio Systems; ii) Green MIMO communications; iii) MIMO relay systems.

\subsection{MIMO Cognitive Radio Systems}
\subsubsection{Mutual Information Design}
W. Zeng \textit{et al.} investigate the spectrum sharing
problem in MIMO cognitive radio systems with finite alphabet inputs \cite{Zeng2012JSAC}.
With the global PCSIT, the linear precoder is designed to maximize the spectral efficiency
between the secondary-user transmitter and the secondary-user receiver with the control
of the interference power to primary-user receivers. This design is extended
to the secure communication of MIMO cognitive radio systems in \cite{Zeng2016TWC}. However,
these algorithms only apply for the equip-probable finite alphabet inputs.
By exploiting the I-MMSE relationship in \cite{Palomar2006TIT},
R. Zhu \textit{et al.} propose a gradient based iterative algorithm to
maximize the secondary users' spectrum efficiency for arbitrary inputs \cite{Zhu2013EURASIP}.

\subsubsection{MSE Design}
B. Seo \textit{et al.} investigate the transceiver design
to minimize the MSE of the primary user subject to the secondary user's interference power constraint \cite{Seo2011SP}.
For MIMO cognitive networks,  X. Gong \textit{et al.} jointly design the precoder and decoder
to minimize the MSE of the secondary network subject to both transmit and interference power constraints \cite{Gong2011WCNC}.
The transceiver designs for ad hoc MIMO cognitive radio networks based on total MSE minimization
are proposed in \cite{Gharavol2011TWC,Zhang2012TSP}. For full-duplex MIMO cognitive radio networks,
A. C. Cirik \textit{et al.} minimize both the total MSE and the maximum per secondary users' MSE \cite{Cirik2015TCOM}.

\subsubsection{Diversity Design}
For cognitive radio networks, by considering multiple cognitive radios as a virtual
antenna array, W. Zhang \textit{et al.} design
space-time codes and space-frequency codes
for cooperative spectrum sensing
over flat-fading and frequency-selective fading
channels, respectively \cite{Zhang2008TWC_Dec}. X. Liang \textit{et al.} design a flexible
STBC scheme for cooperative spectrum sensing
by dynamically selecting users to form the clusters \cite{Liang2012}.
A robust STBC scheme based on dynamical clustering
is further proposed in \cite{Wang2013}. The differential
STBCs for cooperative spectrum sensing without CSI knowledge
are also proposed in \cite{Chen2012JICS,Abdulkadir2015}. J.-B. Kim \textit{et al.}
analyze the outage probability and the diversity order
of various spectrum sensing methods
for cognitive radio networks \cite{Kim2012TCom}. K. B. Letaief \textit{et al.}
design a space-time-frequency block code to exploit spatial diversity
for cognitive relay networks with cooperative spectrum sharing \cite{Letaief2009Pro}.
A distributed space-time-frequency block code is further proposed
to reduce the detection complexity \cite{Vien2014IETC}.

\subsection{Green MIMO Communications}

\subsubsection{Mutual Information Design}
M. Gregori \textit{et al.} study a point-to-point MIMO systems with finite alphabet inputs
where the transmitter is equipped with energy harvesters \cite{Gregori2013TCOM}. A total number of $N$ channels are used and
the non-causal knowledge of the channel state and harvested energy are
available at the transmitter. By obtaining the optimal left singular matrix and setting
the right singular matrix to be identity matrix,  M. Gregori \textit{et al.} propose an
optimal offline power allocation solution to maximize the spectral efficiency along $N$ channel uses \cite{Gregori2013TCOM}.
This design is extended to the SCSIT case in \cite{Zeng2016TVT}, where an iterative algorithm to find the entire
optimal precoder is proposed. The further exploitation of partial instantaneous CSI along with
the SCSI is studied in \cite{Zhu2016IET}.  An resource allocation algorithm for the downlink transmission of multiuser
MIMO systems with finite alphabet inputs and energy harvest process at the transmitter is proposed \cite{Zeng2015TWC}.
With the SCSIT and the statistical energy arrival knowledge, the precoders, the MCR selection, the subchannel allocation,
and the energy consumption are jointly optimized by decomposing the original problem into an
equivalent three-layer optimization problem and solving each layer separately.

\subsubsection{Diversity Design}
Y. Tang \textit{et al.} propose a STBC for geographically separated base stations,
which improves the energy efficiency at remote locations \cite{Tang2001VTC}. Y. Zhu \textit{et al.}
design a low energy consumption turbo-based STBC \cite{Zhu2003SIPS}.  A cooperative balanced STBC
is proposed for dual-hop AF wireless sensor networks to reduce
the energy consumption \cite{Eksim2009LNCS}.  M. Tomio \textit{et al.} compare the energy efficiency of transmit beamforming scheme
and transmit antenna selection scheme \cite{Kakitani2013Cletter}. It is revealed that transmit antenna selection scheme
is more efficient for most transmit distances since only a single radio-frequency chain is used at
the transmitter.
Novel non-coherent energy-efficient collaborative
Alamouti codes are constructed  by uniquely factorizing
a pair of energy-efficient cross QAM constellations and
carefully designing an energy scale \cite{Gong2012WCL,Xia2013TIT}.

\subsection{MIMO Relay Systems}

\subsubsection{Mutual Information Design}
W. Zeng \textit{et al.} investigate the linear precoder
design for dual hop amplify-and-forward (AF) MIMO relay systems
with finite alphabet inputs \cite{Zeng2012TWC}.  By exploiting
the optimal precoder structure similar as the point-to-point MIMO case,
a two-step algorithm is proposed to maximize the spectral efficiency.
The obtained precoder in \cite{Zeng2012TWC} also achieves a good BER performance.
Since the publication of \cite{Zeng2012TWC}, the transmission design for the spectral efficiency maximization
under finite alphabet constraints has been investigated for various
MIMO relays systems including:  two-hop nonregenerative three-node MIMO relay systems \cite{Liang2012ICC},
non-orthogonal AF half-duplex single relay systems \cite{Syed2013}, MIMO two-way relay systems \cite{Feng2014},
and  MIMO DF relay secure systems \cite{Vishwakarma2015}.

\subsubsection{MSE Design}
Linear transceiver designs via MMSE
criterion for MIMO AF relay systems are proposed in \cite{Mo2009TWC_2,Xing2010TSP,Tseng2010TVT,Lee2010TWC_2,Fu2011TSP,Wang2012TSP,Xing2013TSP,Shen2013TSP}.
In addition, non-linear THP/DFE transceiver structures based on the optimization of MSE function
are designed for MIMO AF relay systems in \cite{Tseng2011TVT,Wang2012TVT,Xing2012JSAC,Zhang2014TCOM,Ahn2015TWC}. The average error
probability of AF relay networks employing MMSE based precoding schemes
is derived in \cite{Song2011TCOM}. At the same time, the MMSE based transceiver designs for non-regenerative
MIMO relay networks are provided in \cite{Rong2009TSP,Rong2010TWC,Song2010TWC,Xing2010TSP_2,Rong2012TSP}. The MMSE transceiver designs
are further investigated in multiuser MIMO relay networks \cite{Jang2010Cletter,Wang2012TWC,Nguyen2015TWC,Yang2016TVT,Nguyen2016SPletter}.
Two MMSE transceiver designs for filter-and-forward
MIMO relay systems and single-carrier frequency-domain equalization based MIMO relay systems are given
in \cite{Liang2011TWC} and \cite{Wu2013TWC}, respectively.

\subsubsection{Diversity Design}
For AF relay systems with multiple antennas and finite alphabet inputs, various transmission schemes are provided
to optimize the diversity gain of the system in \cite{Ding2007TSP,Safari2008TWC,Cui2009TSP,Ding2009TWC,Maham2009TWC}. Moreover, the diversity order
and array gains for AF relay MIMO systems  employing OSTBCs are derived analytically in \cite{Dharmawansa2010TCom,Chen2010TVT}.
For DF relay systems with multiple antennas and finite alphabet inputs, various transmission schemes are proposed based on
minimizing the error probability of the system in \cite{Cui2009TSP,Zhou2010TCom,You2012TVT}. Meanwhile, the error performance of
DF relay MIMO systems
are analyzed in \cite{Shi2013WCL,Chen2015TCom}. Further transmission schemes to
increase the reliability of MIMO relay networks with finite alphabet inputs are designed
and analyzed in \cite{Scutari2005TWC,Jing2005TWC,Shang2006TIT,Jing2007TIT,Koyuncu2008JSAC,Guo2008TWC,Jing2009TWC,Yan2015TVT}.
The diversity designs for the non-coherent MIMO relay networks are investigated in \cite{Nguyen2009TVT,Lai2015TCom,Avendi2015TWC}

\subsubsection{Adaptive Transmission}
I. Y. Abualhaol \textit{et al.} propose an adaptive transmission
scheme for MIMO-OFDM systems with a cooperative DF relay \cite{Abualhaol2008}.
The bit and power are adaptively allocated among different subchannels to
maximize the receive signal power under the constraint that each subchannel is
assigned either to the direct link or the relay link.
D. Munoz \textit{et al.} propose an adaptive transmission
scheme for MIMO-OFDMA systems with multiple relay nodes \cite{Munoz2012WCL}. Based
on the channel conditions, the transmission mode
adaptively switches between joint relay and antenna selection and space-frequency block coding to minimize the sum BER of the direct link
 and the relay link subject to a total power constraint.

\subsection{Others}
An optimal precoding technique for coded MIMO
inter-symbol interference  MAC with joint linear MMSE detection
and forward-error-correction codes is proposed in \cite{Yuan2009JSTSP}.
The geometric mean method, which obtains an equal SINR for all transmit streams, is applied
to single user and multiuser MIMO systems \cite{Jiang2005TSP,Liu2008TVT,Lin2008TWC,Liu2010TVT}.
There are some precoder designs  for multiuser MIMO systems based on the receiver side's SINR \cite{Schubert2004TVT,Wiesel2006TSP,Luo2008JSTSP,Chiu2010TWC,Liu2011TSP,Mukherjee2011TSP,Liao2011TSP,Wang2013TWC}.
There are some transmission designs for MIMO X channel based on the diversity performance \cite{Li2011,Shi2013TCom,Li2013TIT,Ganesan2014TIT,Ganesan2014,Ganesan2015}.
The diversity designs for cooperative systems, Ad Hoc networks, and distributed antenna systems
are given in \cite{Li2007SPL}, \cite{Zhang2008TWC}, and \cite{Liu2014TWC}, respectively.

\section{Conclusions}
This paper has provided a first comprehensive summary for the research on MIMO transmission design with practical
input signals.  We  introduced the existing fundamental understanding
for discrete input signals. Then, to facilitate a better clarification
of practical MIMO transmission techniques, we provided a detailed discussion of the transmission
designs based on the criterions of mutual information, MSE, diversity, and the adaptive transmission
switching among these criterions. These transmission schemes were designed for point-to-point MIMO systems, multiuser MIMO
systems, MIMO cognitive radio systems, green MIMO communication systems, MIMO relay systems, etc.
Valuable insights and lessons were extracted from these existing designs.

For the forthcoming massive MIMO systems in 5G wireless networks, the transmission designs with discrete input signals
will become more challenging. A unified framework for the mutual information maximization design
for point-to-point massive MIMO systems with discrete input signals  was introduced in this paper.
However, there are a variety of open research issues currently. The tractable
mutual information maximization design for multiuser massive MIMO systems with discrete input signals
is still unknown. The robust MSE design for large scale antenna arrays with channel uncertainty due to
pilot contamination and other channel variation effects is less studied. The high rate low detection
complexity STBCs for large MIMO systems need to be designed. Most importantly, an integrity and efficient
adaptive transmission scheme for practical massive MIMO systems is still missing yet. This problem is
extremely challenging since it needs to determine the optimal resource allocation (time/space/frequency), the user scheduling,
the transmission mode selection, etc, in a significant large dimension. The characteristics of massive MIMO channels should be exploited
and heuristic methods from other research areas such as machine learning may be helpful.


\end{document}